\documentclass{mn2e}
\usepackage{epsfig}
\usepackage{rotating}
\usepackage{longtable}
\usepackage{graphics}
\usepackage{amssymb}
\usepackage{supertabular}
\usepackage{times}
\title[A deep survey of heavy element lines in PNe - II]
{A deep survey of heavy element lines in planetary nebulae -- II.
Recombination line abundances and evidence for ultra-cold plasma}
\author[Y. G. Tsamis et al.]
{Y. G. Tsamis$^{1,2}$, M. J.\,Barlow$^1$, X.-W. Liu$^{1,3}$, P. J.
Storey$^1$ and I. J. Danziger$^4$ \\
$^1$Department of Physics and Astronomy, University College London,
      Gower Street, London WC1E 6BT, U. K.\\
$^2$Current address: LUTH, Laboratoire l'Univers et ses Th\'eories,
associ\'e au CNRS (FRE 2462) et \'a l'Universit\'e Paris 7, \\
~~~~~Observatoire de Paris-Meudon, F-92195 Meudon C\'edex, France;
yiannis.tsamis@obspm.fr \\
$^3$Current address: Department of Astronomy, Peking University, 
Beijing, China\\
$^4$Osservatorio Astronomico di Trieste, Via G. B. Tiepolo 11, I-34131
Trieste, Italy}

\date{Received:}
\newcommand{\eld}{$N_{\rm e}$}
\newcommand{\crd}{$N_{\rm cr}$}
\newcommand{\elt}{$T_{\rm e}$}
\newcommand{\exe}{$E_{\rm ex}$}
\newcommand{\cmt}{cm$^{-3}$}
\newcommand{\cp}{C$^+$}
\newcommand{\cpp}{C$^{2+}$}
\newcommand{\cppp}{C$^{3+}$}
\newcommand{\cfp}{C$^{4+}$}
\newcommand{\op}{O$^+$}
\newcommand{\opp}{O$^{2+}$}
\newcommand{\oppp}{O$^{3+}$}
\newcommand{\ofp}{O$^{4+}$}
\newcommand{\np}{N$^+$}
\newcommand{\npp}{N$^{2+}$}
\newcommand{\nppp}{N$^{3+}$}
\newcommand{\nfp}{N$^{4+}$}
\newcommand{\nepp}{Ne$^{2+}$}

\newcommand{\Hb}{H$\beta$}
\newcommand{\foiii}{[O~{\sc iii}]}
\newcommand{\fniii}{[N~{\sc iii}]}

\newcommand{\fsii}{[S~{\sc ii}]}

\newcommand{\fariv}{[Ar~{\sc iv}]}
\newcommand{\fcliii}{[Cl~{\sc iii}]}
\newcommand{\fneiii}{[Ne~{\sc iii}]}

\newcommand{\oiii}{O~{\sc iii}}
\newcommand{\oiv}{O~{\sc iv}}

\newcommand{\nii}{N~{\sc ii}}
\newcommand{\niii}{N~{\sc iii}}
\newcommand{\niv}{N~{\sc iv}}
\newcommand{\oii}{O~{\sc ii}}
\newcommand{\cii}{C~{\sc ii}}

\newcommand{\ciii}{C~{\sc iii}}
\newcommand{\civ}{C~{\sc iv}}

\newcommand{\hi}{H\,{\sc i}}
\newcommand{\hii}{H\,{\sc ii}}

\newcommand{\tmf}{10$^{-4}$}
\newcommand{\tmfi}{10$^{-5}$}
\newcommand{\hp}{H$^+$}
\newcommand{\hep}{He$^+$}
\newcommand{\hepp}{He$^{2+}$}

\begin{document}
\maketitle

\begin{abstract}
\noindent Deep optical observations of the spectra of 12
Galactic planetary nebulae (PNe) and 3 Magellanic Cloud PNe were presented
in Paper~I by Tsamis et al. (2003b), who carried out an abundance analysis
using the collisionally excited forbidden lines. Here, the relative
intensities of faint optical recombination lines (ORLs) from ions of
carbon, nitrogen and oxygen are analysed in order to derive the abundances
of these ions relative to hydrogen. The relative intensities of four
high-$l$ C~{\sc ii} recombination lines with respect to the well-known
3d--4f $\lambda$4267 line are found to be in excellent agreement with the
predictions of recombination theory, removing uncertainties about whether
the high C$^{2+}$ abundances derived from the $\lambda$4267 line could be
due to non-recombination enhancements of its intensity.

We define an abundance discrepancy factor (ADF) as the ratio of the
abundance derived for a heavy element ion from its recombination lines to
that derived for the same ion from its ultraviolet, optical or infrared
collisionally excited lines (CELs). All of the PNe in our sample are found
to have ADF's that exceed unity. Two of the PNe, NGC\,2022 and LMC~N66,
have O$^{2+}$ ADF's of 16 and 11, respectively, while the remaining 13 PNe
have a mean O$^{2+}$ ADF of 2.6, with the smallest value being 1.8. 

%The ADF's derived for a range of ions are found to be fairly similar to
%those for O$^{2+}$, so that although ORLs and CELs yield different ionic
%abundances, the abundance ratios derived from ORLs for a pair of heavy
%element ions are similar to the abundance ratios derived for the same two
%ions using CELs. Thus reliable abundance ratios may be obtained for two
%heavy elements provided that both abundances are based upon the same type
%of line (e.g. ORL/ORL or CEL/CEL).

Garnett \& Dinerstein (2001a) found that for a sample of about a dozen PNe the
magnitude of the O$^{2+}$ ADF was inversely correlated with the nebular Balmer
line surface brightness. We have investigated this for a larger sample of 20
PNe, finding weak correlations with decreasing surface brightness for the
ADF's of O$^{2+}$ and C$^{2+}$. The C$^{2+}$ ADF's are well correlated with
the absolute radii of the nebulae, though no correlation is present for the
O$^{2+}$ ADF's. We also find both the C$^{2+}$ and O$^{2+}$ ADF's to be
strongly correlated with the magnitude of the difference between the nebular
[O~{\sc iii}] and Balmer jump electron temperatures ($\Delta T$),
corroborating a result of Liu et al. (2001b) for the O$^{2+}$ ADF. $\Delta T$
is found to be weakly correlated with decreasing nebular surface brightness
and increasing absolute nebular radius.

There is no dependence of the magnitude of the ADF upon the excitation energy
of the UV, optical or IR CEL transition used, indicating that classical
nebular temperature fluctuations---i.e. in a chemically homogeneous
medium---are not the cause of the observed abundance discrepancies. Instead,
we conclude that the main cause of the discrepancy is enhanced ORL emission
from cold ionized gas located in hydrogen-deficient clumps inside the main
body of the nebulae, as first postulated by Liu et al. (2000) for the high-ADF
PN NGC\,6153. We have developed a new electron temperature diagnostic, based
upon the relative intensities of the O~{\sc ii} 4f--3d $\lambda$4089 and
3p--3s $\lambda$4649 recombination transitions. For six out of eight
PNe for which both transitions are detected, we derive O$^{2+}$ ORL electron
temperatures of $\le$300\,K, very much less than the O$^{2+}$
forbidden-line and H$^+$ Balmer jump temperatures derived for the same
nebulae. These results provide direct observational evidence for the presence
of cold plasma regions within the nebulae, consistent with gas cooled largely
by infrared fine structure and recombination transitions; at such low
temperatures recombination transition intensities will be significantly
enhanced due to their inverse power-law temperature dependence, while UV and
optical CELs will be significantly suppressed.


\end{abstract}

\section{Introduction}

This is the second of two papers devoted to the study of elemental
abundances in a sample of Galactic and Magellanic Cloud planetary nebulae
(PNe). In a companion paper, Tsamis et al. (2003a) have presented a
similar analysis of a number of Galactic and Magellanic Cloud {\hii}
regions. The main focus of these papers is on the problem of the optical
recombination-line emission from heavy element ions (e.g. {\cpp}, {\npp},
{\opp}) in photoionized nebulae.  The main manifestation of this problem
is the observed discrepancy between nebular elemental abundances derived
from weak, optical recombination lines (ORLs; such as {\cii}
$\lambda$4267, {\nii} $\lambda$4041, {\oii} $\lambda\lambda$4089, 4650) on
the one hand and the much brighter collisionally-excited lines (CELs;
often collectively referred to as forbidden lines) on the other (Kaler
1981; Peimbert, Storey \& Torres-Peimbert 1993; Liu et al. 1995, 2000,
2001b; Garnett \& Dinerstein 2001a; Tsamis 2002; Tsamis et al. 2003a),
with
ORLs typically being found to yield ionic abundances that are factors of
two or more larger than those obtained from CELs emitted by the same ions.
A closely linked problem involves the observed disparity between the
nebular electron temperatures derived from the traditional {\foiii}
($\lambda$4959+$\lambda$5007)/$\lambda$4363 CEL ratio and the {\hi} Balmer
discontinuity diagnostic: the latter yields temperatures that are in most
cases lower than those derived from the {\foiii} ratio (Peimbert 1971; Liu
\& Danziger 1993b; Liu et al. 2001b; Tsamis 2002).

The ORL analysis of the current paper is based upon deep optical spectra
of twelve galactic and three Magellanic Cloud PNe that were acquired by
Tsamis et al. (2003b; hereafter Paper~I). Paper~I describes how the
observations were obtained and reduced and presents tabulations of
observed and dereddened relative intensities for the detected lines.
Collisionally excited lines (CELs) in the spectra were used to derive
nebular electron temperatures and densities from a variety of diagnostic
ratios. They also derived CEL-based abundances for a range of heavy
elements, using standard ionization correction factor (icf) techniques to
correct for unobserved ion stages.

In the current paper we analyze the ORL data that were presented in Paper~I.
In Section~2 we derive recombination-line ionic abundances for a number of
carbon, nitrogen and oxygen ions. Section~3 presents a comparison between
total C, N and O abundances derived from ORLs and from ultraviolet, optical
and infrared CELs and derives abundance discrepancy factors (ADF's; the ratio
of the abundances derived for the same ion from ORLs and from CELs) for a
range of carbon, nitrogen and oxygen ions. In Section~4 we investigate how
ORL/CEL ADF's correlate with other nebular parameters, such as the difference
between [O~{\sc iii}] forbidden line and H~{\sc i} Balmer jump temperatures;
the H$\beta$ nebular surface brightness; and the nebular absolute radius.
Section~5 looks at whether the observational evidence provides support for the
presence of classical Peimbert-type temperature fluctuations within the
nebulae, and whether the observational evidence points to strong density
variations within the nebulae. In Section~6 we present evidence for the
presence of cold plasma ({\elt} $\leq$ 2000~K) in a number of nebulae in our
sample, making use of the fact that the strengths of several well observed
O~{\sc ii} and He~{\sc i} recombination lines have sufficiently different
temperature dependences for the relative intensities of two O~{\sc ii} lines,
or two He~{\sc i} lines, to be used as diagnostics of the electron
temperatures prevailing in their emitting regions. Section~7 summarizes our
conclusions.

\section{Recombination-line abundances}
\subsection{Carbon ions: C$^{2+}$/H$^+$, C$^{3+}$/H$^+$ and
C$^{4+}$/H$^+$}

We have detected recombination lines of carbon from all of the PNe, with the
exception of LMC N66. {\cii} lines in particular were detected from all
galactic nebulae, as well as from SMC N87 and LMC N141; {\ciii} lines were
detected from the majority of them, excluding NGC\,3132 and My\,Cn\,18 only.
The strongest observed {\cii} optical recombination line is the $\lambda$4267
(V\,6) 3d--4f transition, which was consistently recorded with a high
signal-to-noise (S/N) ratio in our high resolution deep spectra (see Fig.\,~1 in Paper~I).
We derived C$^{2+}$/H$^+$ abundance ratios from it using the recently calculated
effective recombination coefficients of Davey, Storey \& Kisielius (2000),
which include both radiative and dielectronic processes; it should be noted
that abundance ratios derived using $\lambda$4267 are insensitive to the
adoption of Case A or B. The {\cii} $\lambda$4267 line has been used in the
past in abundance analyses of galactic PNe (e.g. Kaler 1986, Rola \&
Stasi{\'n}ska 1994) with conflicting results. It has also been detected
from a
number of LMC and SMC planetary nebulae (Barlow 1987; Meatheringham \& Dopita
1991; and from SMC~N42 by Vassiliadis et al. 1992). No detections of this line
have been reported however for LMC N141 and SMC N87 and carbon ORLs have not
been used to date in abundance studies of Magellanic Cloud PNe.

Due to the structure of the C$^+$ ion, the configuration of the valence
orbital gives rise to only one atomic term, compared to three atomic terms for
N$^+$ and O$^+$ (e.g. Kuhn 1969; Allen 1973). As a result there are fewer
C~{\sc ii} recombination lines than of O~{\sc ii} or N~{\sc ii}, so they are
of greater intensity. As an observational consequence of this fact, in several
PNe of our sample we have also detected C~{\sc ii} recombination lines from
higher principal quantum numbers, originating from states above the
4f\,$^2$F$^{\rm o}$ level. In Table~1 we compare the observed intensities of
these lines of high excitation energy, normalized such that
\emph{I}($\lambda$4267) = 1.00, against the recombination theory predictions
of Davey et al. (2000). This permits us to check whether the 4f\,$^2$F$^{\rm
o}$--\emph{n}g\,$^2$G transitions which populate the upper level of 3d--4f
$\lambda$4267 can be safely attributed to recombination only, or whether
unidentified processes contribute as well. This is of importance in the light
of results from this and previous works (e.g. the PN studies of Kaler 1981,
1986; Rola \& Stasi{\'n}ska 1994 and Liu et al. 1995, 2000 as well as the
{\hii} region analysis of Tsamis et al. 2003a), which found that the
C$^{2+}$/H$^+$ abundances derived from the $\lambda$4267 recombination line
are often significantly higher than those derived from the collisionally
excited C~{\sc iii}] $\lambda$1908 line. This fact had sometimes been
attributed in the past to erroneous recombination coefficients, inaccurate
line detections, or blending of the $\lambda$4267 transition with a line from
an unknown ionic species.

In the case of NGC\,3242, Table~1 shows that the agreement between
observations and theory is excellent for all detected lines. The
4f\,$^2$F$^{\rm o}$--7g\,$^2$G $\lambda$5342 line is blended with a feature
identified as [Kr~{\sc iv}] $\lambda$5345.9 (cf. Hyung, Aller \& Feibelman
1999, their Table 6) and its intensity was retrieved though Gaussian profile
fitting. For NGC\,5315 the agreement is very good for both detected lines of
high excitation. For NGC\,5882 the $\lambda$6462 line is stronger by 49\,per
cent and the $\lambda$4802 line weaker by 19\,per cent than predicted,
relative to $\lambda$4267. For IC\,4191, as measured on the fixed-slit
spectrum, the former line is stronger than predicted by 40\,per cent, while
the $\lambda$5342 line is within 8\,per cent of the predicted value. It should
be noted that apart from $\lambda$4802, the other high-level C~{\sc ii} lines
are covered in our lower resolution 4.5\,\AA~FWHM spectra only and the modest
discrepancies for NGC\,5882 and IC\,4191 regarding the $\lambda$6462 line are
within the estimated error margins. In the above cases the $\lambda$4802 line
was deblended from the N~{\sc ii} $\lambda$4803.3 line via Gaussian line
fitting.

In conclusion, the good consistency among the observed and predicted relative
intensities of C~{\sc ii} 4f\,$^2$F$^{\rm o}$--\emph{n}g\,$^2$G transitions in
this PN sample suggests that there is no other mechanism competing with
recombination that could contribute to the excitation of the 3d--4f
$\lambda$4267 line. Thus the high S/N ratio detections of the other {\cii}
lines from the current PN sample indicate that C~{\sc ii} $\lambda$4267 is a
reliable {\cpp} abundance diagnostic. A similar conclusion can be drawn from
the C~{\sc ii} recombination lines found in the spectrum of the Orion Nebula
(M\,42). In Table~1 we extend the above comparison to 4f--\emph{n}g C~{\sc ii}
transitions from M\,42, whose intensities were presented recently by Baldwin
et al. (2000). These authors did not mark the listed lines as C~{\sc ii}
transitions, but left them unidentified (cf. their Table~1). However, the
measured wavelengths and our examination of their relative intensities leave
no doubt as to their identity. The agreement with theory is excellent in this
case as well, confirming beyond reasonable doubt the interpretation of
recombination excitation for these lines.

\setcounter{table}{0}
\begin{table*}
%\begin{center}
\centering
\begin{minipage}{150mm}
\caption{High-excitation C~{\sc ii} recombination line relative intensities.}
\begin{tabular}{lccccccc}
\noalign{\vskip3pt} \noalign{\hrule}\noalign{\vskip3pt}
$\lambda_{\rm 0}$(\AA)  &Trans.     &NGC\,3242      &NGC\,5315      &NGC\,5882      &IC\,4191              &Orion (M\,42)   &Theory\\
                        &           &$I_{\rm obs}$  &$I_{\rm obs}$  &$I_{\rm obs}$  &$I_{\rm obs}$         &$I_{\rm obs}$  &$I_{\rm pred}$ \\
\noalign{\vskip3pt} \noalign{\hrule} \noalign{\vskip3pt}
4267.15                 &3d--4f     &1.000          &1.000          &1.000          &1.000                 &1.000     &1.000 \\
%6258.78                 &4p--5d     &*              &*              &*              &*                     &*         &0.012 \\
6151.43                 &4d--6f     &0.030          &*              &*              &~~~~~*$\mid$0.055$^a$ &0.038     &0.040 \\
6461.95                 &4f--6g     &0.102          &0.095          &0.153          &0.166$\mid$0.144      &0.087     &0.103 \\
5342.38                 &4f--7g     &0.065          &*              &*              &~~~~~*$\mid$0.056     &0.049     &0.053 \\
4802.23                 &4f--8g     &0.031          &0.020          &0.025          &*                     &0.034     &0.031 \\
\noalign{\vskip3pt} \noalign{\hrule} \noalign{\vskip3pt}
\end{tabular}
\begin{description}
\item[$^a$] The value before the dash is for the entire nebula; those after are from a fixed-slit observation.
\end{description}
\end{minipage}
%\end{center}
\end{table*}
\normalsize

The C$^{3+}$/H$^+$ abundance ratios were derived from the $\lambda$4187
(V\,18) line and from the $\lambda$4650 (V\,1) multiplet, using the
effective radiative and dielectronic recombination coefficients of
P\'{e}quignot, Petitjean \& Boisson (1991) and Nussbaumer \& Storey
(1984), respectively; they are insensitive to the assumption of Case A or
B. The {\ciii} (V\,16) triplet at $\lambda$4069 was also detected from
NGC\,2022, NGC\,3242, NGC\,5882 and NGC\,6818. This multiplet is primarily
excited by radiative recombination and is usually seriously blended with
{\fsii} and {\oii} V\,10 lines in medium resolution spectra; its overall
intensity was retrieved via multiple Gaussian fitting. As was noted by Liu
(1998), the observed relative intensities of {\ciii} V\,16, V\,18 and V\,1
multiplets are not in accord with theoretical predictions; in an analysis
of the spectrum of NGC\,4361 he found that the {\cppp}/{\hp} abundance
ratio derived from these three {\ciii} multiplets spans a range of
0.4\,dex, which he attributed to probable uncertainties in the effective
recombination coefficients. The four PNe for which all V\,16, V\,18 and
V\,1 multiplets have been detected offer a further testing ground for the
reliability of current ORL {\ciii} recombination coefficients. An
instructive case is that of the high excitation nebula NGC\,2022. As is
apparent from our high resolution 1.5\,\AA~FWHM spectra of this object,
the auroral {\fsii} $\lambda\lambda$4068, 4075 lines are virtually absent
and the observed intensity of {\ciii} V\,16 is most accurately recorded,
since it is less affected by blending effects. In this nebula, the
observed {\ciii} $\lambda\lambda$4069, 4187 and $\lambda4650$ multiplets
have intensity ratios of 1.80 : 0.34 : 1.00, versus the theoretical ratios
of 0.59 : 0.21 : 1.00 calculated using the effective radiative and
dielectronic recombination coefficients of P\'{e}quignot et al. (1991) and
Nussbaumer \& Storey (1984), respectively. In the case
of NGC\,5882, the observed multiplet intensities show ratios of 2.38 :
0.22 : 1.00. Similarly for IC\,4191, where {\ciii} $\lambda$4069 is not
detected, the observed intensities of $\lambda$4187 (V\,18) and
$\lambda$4650 (V\,1) have a ratio of 0.20 : 1.00. We see that the
theoretical predictions for the relative intensities of $\lambda$4187 and
$\lambda$4650 appear to be more secure than that for $\lambda$4069. It
would seem that the effective recombination coefficient for the
$\lambda$4069 V\,16 multiplet has been underestimated by a factor of
$\sim$2--4, or else that it is blended with an unknown line from an ion of
similar excitation.

As a result of the above discussion, in our ORL abundance analysis for {\cppp}
we have discarded values derived from the $\lambda$4069 triplet; instead, we
have adopted {\cppp}/{\hp} abundance ratios as derived from an
intensity-weighted mean of the {\ciii} $\lambda$4187 and $\lambda$4650 lines.

Regarding {\cfp}, we have derived C$^{4+}$/H$^+$ fractions for a number of
galactic PNe from the {\civ} $\lambda$4658 line, using the Case A
effective recombination coefficients of P\'{e}quignot et al. (1991). The
ionic and total carbon abundances derived from
recombination lines are presented in Tables~2 and 3 for the Galactic and
Magellanic Cloud PNe, respectively (see Appendix A for a discussion of the
adopted \emph{icf} scheme in the derivation of total C abundances).

\setcounter{table}{1}
\begin{table*}
\begin{center}
\caption{Recombination line carbon abundances for Galactic PNe.}
\begin{tabular}{lcccccc}
\noalign{\vskip3pt}\noalign{\hrule}\noalign{\vskip3pt}
                                            &\bf{NGC\,3242}&\bf{NGC\,5882}  &\bf{NGC\,5315}  &\bf{NGC\,3918}  &\bf{NGC\,2022}  &\bf{NGC\,6818}    \\
\noalign{\vskip3pt} \noalign{\hrule} \noalign{\vskip3pt}
$I(\lambda4267)$                            &0.620      &0.399          &0.706          &0.497      &0.820      &0.449       \\
10$^4$\,$\times$\,{\cpp}/{\hp}              &6.15       &3.77           &6.60           &5.02       &8.83       &4.69       \\
                                            &           &               &               &           &           &           \\
$I(\lambda4069)$                            &0.755      &0.412          &*              &*          &1.75       &0.957      \\
10$^4$\,$\times$\,{\cppp}/{\hp}             &3.64       &2.11           &*              &*          &7.49       &3.97       \\
$I(\lambda4187)$                            &0.204      &0.0376         &0.0107         &0.152      &0.332      &0.118       \\
10$^4$\,$\times$\,{\cppp}/{\hp}             &2.83       &0.552          &0.161          &2.03       &4.07       &1.41        \\
$I(\lambda4650)$                            &0.590      &0.173          &*              &0.418      &0.974      &0.389        \\
10$^4$\,$\times$\,{\cppp}/{\hp}             &1.75       &0.522          &*              &1.22       &2.77       &1.10            \\
\emph{Adopted}                              &           &               &               &           &           &             \\
10$^4$\,$\times$\,{\cppp}/{\hp}             &2.03       &0.527          &0.161          &1.44       &3.10       &1.37          \\
                                            &           &               &               &           &           &           \\
$I(\lambda4658)$                            &0.0776     &*              &*              &0.109      &1.03       &0.180         \\
10$^4$\,$\times$\,{\cfp}/{\hp}              &0.188      &*              &*              &0.262      &2.62       &0.447         \\
icf(C)                                      &1.010      &1.030          &1.078          &1.116      &1.020      &1.185      \\
%icf(C)                                      &1.01       &1.03           &*              &*          &1.02       &*             \\
\bf{10$^4$\,$\times$\,C/H}           &\bf{8.45}       &\bf{4.43}      &\bf{7.29}    &\bf{7.50}    &\bf{14.84}  &\bf{7.71}         \\
\noalign{\vskip3pt}
                                           &\bf{NGC\,3132}  &\bf{NGC\,2440}&\bf{NGC\,6302} &\bf{IC\,4406}  &\bf{IC\,4191}&\bf{My\,Cn\,18}    \\
$I(\lambda4267)$                           &0.697              &0.403      &0.163          &0.805          &0.546      &0.252         \\
10$^4$\,$\times$\,{\cpp}/{\hp}             &6.60               &4.47       &2.00           &7.72           &5.16       &3.80          \\
                                            &           &               &               &           &           &           \\
$I(\lambda4187)$                           &*                  &0.222      &*              &0.0912         &0.0711     &*             \\
10$^4$\,$\times$\,{\cppp}/{\hp}            &*                  &2.58       &*              &1.32           &1.04       &*             \\
$I(\lambda4650)$                           &*                  &0.337      &0.0754$^a$     &0.238          &0.359      &*             \\
10$^4$\,$\times$\,{\cppp}/{\hp}            &*                  &1.06       &*              &0.715          &1.08       &*             \\
\emph{Adopted}                             &                   &           &               &               &           &              \\
10$^4$\,$\times$\,{\cppp}/{\hp}            &*                  &1.66       &*              &0.883          &1.07       &*             \\
                                           &           &               &               &           &           &           \\
$I(\lambda4658)$                           &*                  &0.722      &*              &*              &*          &*             \\
10$^4$\,$\times$\,{\cfp}/{\hp}             &*                  &1.82       &*              &*              &*          &*             \\
icf(C)                                     &1.930              &1.284      &2.725          &1.310          &1.040      &1.589  \\
%icf(C)                                     &1.93               &1.28       &2.72           &1.31           &1.04       &1.59          \\
\bf{10$^4$\,$\times$\,C/H}           &\bf{12.74}              &\bf{10.21} &\bf{5.45}     &\bf{11.27}  &\bf{6.48}       &\bf{6.04}          \\
\noalign{\vskip3pt} \noalign{\hrule} \noalign{\vskip3pt}

\end{tabular}
\begin{description}
\item[$^a$] From {\ciii} $\lambda$4647.
\end{description}
\end{center}
\end{table*}

\setcounter{table}{2}
\begin{table}
\begin{center}
\caption{Carbon and oxygen recombination-line abundances for
Magellanic Cloud PNe.}
\begin{tabular}{lccc}
\noalign{\vskip3pt}\noalign{\hrule}\noalign{\vskip3pt}
                                      &\bf{SMC N87}   &\bf{LMC N66}  &\bf{LMC N141}       \\
\noalign{\vskip3pt} \noalign{\hrule} \noalign{\vskip3pt}
$I(\lambda4267)$                      &0.672            &*              &0.683          \\
10$^4$\,$\times$\,{\cpp}/{\hp}        &6.73             &*              &6.79           \\
$I(\lambda4187)$                      &0.108            &*              &0.0522        \\
10$^4$\,$\times$\,{\cppp}/{\hp}       &1.46             &*              &0.733         \\
$I(\lambda4650)$                      &0.141            &*                &0.181       \\
10$^4$\,$\times$\,{\cppp}/{\hp}       &0.413            &*                &0.538       \\
\emph{Adopted}                        &                 &                 &            \\
10$^5$\,$\times$\,{\cppp}/{\hp}       &6.75             &*                &5.82        \\
$I(\lambda4658)$                      &*                &*                &*          \\
10$^4$\,$\times$\,{\cfp}/{\hp}        &*                &*                &*          \\
icf(C)                               &1.020            &*                &1.023      \\
\bf{10$^4$\,$\times$\,C/H}            &\bf{7.55}             &\bf{*}              &\bf{7.54}    \\
\noalign{\vskip2pt}
$I(\lambda4069)$$\mid${\opp}          &*              &*              &0.188$\mid$7.27    \\
$I(\lambda4072)$$\mid${\opp}          &*              &*              &0.124$\mid$5.15   \\
$I(\lambda4075)$$\mid${\opp}          &*              &*              &0.254$\mid$7.30   \\
$I(\lambda4638)  $$\mid${\opp}        &.0832$\mid$7.73&*              &.0255$\mid$2.38      \\
$I(\lambda4641)  $$\mid${\opp}        &.0831$\mid$3.06&*              &0.120$\mid$4.45      \\
$I(\lambda4649)  $$\mid${\opp}        &.0781$\mid$1.51&0.227$\mid$3.98&0.127$\mid$2.47      \\
$I(\lambda4650)  $$\mid${\opp}        &.0781$\mid$7.26&0.227$\mid$19.1&.0836$\mid$7.81      \\
$I(\lambda4661)  $$\mid${\opp}        &.0243$\mid$1.76&*              &.0518$\mid$3.79      \\
$I(\lambda4676)  $$\mid${\opp}        &.0248$\mid$2.15&*              &.0465$\mid$4.04      \\
\emph{Adopted}                        &                 &                 &            \\
10$^4$\,$\times$\,{\opp}/{\hp}         &2.95          &8.53           &4.96     \\
CEL {\op}/{\opp}                       &0.018         &0.076          &0.023      \\
icf(O)                               &1.000         &3.770          &1.000      \\
\bf{10$^4$\,$\times$\,O/H}             &\bf{3.00}   &\bf{34.6}      &\bf{5.08}       \\
\noalign{\vskip3pt} \noalign{\hrule} \noalign{\vskip3pt}
\end{tabular}
\end{center}
\end{table}

\subsection{Nitrogen ions: {\npp}/{\hp}, {\nppp}/{\hp} and {\nfp}/{\hp}}

We have detected recombination lines from all galactic PNe and from up to three
ionization stages of nitrogen. Nitrogen ORLs were not detected from any of the
three Cloud nebulae. Both 3s--3p as well as 3d--4f transitions were recorded.
The derived recombination-line {\npp}, {\nppp}, {\nfp} and total N abundances
are presented in Table 4 (see also Appendix A).

The {\npp}/{\hp} fractions were derived using effective recombination
coefficients from Escalante \& Victor (1990), assuming Case A for singlets and
Case B for triplets. The adopted {\npp}/{\hp} abundances are derived for each
nebula by averaging the values obtained from each {\nii} line, weighted
according to the predicted relative intensity of the transition. The
$\lambda$4379 (V\,18) {\niii} recombination line was detected from all
galactic PNe, except NGC\,5315 and My\,Cn\,18. We derived {\nppp}/{\hp}
abundances from it using the effective radiative and dielectronic
recombination coefficients of P\'{e}quignot et al. (1991) and Nussbaumer \&
Storey (1984) respectively. The {\niv} 5g--6h $\lambda$4606 line has also been
detected from the high excitation nebulae NGC\,2022, NGC\,2440, NGC\,3918,
NGC\,6302 and NGC\,6818; it was used for the derivation of {\nfp}/{\hp}
abundances employing effective recombination coefficients as in the case of
{\niii} $\lambda$4379.

\setcounter{table}{3}
\begin{table*}
\begin{center}
\caption{Recombination-line {\npp}/{\hp} abundances for Galactic PNe.}
\begin{tabular}{lclclclclc}
\noalign{\hrule}\noalign{\vskip3pt}
$\lambda_{0}$ &Mult.&$I_{\rm obs}$ &$\displaystyle\frac{\rm{~N}^{2+}}{\rm{H}^+}$ &$I_{\rm obs}$ &$\displaystyle\frac{\rm{~N}^{2+}}{\rm{H}^+}$ &$I_{\rm obs}$ &$\displaystyle\frac{\rm{~N}^{2+}}{\rm{H}^+}$ &$I_{\rm obs}$ &$\displaystyle\frac{\rm{~N}^{2+}}{\rm{H}^+}$ \\
\noalign{\vskip2pt}
(\AA)         &     &          &(\tmf)                                       &          &(\tmf)                                       &          &(\tmf)                                       &          &(\tmf)   \\
\noalign{\vskip3pt} \noalign{\hrule} \noalign{\vskip3pt}
& &\multicolumn{2}{c}{\bf{NGC\,3242}}           &\multicolumn{2}{c}{\bf{NGC\,5882}}        &\multicolumn{2}{c}{\bf{NGC\,5315}} &\multicolumn{2}{c}{\bf{NGC\,3918}}\\
\noalign{\vskip2pt}
\multicolumn{10}{l}{\bf{V\,3 3s\,$^{3}$P$^\mathrm{{o}}$--3p\,$^{3}$D}} \\
5666.63    &V3      &.0321         &2.34     &*                &*       &*      &*                      &*    &*     \\
5676.02    &V3      &.0134         &2.20     &.0189            &3.10    &*      &*                      &*    &*     \\
5679.56    &V3      &.0428         &1.68     &.0604            &2.36    &*      &*                      &*    &*    \\
\multicolumn{10}{l}{\bf{V\,5 3s\,$^{3}$P$^\mathrm{{o}}$--3p\,$^{3}$P}} \\
4601.48    &V5      &.0050         &1.19     &*                &*      &*        &*                         &*&*  \\
4607.16    &V5      &*             &*        &*                &*      &*        &*                         &*&*   \\
4621.39    &V5      &*             &*        &*                &*      &*        &*                         &*&*  \\
4630.54    &V5      &.0179         &1.43     &*                &*      &*        &*                         &*&*  \\
\multicolumn{10}{l}{\bf{3d--4f}} \\
4041.31    &V39b    &.0128         &0.88     &.0273            &1.79   &.0573    &3.71                      &.0134     &0.94   \\
4237.05    &V48b    &.0173         &1.71     &.0139            &1.38   &.0166    &1.64                      &.0163     &1.65   \\
4241.78    &V48a    &*             &*        &*                &*      &.0422    &4.16                      &.0163     &1.75   \\
4530.41    &V58b    &*             &*        &.0180            &1.75   &*        &*                         &.0148     &1.55   \\
4678.14    &V61b    &*             &*        &*                &*      &.0312    &4.80                      &*        &*      \\
\multicolumn{1}{l}{\bf{Sum}} & &\bf{0.141}   &\bf{1.71}     &\bf{0.138}   &\bf{2.24}   &\bf{0.147}         &\bf{3.43}  &\bf{.0607}   &\bf{1.42}  \\
\noalign{\vskip3pt}
      & &\multicolumn{2}{c}{\bf{NGC\,6302}}  &\multicolumn{2}{c}{\bf{IC\,4191}}  &\multicolumn{2}{c}{\bf{NGC\,6818}}        &\multicolumn{2}{c}{\bf{NGC\,2022}} \\
\noalign{\vskip2pt}
\multicolumn{10}{l}{\bf{V\,3 3s\,$^{3}$P$^\mathrm{{o}}$--3p\,$^{3}$D}} \\
5666.63    &V3      &*             &*        &.0921            &6.90    &*      &*         &*      &*      \\
5676.02    &V3      &*             &*        &.0400            &6.58    &.0322  &5.29      &.0288  &4.73   \\
5679.56    &V3      &*             &*        &.1282            &5.02    &.1029  &4.03      &.0921  &3.61   \\
\multicolumn{10}{l}{\bf{V\,5 3s\,$^{3}$P$^\mathrm{o}$--3p\,$^{3}$P}} \\
4630.54    &V5      &.0743         &5.55     &.0549            &4.50    &*      &*         &*    &*   \\
\multicolumn{10}{l}{\bf{V\,28 3s\,$^{3}$P$^\mathrm{{o}}$--3p\,$^{3}$D$^\mathrm{o}$}} \\
5927.81    &V28     &.0205         &8.26     &*                &*          &* &*               &*    &*   \\
5931.78    &V28     &.0361         &6.47     &.0460            &5.73       &* &*               &*    &*   \\
\multicolumn{10}{l}{\bf{3d--4f}} \\
4041.31    &V39b    &*             &*        &.0639            &4.26    &*      &*         &.0580  &4.27  \\
\multicolumn{1}{l}{\bf{Sum}}  & &\bf{0.131}  &\bf{5.64}     &\bf{0.425}     &\bf{5.50}     &\bf{0.135}  &\bf{4.27} &\bf{0.179}     &\bf{4.38} \\
\noalign{\vskip3pt}
           &     &\multicolumn{2}{c}{\bf{NGC\,3132}}  &\multicolumn{2}{c}{\bf{NGC\,2440}}     &\multicolumn{2}{c}{\bf{IC\,4406}}  &\multicolumn{2}{c}{\bf{My\,Cn\,18}}  \\
\noalign{\vskip2pt}
\multicolumn{10}{l}{\bf{V\,3 3s\,$^{3}$P$^\mathrm{{o}}$--3p\,$^{3}$D}} \\
5676.02    &V3      &.0249         &4.10     &*                &*       &*      &*         &*      &*      \\
5679.56    &V3      &.0799         &3.13     &*                &*       &*      &*         &*      &*      \\
\multicolumn{10}{l}{\bf{V\,5 3s\,$^{3}$P$^\mathrm{o}$--3p\,$^{3}$P}}\\
4607.16    &V5      &*             &*        &*                &*           &*       &*    &.0842  &26.7     \\
4621.39    &V5      &*             &*        &*                &*           &*       &*    &.0731     &23.3 \\
4630.54    &V5      &.0411         &3.40     &*                &*           & .2318  &1.90    &.2099  &17.9   \\
\multicolumn{10}{l}{\bf{3d--4f}}  \\
4041.31    &V39b    &*             &*        &.0618            &4.28                &* &*        &*&* \\
4237.05    &V48b    &*             &*        &.0495            &4.90                &* &*        &*&* \\
4241.78    &V48a    &*             &*        &.0558            &5.89                &* &*        &*&* \\
4530.41    &V58b    &*             &*        &.0415            &4.28                &* &*        &*&* \\
\multicolumn{1}{l}{\bf{Sum}} & &\bf{0.146}   &\bf{3.40}    &\bf{0.209}            &\bf{4.77}        &\bf{0.232}  &\bf{1.90}   &\bf{0.283}&\bf{20.4}\\
\noalign{\vskip3pt} \noalign{\hrule}
\end{tabular}
\end{center}
\end{table*}

\setcounter{table}{4}
\begin{table*}
\begin{center}
\caption{Overall recombination-line nitrogen abundances for Galactic PNe.}
\begin{tabular}{lcccccc}
\noalign{\vskip3pt} \noalign{\hrule}\noalign{\vskip3pt}

                                            &\bf{NGC\,3242}&\bf{NGC\,5882}&\bf{NGC\,5315}&\bf{NGC\,3918}&\bf{NGC\,2022}&\bf{NGC\,6818}    \\
\noalign{\vskip3pt} \noalign{\hrule} \noalign{\vskip3pt}
10$^4$\,$\times$\,{\npp}/{\hp}              &1.71       &2.24           &3.43           &1.42       &4.38       &4.27        \\
\noalign{\vskip2pt}
$I(\lambda4379)$                            &0.158      &0.136          &*              &0.146      &0.472      &0.203       \\
10$^4$\,$\times$\,{\nppp}/{\hp}             &0.685      &0.575          &*              &0.641      &2.080      &0.887       \\
\noalign{\vskip2pt}
$I(\lambda4606)$                            &*          &*              &*              & .0660     &0.112      & .0460      \\
10$^4$\,$\times$\,{\nfp}/{\hp}              &*          &*              &*              &0.207      &0.368      &0.146       \\
\noalign{\vskip2pt}
CEL {\np}/{\npp}                            &0.008      &0.028          &0.163          &0.203      &0.013      &0.241       \\
\noalign{\vskip2pt}
\bf{10$^4$\,$\times$\,N/H}              &\bf{2.42}     &\bf{2.88}    &\bf{5.93}     &\bf{2.56}     &\bf{6.88} &\bf{6.33}     \\

\noalign{\vskip3pt} \noalign{\hrule} \noalign{\vskip3pt}
                                            &\bf{NGC\,3132}  &\bf{NGC\,2440}&\bf{NGC\,6302}      &\bf{IC\,4406} &\bf{IC\,4191} &\bf{My\,Cn\,18}     \\
\noalign{\vskip3pt} \noalign{\hrule} \noalign{\vskip3pt}
10$^4$\,$\times$\,{\npp}/{\hp}              &3.40       &4.77       &5.64           &1.90           &5.00       &20.35       \\
\noalign{\vskip2pt}
$I(\lambda4379)$                            &.0571      &0.324      &0.182          &0.172          &0.270      &*           \\
10$^4$\,$\times$\,{\nppp}/{\hp}             &0.247      &1.430      &0.807          &0.731          &1.150      &*           \\
\noalign{\vskip2pt}
$I(\lambda4606)$                            &*          &0.475      &0.668          &*              &*          &*           \\
10$^4$\,$\times$\,{\nfp}/{\hp}              &*          &1.70       &2.28           &*              &*          &*           \\
\noalign{\vskip2pt}
CEL {\np}/{\npp}                            &1.296      &0.475      &0.435          &0.536          &0.393      &*           \\
\noalign{\vskip2pt}
\bf{10$^4$\,$\times$\,N/H}                  &\bf{8.05}  &\bf{10.17} &\bf{11.18}     &\bf{3.65}      &\bf{8.12}  &\bf{$\geq$20.35}     \\
\noalign{\vskip3pt} \noalign{\hrule}
\end{tabular}
\end{center}
\end{table*}

\subsection{Oxygen ions: O$^{2+}$/H$^+$, O$^{3+}$/H$^+$ and
O$^{4+}$/H$^+$}

Our observations provide us with one the most extensive records of O~{\sc ii}
recombination spectra in gaseous nebulae thus far, obtained for PNe possessing
a wide range of physical conditions. For the first time also to our knowledge,
O~{\sc ii} lines have been detected and measured from Magellanic Cloud
planetary nebulae (SMC N87, LMC N66, and LMC N141). This allowed us to obtain
accurate ORL O$^{2+}$/H$^+$ abundances for all our nebulae and permitted a
comprehensive investigation on the occurrence of the discrepancy between
recombination-line {\opp} abundances and those derived in the usual manner
from forbidden lines of O$^{2+}$. Furthermore, having such an extensive
inventory of O~{\sc ii} line intensities, we are able to perform a thorough
comparison with current theoretical predictions of the O~{\sc ii}
recombination spectrum (see Appendix B).

In Table~5 we present ORL O$^{2+}$/H$^+$ ionic ratios for the complete sample
(for the Magellanic Cloud objects, these ratios and total O abundances are
listed in Table 3). Effective recombination coefficients are taken from Storey
(1994) for 3s--3p transitions (under \emph{LS}-coupling) and from Liu et al.
(1995, hereafter LSBC) for 3p--3d and 3d--4f transitions (under intermediate
coupling), assuming Case A for doublet and Case B for quartet lines. For
several nebulae, we have also detected lines arising from doubly excited
spectral terms, such as multiplet V\,15
3s$^{'}$\,$^{2}$D--3p$^{'}$\,$^{2}$F$^{\rm o}$ at 4590\,\AA~(cf. Paper~I). The
excitation of this multiplet is dominated by dielectronic recombination, but
we have not derived abundance ratios from it since the existing recombination
coefficients are not of the desired accuracy.\footnote{Garnett \& Dinerstein
(2001a) noted that for a sample of about 12 PNe, O$^{2+}$ abundances derived
from multiplet V~15 of O~{\sc ii} were anomalously high compared to those
derived from other O~{\sc ii} multiplets, owing either to an underestimated
dielectronic recombination coefficient for this multiplet, or to additional
contributions, e.g. high-temperature dielectronic recombination originating
from very hot regions, which might also contribute to the strengths of other
O~{\sc ii} lines. Liu et al. (2001b) looked at this issue in detail and
concluded that the discrepancy was most likely due to the lower accuracy of
the recombination coefficients available for the V~15 multiplet.}~For each PN,
the mean {\opp}/{\hp} fractions derived by averaging the values from all 3--3
multiplets and the co-added 3d--4f transitions have been adopted as the
recombination line values in the subsequent discussion. The {\opp}/{\hp}
values listed for the 3d--4f lines in Table~6 were obtained by summing all the
intensities and dividing by the sum of all the recombination coefficients,
allowing for weak unobserved or blended multiplet components, as listed in
Table~4(a) of LSBC.

In Table~7 we
present a summary of the oxygen recombination-line ionic and total
abundances, including the {\oppp}/{\hp} fractions for several PNe derived
by Liu \& Danziger (1993a) from the {\oiii} V\,8 multiplet at 3265\,\AA;
we also list {\ofp}/{\hp} fractions derived from the {\oiv} $\lambda$4632
line which was detected from the high excitation nebulae NGC\,3918,
NGC\,2022, NGC\,6818, NGC\,2440 and NGC\,6302. For the latter calculation
we used the Case A dielectronic recombination coefficients of Nussbaumer
\& Storey (1984) and the radiative recombination coefficients of
P\'{e}quignot et al. (1991). For a discussion on the derivation of total O
abundances the reader is referred to Appendix~A.

\setcounter{table}{5}
\begin{table}
\begin{minipage}{75mm}
\centering \caption{Recombination-line {\opp}/{\hp} abundances for galactic PNe.}
\begin{tabular}{lc@{\hspace{2.8mm}}c@{\hspace{2.8mm}}c@{\hspace{2.8mm}}c@{\hspace{2.8mm}}c@{\hspace{2.8mm}}}
\noalign{\vskip3pt} \noalign{\hrule} \noalign{\vskip3pt} $\lambda_{0}$ &Mult.
&$I_{\rm obs}$ &$\displaystyle\frac{\rm{~O}^{2+}}{\rm{H}^+}$ &$I_{\rm obs}$
&$\displaystyle\frac{\rm{~O}^{2+}}{\rm{H}^+}$ \\
\noalign{\vskip3pt}
(\AA)         &     &          &(\tmf)     &          &(\tmf) \\

\noalign{\vskip3pt} \noalign{\hrule} \noalign{\vskip3pt}
\multicolumn{2}{c}{}&\multicolumn{2}{c}{\bf{NGC\,3242}}&\multicolumn{2}{c}{\bf{NGC\,5882}}\\

4638.86   & V1    &.1692   &15.9:  &.1690   &16.2: \\
4641.81   & V1    &.2152   &8.00   &.2522   &9.59 \\
4649.13   & V1    &.2231   &4.36   &.3447   &6.89 \\
4650.84   & V1    &.0979   &9.18   &.1141   &10.9 \\
4661.63   & V1    &.0870   &6.38   &.1188   &8.92 \\
4673.73   & V1    &.0304   &13.1:  &.0188   &9.10 \\
4676.24   & V1    &.0574   &5.01   &.0906   &8.10 \\

\bf{V\,1 3s\,$^{4}$P--3p\,$^{4}$D$^\mathrm{{o}}$}  & &\bf{0.681} &\bf{6.02} &\bf{0.939} &\bf{8.29} \\

4317.14   &V2     &.0454  &5.39   &.0638   &8.43   \\
4319.63   &V2     &.0227  &2.74   &.0333   &4.08   \\
4325.76   &V2     &.0311  &20.3:  &*        &*      \\
4345.56   &V2     &.0881  &11.1:  &.1100   &14.6:   \\
4349.43   &V2     &.0712  &3.71   &.0911   &4.82      \\
4366.89   &V2     &.0909  &10.6:  &*        &*        \\

\bf{V\,2 3s\,$^{4}$P--3p\,$^{4}$P$^\mathrm{{o}}$}  & &\bf{0.139} &\bf{3.88}  &\bf{0.188} &\bf{5.43} \\

4414.90   &V5     &.0349  &6.20   &.0439   &8.51  \\
4416.97   &V5     &.0331  &10.6   &.0399   &13.9  \\
4452.37   &V5     &.0179  &28.8:  &*        &*  \\

\bf{V\,5 3s\,$^{2}$P--3p\,$^{2}$D$^\mathrm{{o}}$}  & &\bf{0.068} &\bf{7.77} &\bf{0.084} &\bf{10.4}   \\

4069.89   &V10    &.1828  &7.08   &.1361   &5.27 \\
4072.16   &V10    &.2153  &8.95   &.2385   &9.91 \\
4075.86   &V10    &.1664  &4.79   &.2754   &7.92 \\
4078.84   &V10    &.0213  &5.81   &*       &*     \\
4085.11   &V10    &.0351  &7.81   &.0290   &6.45 \\
4092.93   &V10    &.0240  &7.32   &.0212   &6.45 \\

\bf{V\,10 3p\,$^{4}$D$^\mathrm{{o}}$}--3d\,$^{4}$F  & &\bf{0.645} &\bf{6.71}  &\bf{0.700} &\bf{7.58}  \\

4121.46   &V19    &.0283  &9.81   &.0289   &10.31 \\
4129.32   &V19    &.0517  &79.2:  &*       &*     \\
4132.80   &V19    &.0210  &3.87   &.0535   &9.77  \\
4153.30   &V19    &.0304  &3.91   &.0955   &12.2  \\
4156.53   &V19    &.0373  &30.1:  &.0728   &58.3: \\
4169.22   &V19    &.0126  &4.76   &.0279   &10.5  \\

\bf{V\,19 3p\,$^{4}$P$^\mathrm{{o}}$}--3d\,$^{4}$P  & &\bf{0.091} &\bf{4.90}  &\bf{0.206} &\bf{10.9}   \\

4110.78   &V20     &*       &*      &.0359   &14.9      \\
4119.22   &V20     &*       &*      &.0919   &10.3     \\

\bf{V\,20 3p\,$^{4}$P$^\mathrm{{o}}$}--3d\,$^{4}$D  & &\bf{*} &\bf{*} &\bf{0.128} &\bf{11.3}   \\

4890.86   &V28     &*&*             &.0086   &7.39      \\
4906.83   &V28     &*&*             &.0362   &14.5      \\
4924.53   &V28     &*&*             &.0656   &15.4      \\

\bf{V\,28 3p\,$^{4}$P$^\mathrm{{o}}$}--3d\,$^{4}$D  & &\bf{*} &\bf{*} &\bf{0.110} &\bf{13.9}   \\

4083.90   &V48b   &.0211  &6.88   &.0210   &6.58     \\
4087.15   &V48c   &.0204  &7.04   &.0359   &11.8     \\
4089.29   &V48a   &.1113  &10.4   &.1452   &12.9     \\
4275.55   &V67a   &.0310  &4.47   &.0579   &7.99     \\
4276.75   &V67b   &.0276  &8.36   &.0236   &7.65     \\
4277.43   &V67c   &.0172  &6.90   &.0265   &10.2      \\
4282.96   &V67c   &.0086  &5.24   &*  &*    \\
4283.73   &V67c   &.0095  &9.26   &* &*     \\
4285.69   &V78b   &.0120  &5.90   &* &*     \\
4288.82   &V53c   &.0055  &5.11   &* &*     \\
4291.25   &V55    &.0162  &9.27   &.0221   &12.2     \\
4292.21   &V78c   &.0086  &8.67   &.0078    &7.56     \\
4294.78   &V53b   &.0314  &10.2   &.0156    &4.88     \\
4303.83   &V53a   &.0582  &11.6   &.0573    &10.9     \\
4307.23   &V53b   &.0087  &7.71   &*         &*        \\
4313.44   &V78a   &.0086  &6.59   &*          &*         \\
4353.59   &V76c   &.0098  &8.98   &.0150    &13.2        \\
4357.25   &V63a   &.0182  &29.3:  &*        &*                  \\
4366.53   &V75a   &*       &*     &.0605    &7.13     \\
4466.42   &V86b   &.0260  &23.7:  &.0305    &26.7:      \\
4477.90   &V88    &.0145  &15.7   &*        &*                  \\

\end{tabular}
\end{minipage}
\end{table}

\setcounter{table}{5}
\begin{table}
\begin{minipage}{75mm}
\centering \caption{{\it --continued}}
\begin{tabular}{lc@{\hspace{2.8mm}}c@{\hspace{2.8mm}}c@{\hspace{2.8mm}}c@{\hspace{2.8mm}}c@{\hspace{2.8mm}}}
\noalign{\vskip3pt} \noalign{\hrule} \noalign{\vskip3pt} $\lambda_{0}$ &Mult.
&$I_{\rm obs}$ &$\displaystyle\frac{\rm{~O}^{2+}}{\rm{H}^+}$ &$I_{\rm obs}$
&$\displaystyle\frac{\rm{~O}^{2+}}{\rm{H}^+}$ \\
\noalign{\vskip3pt}
(\AA)         &     &          &(\tmf)     &          &(\tmf) \\

\noalign{\vskip3pt} \noalign{\hrule} \noalign{\vskip3pt}
\multicolumn{2}{c}{}&\multicolumn{2}{c}{\bf{NGC\,3242}}&\multicolumn{2}{c}{\bf{NGC\,5882}}\\

4489.49   &V86b   &.0055  &7.86   &*        &*                  \\
4491.23   &V86a   &.0235  &16.0   &.0203    &13.3        \\
4609.44   &V92a   &.0439  &7.26   &.0609    &9.64        \\
4669.27   &V89b   &.0119  &27.6:  &*        &*                 \\

\bf{3d--4f}  & &\bf{0.493} &\bf{8.40} &\bf{0.569} &\bf{9.67} \\

\bf{Adopted} & & &\bf{6.28}     &   &\bf{9.70} \\
\noalign{\vskip3pt}

\multicolumn{2}{c}{}&\multicolumn{2}{c}{\bf{NGC\,5315}}&\multicolumn{2}{c}{\bf{NGC\,3918}}\\

4638.86   &V1      &.1740   &16.5:    &.2200   &20.4:   \\
4641.81   &V1      &.2367   &9.86     &.2168   &7.96    \\
4649.13   &V1      &.3574   &7.08     &.2108   &4.07    \\
4650.84   &V1      &.0900   &8.56     &.0415   &3.84    \\
4661.63   &V1      &.0897   &6.67     &.0743   &5.39    \\
4673.73   &V1      &.0137   &6.56     &.0310   &14.5:   \\
4676.24   &V1      &.0789   &6.99     &.0656   &5.66    \\

\bf{V\,1 3s\,$^{4}$P--3p\,$^{4}$D$^\mathrm{{o}}$}  & &\bf{0.866} &\bf{7.58} &\bf{0.609} &\bf{5.28} \\

4317.14   &V2      &.0694   &8.23     &.0417   &4.93 \\
4319.63   &V2      &.0491   &5.94     &.0170   &2.04 \\
4325.76   &V2      &.0191   &11.8:    &.0141   &9.2:  \\
4345.56   &V2      &.0635   &8.34     &*&* \\
4349.43   &V2      &.0744   &9.77     &.0638   &3.31 \\
4366.89   &V2      &.0589   &6.87     &*&* \\

\bf{V\,2 3s\,$^{4}$P--3p\,$^{4}$P$^\mathrm{{o}}$}  & &\bf{0.315} &\bf{6.06} &\bf{0.123} &\bf{3.39} \\

4414.90   &V5      &.0721   &13.0     &.0518   &8.88        \\
4416.97   &V5      &.0620   &20.2     &.0290   &8.96        \\
4452.37   &V5      &.0146   &29.8:    &.0653   &100.:       \\

\bf{V\,5 3s\,$^{2}$P--3p\,$^{2}$D$^\mathrm{{o}}$}  & &\bf{0.134} &\bf{15.55} &\bf{0.081} &\bf{8.91} \\

4069.62   &V10     &.4263   &16.5:    &*&*  \\
4072.16   &V10     &.1879   &7.78     &*&*  \\
4075.86   &V10     &.1412   &4.05     &*&*  \\
4085.11   &V10     &.0437   &9.69     &*&*  \\

\bf{V\,10 3p\,$^{4}$D$^\mathrm{{o}}$}--3d\,$^{4}$F  & &\bf{0.329} &\bf{5.58} &*&* \\

4121.46   &V19     &*&*     &.0920   &33.1:     \\
4132.80   &V19     &*&*     &.0180   &3.32      \\
4153.30   &V19     &*&*     &.0260   &3.35      \\
4156.53   &V19     &*&*     &.0370   &29.9:     \\
4169.22   &V19     &*&*     &.0045   &1.72      \\

\bf{V\,19 3p\,$^{4}$P$^\mathrm{{o}}$}--3d\,$^{4}$P  & &*&* &\bf{0.049} &\bf{3.07}  \\

4083.90   &V48b    &.0241   &7.80     &*&* \\
4087.15   &V48c    &.0319   &10.9     &*&* \\
4089.29   &V48a    &.0899   &8.30     &*&* \\
4275.55   &V67a    &.1005   &8.11     &*&* \\
4303.82   &V53a    &.0331   &6.50     &*&* \\
4609.44   &V92a    &.0476   &7.79     &*&* \\

\bf{3d--4f}  & &\bf{0.332} &\bf{8.08} &*&*\\

\bf{Adopted} &  &&\bf{8.57} & &\bf{5.36}   \\

\noalign{\vskip3pt}
\multicolumn{2}{c}{}&\multicolumn{2}{c}{\bf{NGC\,2022}}&\multicolumn{2}{c}{\bf{NGC 6818}}\\

4638.86   &V1      &.1051   &9.41:   &.2160    &19.8:   \\
4641.81   &V1      &.2538   &9.01:   &.4956    &18.0:   \\
4649.13   &V1      &.3674   &6.86    &.1687    &3.23    \\
4650.84   &V1      &.0892   &8.00    &.0262    &2.50   \\
4661.63   &V1      &.1093   &7.67    &.0851    &6.11   \\
4676.24   &V1      &.0772   &6.44    &*&* \\

\bf{V\,1 3s\,$^{4}$P--3p\,$^{4}$D$^\mathrm{{o}}$}  & &\bf{1.00}  &\bf{7.69} &\bf{0.280} &\bf{3.65} \\

4414.90   &V5      &.1247   &19.3    &.0594  &9.90  \\

\bf{V\,5 3s\,$^{2}$P--3p\,$^{2}$D$^\mathrm{{o}}$}  & &\bf{0.125} &\bf{19.3}  &\bf{0.059} &\bf{9.90}   \\

4069.78   &V10     &*&*                 &0.132 &5.11 \\
4072.16   &V10     &.3164   &13.1       &0.175 &7.26 \\
4075.86   &V10     &.2949   &8.43        &0.438 &12.6\\
4085.11   &V10     &*&*                 &.0722: &15.8:\\

\bf{V\,10 3p\,$^{4}$D$^\mathrm{{o}}$}--3d\,$^{4}$F  & &\bf{0.611} &\bf{10.3}  &\bf{0.745} &\bf{6.41} \\

\end{tabular}
\end{minipage}
\end{table}

\setcounter{table}{5}
\begin{table}
\begin{minipage}{75mm}
\centering \caption{{\it --continued}}
\begin{tabular}{lc@{\hspace{2.8mm}}c@{\hspace{2.8mm}}c@{\hspace{2.8mm}}c@{\hspace{2.8mm}}c@{\hspace{2.8mm}}}
\noalign{\vskip3pt} \noalign{\hrule} \noalign{\vskip3pt} $\lambda_{0}$ &Mult.
&$I_{\rm obs}$ &$\displaystyle\frac{\rm{~O}^{2+}}{\rm{H}^+}$ &$I_{\rm obs}$
&$\displaystyle\frac{\rm{~O}^{2+}}{\rm{H}^+}$ \\
\noalign{\vskip3pt}
(\AA)         &     &          &(\tmf)     &          &(\tmf) \\

\noalign{\vskip3pt} \noalign{\hrule} \noalign{\vskip3pt}
\multicolumn{2}{c}{}&\multicolumn{2}{c}{\bf{NGC\,2022}}&\multicolumn{2}{c}{\bf{NGC 6818}}\\

4089.29   &V48a    &.2020   &16.9    &.0984    &9.37   \\
4276.75   &V67b    &.1856   &13.8    &*&*\\
4609.44   &V92a    &.0578   &10.0    &*&*\\

\bf{3d--4f}  & &\bf{0.416} &\bf{14.8 } &\bf{0.098} &\bf{9.37} \\

\bf{Adopted} & & &\bf{13.0}   & &\bf{7.33}   \\
\noalign{\vskip3pt}

\multicolumn{2}{c}{}&\multicolumn{2}{c}{\bf{NGC\,3132}}&\multicolumn{2}{c}{\bf{IC\,4406}}\\

4638.86   &V1      &.1043    &10.0    &.1817   &17.3:  \\
4641.81   &V1      &.1706    &6.49    &.2627   &9.94   \\
4649.13   &V1      &.1649    &3.30    &.2585   &5.14   \\
4650.84   &V1      &.1226    &11.8    &.0959   &9.16   \\
4661.63   &V1      &.1157    &8.70    &.0988   &7.39   \\
4676.24   &V1      &.0762    &6.82    &.0481   &4.28   \\

\bf{V\,1 3s\,$^{4}$P--3p\,$^{4}$D$^\mathrm{{o}}$}  & &\bf{0.754} &\bf{7.06} &\bf{0.764} &\bf{6.84} \\

4069.62   &V10     &.4086    &15.8:   &.4106    &15.9: \\
4072.16   &V10     &.1484    &6.17    &.3776    &15.7: \\
4075.86   &V10     &.3592    &10.3    &.2530    &7.28 \\
4085.11   &V10     &.0433    &9.64    &*&*\\

\bf{V\,10 3p\,$^{4}$D$^\mathrm{{o}}$}--3d\,$^{4}$F  & &\bf{0.551} &\bf{8.70} &\bf{0.253} &\bf{7.28}  \\

\bf{Adopted} & & &\bf{8.15}  & &\bf{7.06}   \\
\noalign{\vskip3pt}

\multicolumn{2}{c}{}&\multicolumn{2}{c}{\bf{My\,Cn\,18}}&\multicolumn{2}{c}{\bf{NGC\,6302}}\\

4649.13   &V1       &.2294      &4.66      &.154   &2.70           \\
4650.84   &V1       &.1531      &14.9      &.0609  &5.13           \\
4661.63   &V1       &*          &*         &.0353  &2.33           \\

\bf{Adopted} & & &\bf{6.43}  & &\bf{3.28}   \\

\noalign{\vskip3pt}
\multicolumn{2}{c}{}&\multicolumn{4}{c}{\bf{IC\,4191}}\\

\multicolumn{2}{c}{}&\multicolumn{2}{c}{entire nebula}&\multicolumn{2}{c}{fixed slit}\\

4638.86   &V1      &.1912   &18.0:   &.2426   &22.8: \\
4641.81   &V1      &.3218   &12.0    &.2020   &12.2  \\
4649.13   &V1      &.5439   &10.7    &.5757   &11.9  \\
4650.84   &V1      &.0997   &9.38    &.1168   &14.5  \\
4661.63   &V1      &.1539   &11.3    &.1442   &10.6  \\
4676.24   &V1      &.1226   &10.8    &.1084   &9.29  \\

\bf{V\,1 3s\,$^{4}$P--3p\,$^{4}$D$^\mathrm{{o}}$}  & &\bf{1.24}  &\bf{10.9}  &\bf{1.15}   &\bf{10.1} \\

4317.14  &V2 &* &*                       &.0676   &8.77   \\
4319.63  &V2 &* &*                       &.0551   &6.62   \\
4345.56  &V2 &* &*                       &.1210   &15.2     \\
4349.43  &V2 &* &*                       &.2324   &12.1     \\
4366.89  &V2 &* &*                       &.1064   &12.4     \\

\bf{V\,2 3s\,$^{4}$P--3p\,$^{4}$P$^\mathrm{{o}}$}  & &\bf{*} &\bf{*} &\bf{0.582} &\bf{11.1}  \\

4414.90   &V5     &.0757   &13.2     &.0546    &9.49     \\
4416.97   &V5     &.0531   &16.6     &.0444    &13.9     \\

\bf{V\,5 3s\,$^{2}$P--3p\,$^{2}$D$^\mathrm{{o}}$}  & &\bf{0.129} &\bf{14.4}  &\bf{0.099} &\bf{11.1}  \\

4069.62   &V10   &.7056  &27.2:   &.5998  &23.2:   \\
4072.16   &V10   &.3654  &15.2    &.3565  &14.8    \\
4075.86   &V10   &.2883  &8.28    &.3452  &9.91       \\
4085.11   &V10   &.0503  &11.0    &.0266  &5.81       \\

\bf{V\,10 3p\,$^{4}$D$^\mathrm{{o}}$}--3d\,$^{4}$F  & &\bf{0.704} &\bf{11.1}  &\bf{0.728} &\bf{11.5}  \\

4083.90   &V48b  &.0418  &13.8   &.0701  &23.2        \\
4087.15   &V48c  &.0580  &18.6   &.0685  &21.9        \\
4089.29   &V48a  &.1475  &13.9   &.2059  &19.1         \\
4275.55   &V67a  &.2096  &16.2   &.2434  &18.8         \\
4282.96   &V67c  &.0587  &12.6   &*      &*              \\
4303.82   &V53a  &.0930  &18.1   &.0684  &13.3         \\
4466.42   &V86b  &*       &*     &.0387: &35.9:         \\
4609.44   &V92a  &.0852  &14.3   &.0603  &10.1           \\

\bf{3d--4f}  & &\bf{0.694} &\bf{15.2}  &\bf{0.717} &\bf{17.6}  \\

%4590.97   &V15    &.07579    &22.97 &.08042    &24.26 \\
%4596.18   &V15    &.05858    &      &.06177    &       \\

\bf{Adopted}         & & &\bf{12.9}      &   &\bf{12.3}  \\
\noalign{\vskip3pt} \noalign{\hrule} \noalign{\vskip3pt}
\end{tabular}
\end{minipage}
\end{table}

\setcounter{table}{5}
\begin{table}
\begin{minipage}{75mm}
\centering \caption{{\it --continued}}
\begin{tabular}{lc@{\hspace{2.8mm}}c@{\hspace{2.8mm}}c@{\hspace{2.8mm}}}
\noalign{\vskip3pt} \noalign{\hrule} \noalign{\vskip3pt} $\lambda_{0}$ &Mult.
&$I_{\rm obs}$ &$\displaystyle\frac{\rm{~O}^{2+}}{\rm{H}^+}$  \\
\noalign{\vskip3pt}
(\AA)         &     &          &(\tmf)    \\

\noalign{\vskip3pt} \noalign{\hrule} \noalign{\vskip3pt}
\multicolumn{2}{c}{}&\multicolumn{2}{c}{\bf{NGC\,2440}}\\

4638.86   &V1      &.2840:  &24.0:  \\
4641.81   &V1      &.2051:  &7.14:  \\
4649.13   &V1      &.1266   &2.32   \\
4650.84   &V1      &.0510   &4.48    \\
4661.63   &V1      &.0346   &2.38   \\
4676.24   &V1      &.0710:  &5.81:  \\

\bf{V\,1 3s\,$^{4}$P--3p\,$^{4}$D$^\mathrm{{o}}$}  & &\bf{0.212}  &\bf{2.63}   \\

4414.90   &V5     &.0472   &6.91          \\
4416.97   &V5     &.0380   &10.0          \\
\bf{V\,5 3s\,$^{2}$P--3p\,$^{2}$D$^\mathrm{{o}}$}  & &\bf{0.085} &\bf{8.02} \\

4153.30   &V19     &.0260   &3.35      \\
\bf{V\,19 3p\,$^{4}$P$^\mathrm{{o}}$}--3d\,$^{4}$P  &&\bf{0.026} &\bf{3.35}  \\
\bf{Adopted}         & & &\bf{5.23}       \\

\noalign{\vskip3pt} \noalign{\hrule} \noalign{\vskip3pt}
\end{tabular}
\end{minipage}
\end{table}

\setcounter{table}{6}
\begin{table*}
\begin{center}
\caption{Recombination-line oxygen abundances for Galactic PNe}
\begin{tabular}{lccccccc}
\noalign{\vskip3pt} \noalign{\hrule}\noalign{\vskip3pt}
                                            &\bf{NGC\,3242}&\bf{NGC\,5882}&\bf{NGC\,5315}&\bf{NGC\,3918}&\bf{NGC\,2022}&\bf{NGC\,6818}    \\
\noalign{\vskip3pt} \noalign{\hrule} \noalign{\vskip3pt}
10$^4$\,$\times$\,{\opp}/{\hp}              &6.28       &9.70           &8.57           &5.36       &13.01      &7.33       \\
\noalign{\vskip2pt}
10$^4$\,$\times$\,{\oppp}/{\hp}             &2.05$^a$   &*              &*              &*          &3.62$^a$   &2.37$^a$   \\
\noalign{\vskip2pt}
$I(\lambda4632)$                            &*          &*              &*              &0.150      &0.334      &0.200      \\
10$^4$\,$\times$\,{\ofp}/{\hp}              &*          &*              &*              &0.474      &1.100      &0.641      \\
\noalign{\vskip2pt}
CEL {\op}/{\opp}                            &0.0091     &0.029          &0.098          &0.086      &0.020      &0.090      \\
\noalign{\vskip2pt}
icf(O)                                      &1.000      &1.010          &1.332          &1.298      &1.000      &1.000      \\
\bf{10$^4$\,$\times$\,O/H}                 &\bf{8.39}   &\bf{10.08}    &\bf{12.53}      &\bf{12.32}  &\bf{17.99} &\bf{11.00} \\
\noalign{\vskip3pt} \noalign{\hrule} \noalign{\vskip3pt}
                                            &\bf{NGC\,3132}&\bf{NGC\,2440}&\bf{NGC\,6302}&\bf{IC\,4406}&\bf{IC\,4191}&\bf{My\,Cn\,18}   \\
\noalign{\vskip3pt} \noalign{\hrule} \noalign{\vskip3pt}
10$^4$\,$\times$\,{\opp}/{\hp}              &8.15       &5.23           &3.28          &7.06           &12.92      &6.43     \\
\noalign{\vskip2pt}
10$^4$\,$\times$\,{\oppp}/{\hp}             &*          &2.56$^a$       &*              &*              &*          &*       \\
\noalign{\vskip2pt}
$I(\lambda4632)$                            &*          &0.300          &0.228          &*              &*          &*       \\
10$^4$\,$\times$\,{\ofp}/{\hp}              &*          &1.01           &0.784          &*              &*          &*       \\
\noalign{\vskip2pt}
CEL {\op}/{\opp}                            &0.894      &0.285          &0.092          &0.471          &0.041      &0.590   \\
\noalign{\vskip2pt}
icf(O)                                      &1.020      &1.000          &1.672          &1.040          &1.070      &1.003   \\
\bf{10$^4$\,$\times$\,O/H}                  &\bf{15.74} &\bf{10.29}     &\bf{7.30}      &\bf{10.80}     &\bf{14.39} &\bf{10.24} \\
\noalign{\vskip3pt} \noalign{\hrule}
\end{tabular}
\begin{description}
\item[$^a$] From the Liu \& Danziger (1993a) analysis of {\oiii} V8 $\lambda$3265.
\end{description}
\end{center}
\end{table*}

\section{Comparison of ORL and CEL abundances}

\subsection{Total C, N and O abundances}

The total elemental abundances of C, N and O obtained for the current PN
sample from CELs (Paper~I) and from ORLs are presented in Table~8, on the
usual log\,$N$(H) = 12.0 scale, where they are compared with average
abundances for Galactic PNe (for both Type I and non-Type I nebulae) derived
by Kingsburgh \& Barlow (1994) and the solar photospheric abundances of
Grevesse, Noels \& Sauval (1996) and Allende Prieto et al. (2001, 2002).

\setcounter{table}{7}
\begin{table}
\centering
\begin{minipage}{80mm}
\caption{Elemental C, N, O abundances by number, derived from CELs and ORLs, in
units where log\,$N$(H) = 12.0.$^a$}
\begin{tabular}{lccccccc}
\noalign{\vskip3pt}\noalign{\hrule}\noalign{\vskip3pt}
 PN                     &\multicolumn{2}{c}{C}          &\multicolumn{2}{c}{N}          &\multicolumn{2}{c}{O}                     \\%         &Ne             &S        &Cl       &Ar   He       \\
                       &ORL        &CEL                &ORL        &CEL                &ORL        &CEL                           \\%         &CEL             &CEL      &CEL      &CEL  ORL         \\
\noalign{\vskip3pt}\noalign{\hrule}\noalign{\vskip3pt}

NGC\,2022              &9.17        &8.33               &8.84       &7.46               &9.26       &8.66                           \\%       &   7.84       &6.57     &4.91     &6.12    &11.02                     \\
NGC\,2440(I)           &9.01        &8.37               &9.01       &8.16               &9.01       &8.39                           \\%       &  8.04        & *       & *       & *      & *          \\
NGC\,3132              &9.11        &8.50               &8.91       &8.37               &9.20       &8.82                           \\%       &  8.49        &7.03     &5.36     &6.73    &11.08                      \\
NGC\,3242              &8.93        &8.14               &8.38       &7.53               &8.92       &8.52                           \\%       &   7.89       &6.38     &4.94     &5.99    &11.00                   \\
NGC\,3918              &8.88        &8.64               &8.41       &8.02               &9.09       &8.86               \\
NGC\,5315              &8.86        &8.33               &8.77       &8.52               &9.10       &8.79                           \\%       &   8.30       &7.31     &5.41     &6.56    &11.08                     \\
NGC\,5882              &8.65        &8.18               &8.46       &8.18               &9.00       &8.69                           \\%       &   8.13       &6.92     &5.26     &6.31    &11.04                     \\
NGC\,6302(I)           &8.74        &7.89               &9.05       &8.52               &8.86       &8.40                           \\%       & 7.88         &6.75     &4.99     &6.34    &??                         \\
NGC\,6818              &8.89        &8.41               &8.80       &7.74               &9.04       &8.71                           \\%       &  8.07        &6.56     &5.09     &5.94    &11.00                    \\
IC\,4191               &8.81        &*                  &8.91       &7.59               &9.16       &8.78                           \\%       & 8.21         &7.10     &5.37     &6.49    &11.08                     \\
IC\,4406               &9.05        &8.56               &8.56       &8.32               &9.03       &8.76                           \\%       & 8.33         &6.33     &5.16     &6.40    &11.09                      \\
My\,Cn\,18             &8.78        &*                  &$\geq$9.31 &8.34               &9.01       &8.75                           \\%       & 8.09         &7.24     &5.39     & *      &10.99                  \\
\noalign{\vskip2pt}
SMC\,N87               &8.88        &8.58               &*          &7.55               &8.48       &8.03                           \\%       & 7.03         & *       & *       &5.32    &10.99             \\
LMC\,N66               &*           &7.52               &*          &7.99               &9.54       &8.50                           \\%       & 7.62         &6.63     & *       &6.10    &11.02                 \\
LMC\,N141              &8.88        &8.30               &*          &7.95               &8.71       &8.29                           \\%       & 7.38         &6.61     & *       &5.88    &11.04                  \\
\noalign{\vskip2pt}
KB94 non-Type~I$^b$    &*           &8.81               &*          &8.14               &*          &8.69                           \\%     &8.10             &6.91  &*    &6.38          &11.05                       \\
KB94 Type~I$^b$        &*           &8.48               &*          &8.72               &*          &8.65                           \\%   &8.09             &6.91  &*    &6.42            &11.11                       \\
Solar$^c$              &*           &8.39               &*          &7.97               &*          &8.69                           \\%     &8.09           &7.21  &5.50  &6.56           &10.99                       \\

\noalign{\vskip3pt} \noalign{\hrule} \noalign{\vskip3pt}
\end{tabular}
\begin{description}
\item[$^a$] All values are for the entire nebulae, except NGC\,6302, NGC\,6818, My\,Cn\,18 and IC\,4406 where values are from fixed-slit
observations; CEL results are from Paper~I.
\item[$^b$] From Kingsburgh \& Barlow (1994).
\item[$^c$] Solar photospheric abundances from Grevesse et al. (1996),
except O and C which are from Allende Prieto et al. (2001, 2002) respectively.
\end{description}
\end{minipage}
%\end{center}
\end{table}

\subsection{Abundance grids and ionic ADF's}

Comparisons between the ionic abundances obtained from the UV, optical and
IR collisionally excited lines (Paper~I) and those from ORLs (Tables~2--7)
are presented in Fig.\,~1 bringing together {\it IUE}, optical, and {\it
ISO} data for 9 galactic PNe, and {\it IUE} and optical data for the 3
Cloud PNe (LWS spectra are not available for NGC\,2440 and My\,CN\,18,
while {\it IUE} spectra are not available for IC\,4191 and My\,CN\,18).  
Note that the plotted {\opp}/{\hp} fractions derived from optical CELs are
those from the {\foiii} $\lambda\lambda$4959, 5007 lines only.

In Paper~I we used the {\it ISO} LWS far-IR fine structure (FS) lines
presented by Liu et al. (2001a) to derive {\npp}/{\hp} and {\opp}/{\hp}
abundances for 8 PNe in our sample. In the meantime we retrieved archived LWS
spectra of NGC\,2022 (TDT~69201604) and NGC\,6818 (TDT~34301005) and 
measured
the fluxes of the {\fniii} 57-$\mu$m and {\foiii} 52-, 88-$\mu$m FS lines. We
scaled these to $I$({\Hb}) = 100 using the total nebular {\Hb} fluxes
dereddened by the $c$({\Hb})$^{radio}$ constants of Table~5, Paper~I. The
respective $I$(52-$\mu$m), $I$(57-$\mu$m), and $I$(88-$\mu$m) intensities are
91.7, 21.5, and 42.2 for NGC\,2022; and 164.6, 45.5 and 69.7 for NGC\,6818.
From the {\foiii} 52-$\mu$m/88-$\mu$m line ratio we derived electron densities
of 800\,{\cmt} for NGC\,2022 and 950\,{\cmt} for NGC\,6818. From the measured
FS line intensities and adopting the {\eld}'s from the {\foiii} FS line ratio
we derived: {\npp}/{\hp} = 2.75$\times${\tmfi} and {\opp}/{\hp} =
1.07$\times${\tmf} for NGC\,2022; and 6.91$\times${\tmfi} and
2.20$\times${\tmf} respectively for NGC\,6818. On the other hand adopting the
mean of the {\eld}'s from the {\fcliii} and {\fariv} optical line ratios
(Paper~I) we derived: {\npp}/{\hp} = 3.94$\times${\tmfi} and {\opp}/{\hp} =
1.45$\times${\tmf} for NGC\,2022; and 1.15$\times${\tmf} and
3.41$\times${\tmf} respectively for NGC\,6818. The latter values for these 
two PNe are plotted in Fig.\,~1.

Fig.\,~1 shows that in \emph{all} nebulae and for all ions where ionic
abundances from both ORLs and CELs have been derived, the values from the
optical recombination lines are consistently \emph{higher} than those
derived from the collisionally excited lines (UV, optical or IR). This
includes the Magellanic Cloud nebulae SMC N87, LMC N66 and LMC N141; no
such comparisons had been reported so far regarding extragalactic PNe. In
Table~9 we present a listing of the ionic abundance discrepancy factors,
ADF(X$^{i+}$), defined as the ratio of the ORL ionic abundance to the UV,
optical (OPT) or IR CEL ionic abundance. For comparison, data for the
previously studied planetary nebulae NGC\,7009 (LSBC), NGC\,6153 (Liu et
al. 2000), the galactic bulge nebulae M\,1-42, M\,2-36 (Liu et al. 2001b),
as well as for NGC\,6644 and NGC\,6572 (from unpublished ESO 1.52\,m
observations) are also included.

%\special{\landscape}
%\begin{rotate}{90}
%\begin{sidewaystable}

Possible exceptions to the general pattern of large ADF(X$^{i+}$) values
include {\cppp}/{\hp} in NGC\,3918, where abundances derived from the {\ciii}
$\lambda\lambda$4187, 4650 ORLs and the C~{\sc iv} $\lambda$1550 resonant
doublet almost coincide; for the same nebula the {\nfp}/{\hp} abundance
obtained from the N~{\sc iv} $\lambda$4606 ORL is 10\,per cent \emph{less}
that the value derived from the N~{\sc v} $\lambda$1240 CEL; also
{\nppp}/{\hp} in NGC\,6302, where identical values are derived from the
{\niii} $\lambda$4379 ORL and the N~{\sc iv}] $\lambda$1486 intercombination
line. This could be due to an underestimation on our part of the actual
electron temperature pertaining to highly ionized species in these two
high-excitation nebulae; adopting a higher {\elt} in order to derive
{\cppp}/{\hp}, {\nppp}/{\hp} and {\nfp}/{\hp} would result in lower CEL
abundances and produce a discrepancy with the ORL values as well.
Alternatively, it could mean that at least in some nebulae the mechanisms that
are responsible for the abundance discrepancies do not affect species of
different ionization degree in the same way.

The current work has revealed two further nebulae, NGC\,2022 and LMC~N66,
which exhibit extreme ORL/CEL ADF's: factors of 16 and 11, respectively, for
{\opp}/{\hp}. These nebulae are added to a rare class of PNe, along with
NGC\,6153, M1-42, and Hf~2-2, that exhibit ORL/CEL ADF's of ten or more. We
also reveal another object, the Type-I bipolar nebula NGC\,2440, whose ORL/CEL
abundance discrepancy factor of 5 is similar to those of NGC\,7009 and M\,2-36
(see Table~9).

Taking into account that in Paper~I we presented both `low-density' (adopting
{\foiii} 88 $\mu$m/52 $\mu$m densities) and `high-density' values (adopting
densities from {\fariv}, {\fcliii}), we note the following regarding the
{\npp}/{\hp} and {\opp}/{\hp} fractions derived from the far-IR FS lines that
were plotted in Fig.\,~1: for NGC\,3132 and IC\,4406, which have a low mean
density, the `low-density' values are plotted for both {\npp} and {\opp},
while for all remaining PNe (including NGC\,2022 and NGC\,6818), which are of
higher mean densities, the `high-density' values are plotted. From our Paper~I
discussion of the results of Rubin (1989) it follows that the low critical
densities of the far-IR {\npp} and {\opp} lines lead to their emission being
biased towards lower-than-average density nebular regions, while this is not
true for the UV and optical lines, whose emissivity ratios relative to
H$\beta$ are less sensitive to variations in nebular density, owing to the
significantly higher critical densities of the lines in question. However, it
is obvious from Fig.\,~1 and Table~9 that the {\npp} and {\opp} abundances
deduced from the IR FS lines agree very well with the values obtained from the
N~{\sc iii}] $\lambda$1750 and {\foiii} $\lambda\lambda$4959, 5007 lines. This
shows that even though density variations are present in the nebulae, {\it no
significant} bias in the inferred CEL abundances is present, since very
satisfactory agreement amongst the derived values is achieved (see also the
relevant discussion in Section~5.1). Alternatively, this could mean that
whatever {\it other} bias there is, it affects all UV, optical and IR CEL
abundances in a rather similar manner.

In Sections 5.1 and 5.2 we will look further into the arguments for or
against the existence of density or temperature fluctuations in
our present PNe sample.

\setcounter{figure}{0}
\begin{figure*}
\begin{center}
\epsfig{figure=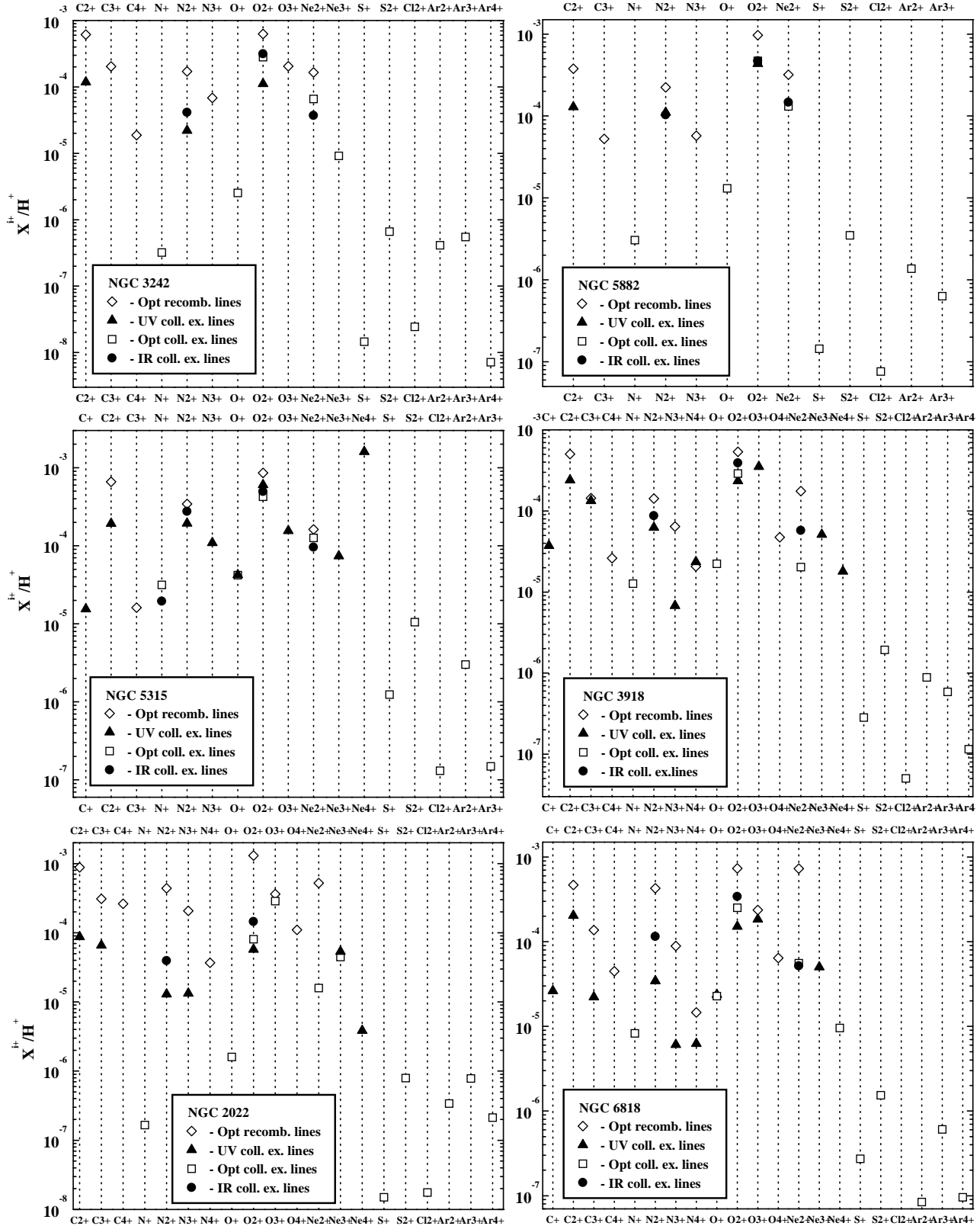, width=17 cm, bbllx=14pt, bblly=50pt,
bburx=581pt, bbury=800pt, clip=, angle=0} \caption{Comparison of ionic
abundances derived from optical recombination lines, and from UV,
optical and IR collisionally excited lines. In these and subsequent graphs the
plotted values are for the entire nebulae originating from ground-based
scanning optical spectroscopy and wide aperture, \emph{IUE}, \emph{ISO} and
\emph{IRAS} observations, except for NGC\,6302, NGC\,6818, IC\,4406 and
My\,Cn\,18, where abundances from ORLs and optical CELs have been derived from
fixed-slit spectra.}
\end{center}
\end{figure*}

\setcounter{figure}{0}
\begin{figure*}
\begin{center} \epsfig{file=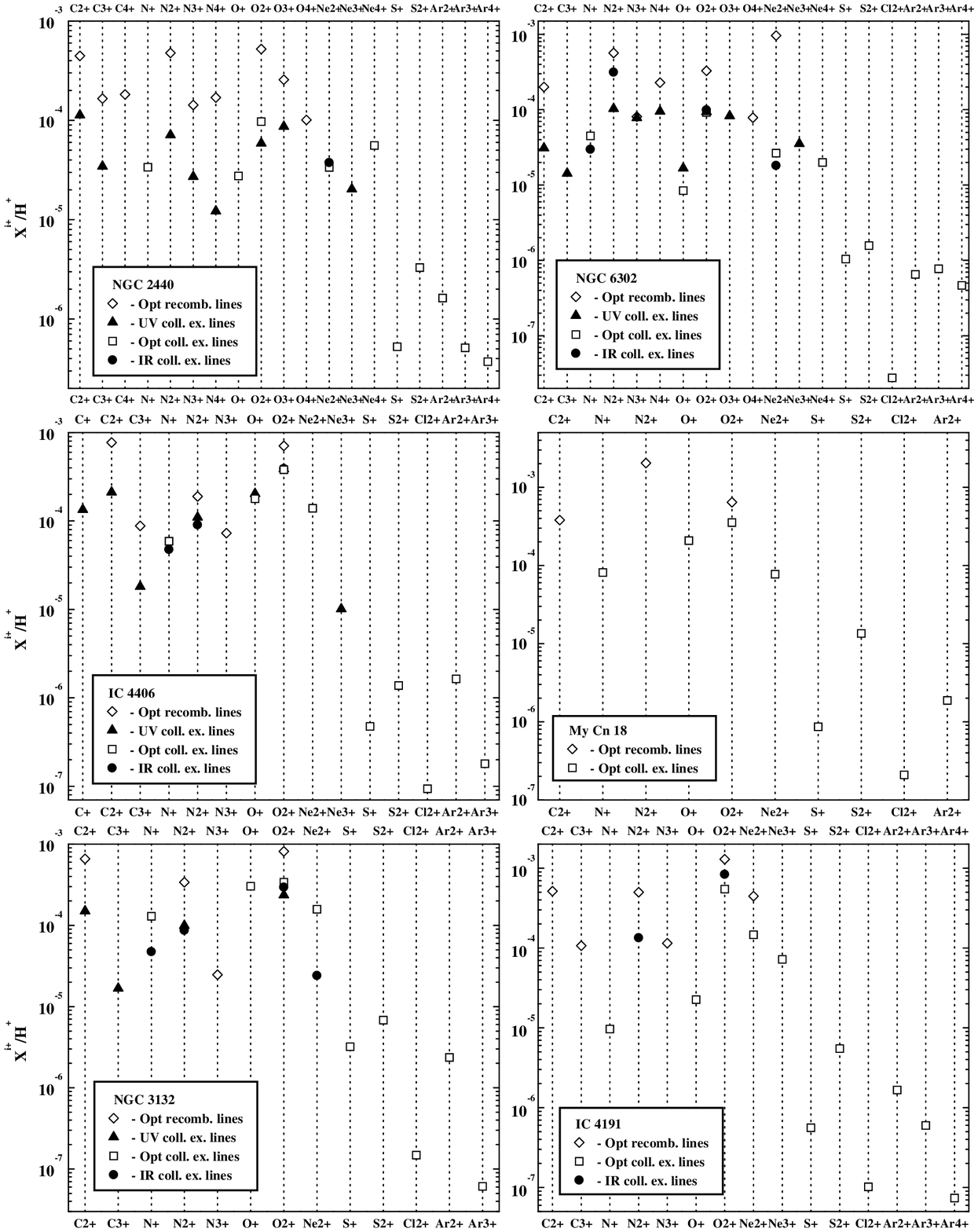, width=17 cm, bbllx=14pt, bblly=50pt,
bburx=581pt, bbury=800pt, clip=, angle=0} \caption{---{\it continued}.}
\end{center}
\end{figure*}

\setcounter{figure}{0}
\begin{figure}
\begin{center} \epsfig{file=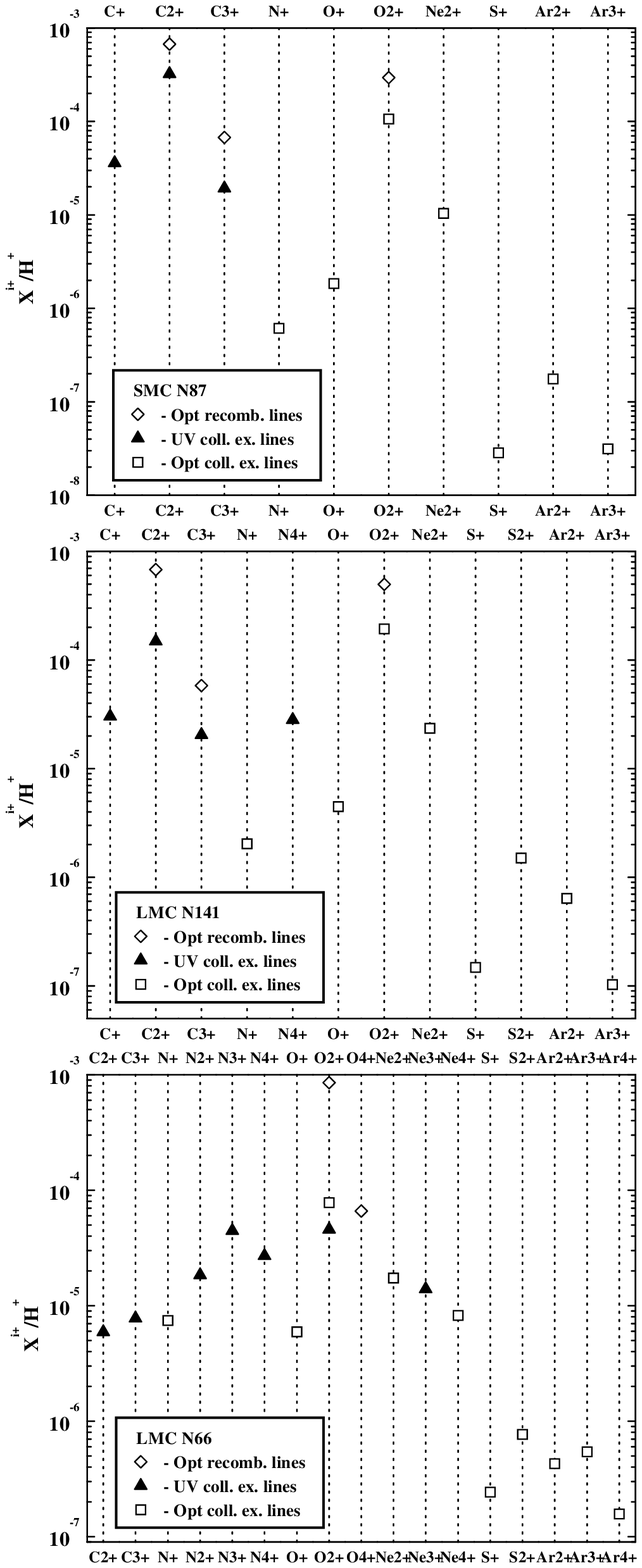, width=10.5 cm, bbllx=150pt, bblly=20pt,
bburx=500pt, bbury=790pt, clip=, angle=0} \caption{---{\it continued}.}
\end{center}
\end{figure}

\setcounter{table}{8}
\begin{table*}
\centering
%\begin{center}
\begin{minipage}{200mm}
\caption{Ionic ORL/CEL abundance discrepancy factors.}

\begin{tabular}{l@{\hspace{2mm}}l@{\hspace{3mm}}c@{\hspace{2mm}}c@{\hspace{2mm}}c@{\hspace{2mm}}c@{\hspace{2mm}}c@{\hspace{2mm}}c@{\hspace{2mm}}c@{\hspace{2mm}}c@{\hspace{2mm}}c@{\hspace{2mm}}c@{\hspace{2mm}}}
\noalign{\vskip3pt}\noalign{\hrule}\noalign{\vskip3pt}
PN                 &ADF({\cpp})&ADF({\cppp})&\multicolumn{2}{c}{\underline{~~~~~~~~ADF({\npp})~~~~~~~}}&ADF({\nppp})&ADF(N$^{4+}$) &\multicolumn{3}{c}{\underline{~~~~~~~~~~~~~~~ADF({\opp})~~~~~~~~~~~~~~~~}} &ADF({\oppp})    &ADF(Ne$^{2+}$)    \\
                    &ORL/UV     &ORL/UV     &ORL/UV   &ORL/IR           &ORL/UV     &ORL/UV     &ORL/OPT &ORL/UV &ORL/IR           &ORL/UV        &ORL/OPT       \\
\noalign{\vskip3pt}\noalign{\hrule}\noalign{\vskip3pt}

NGC\,2022       &10.\,(9.1)$^a$~~~&4.7       &34.~~~~  &11.~~~                &15.~~~           & *     & 16.    & 23.  & 9.0                & 1.3         &33.          \\
NGC\,2440           &4.0\,(2.5)&4.8          &6.7      &*                &5.3           & 14.~~~     & 5.4    &8.9   & *                &  *          &2.9          \\
NGC\,3132           &4.4       & *           &3.4      &3.9              & *            & *       & 2.4    &3.5   & 2.8              &  *          & *           \\
NGC\,3242           &5.2\,(5.5)& *           &7.8      &4.1              & *            &*        & 2.2    &5.7   &2.0               &  *          &2.5          \\
NGC\,3918           &2.1\,(3.4)&1.1          &2.3      &1.6              &9.5           &0.9      & 1.8    &2.3   & 1.4              &  *          &8.7          \\
NGC\,5315           &3.4       & *           &1.8      &1.2              & *            &*        & 2.0    &1.4   & 1.7              &  *          &1.3          \\
NGC\,5882           &2.9       & *           &1.5      &2.2              & *            &*        & 2.1    &2.2   & 2.1              &  *          &2.3    \\
NGC\,6302           &6.5\,(3.3)& *           &5.5      &1.8              &1.0           & 2.4     & 3.6    &3.5   & 3.3              &  *          &36.          \\
NGC\,6818           &2.3\,(1.3)&6.2          &12.~~~~      &3.7              &14.           &2.3      & 2.9    &4.9   & 2.1              & 1.3         &13.          \\
IC\,4191            &*           & *           & *     &3.7              & *            &   *     & 2.4    &*     & 1.5              &  *          &3.0           \\
IC\,4406            &3.7       &4.9          &1.7      &2.1              & *            &   *     & 1.9    &1.8   & 1.8              &   *         &*             \\
My\,Cn\,18          &*           & *           & *       &*                & *            &   *     & 1.8    &*     & *                &  *          &*             \\
\noalign{\vskip2pt}
SMC\,N87            &2.1       &3.5          & *       &*                & *            &   *     & 2.8    &*     & *                &  *          &*             \\
LMC\,N66            &*           & *           & *       &*                & *            &   *     & 11.    &19.   & *                &  *          &*             \\
LMC\,N141           &4.6       &2.9          & *       &*                & *            &   *     & 2.6    &*     & *                &  *          &*              \\
\noalign{\vskip2pt}
\noalign{\vskip2pt}
NGC\,4361           &15.~~~         &4.0          & *       &*                &*             &*        &*       &*     &*                 & *           &*             \\
NGC\,6153           &9.0\,(9.7)&*            &3.3      &9.0              &*             &*        &9.2     &*     &7.3               &*            &11.         \\
NGC\,6572           &2.1\,(1.1)&*            &*        &*                &*             &*        &1.5     &*     &3.9               &*            &*               \\
NGC\,6644           &2.8\,(2.8)&*            &*        &*                &*             &*        &1.9     &*     &*                 &*            &*               \\
NGC\,7009           &4.1\,(5.3)&3.2          &3.2      &7.1              & *            &*        & 5.0    & *    & 5.9              & *           &*            \\
M\,1-42             &23.~~~         &*            &*      &8.2              &*             &*        &22.     &*     &11.               &*            &17.          \\
M\,2-36             &4.8       &*            &*        &5.6              &*             &*        &6.9     &*     &5.6               &*            &7.7         \\

\noalign{\vskip3pt} \noalign{\hrule} \noalign{\vskip3pt}
\end{tabular}
\begin{description}
\item[$^a$] Parenthesized values are from Rola \& Stasi\'{n}ska (1994).
\end{description}
\end{minipage}
\end{table*}

\subsection{C/O and N/O elemental ratios}

\setcounter{figure}{1}
\begin{figure}
\begin{center} \epsfig{file=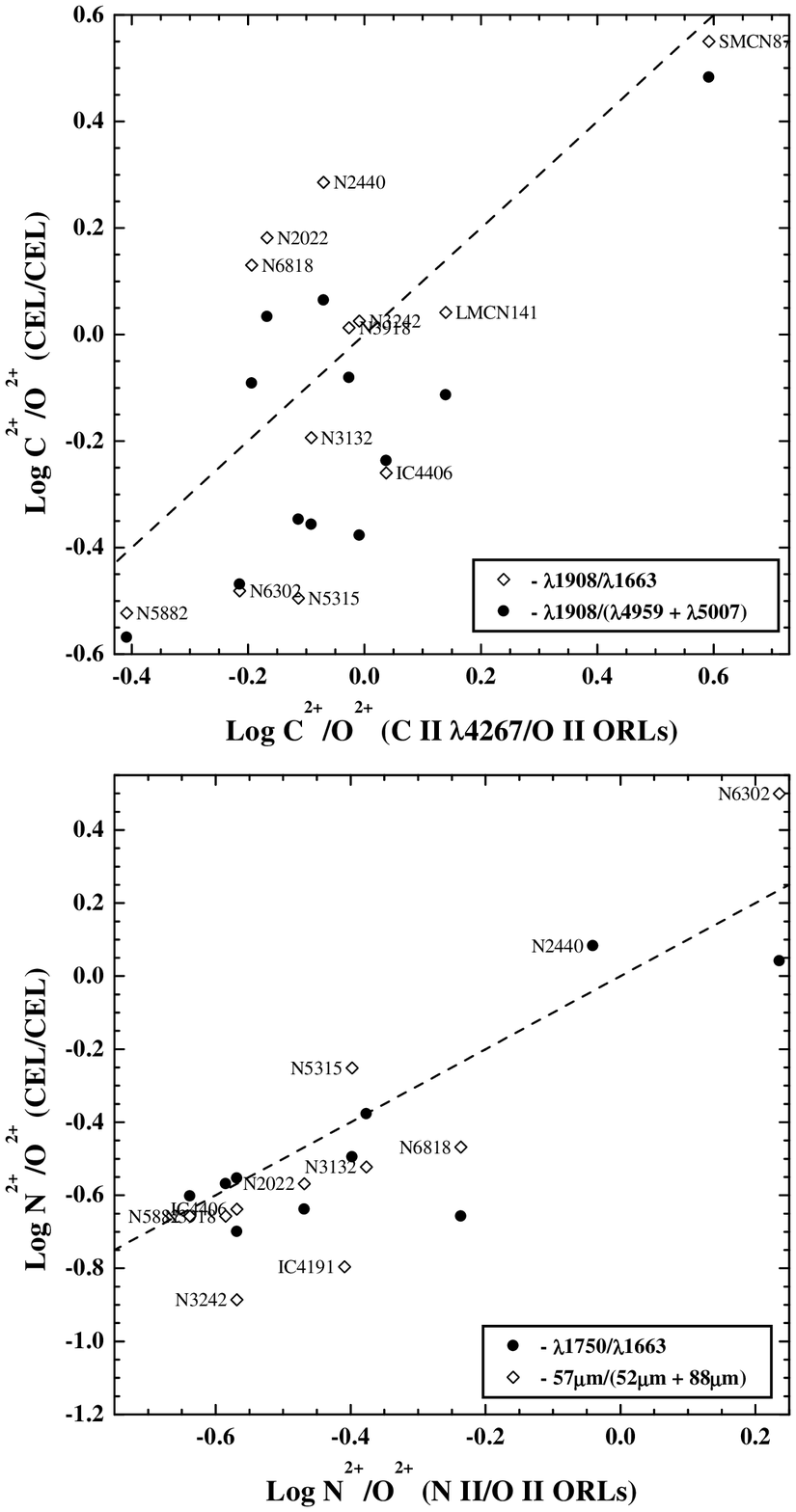, width=12 cm, bbllx=90pt,
bblly=150pt, bburx=581pt, bbury=829pt, clip=, angle=0}
\caption{\label{fig:c2o2-n2o2}Comparison of the {\cpp}/{\opp} (\emph{top}) and
{\npp}/{\opp} (\emph{bottom}) ionic abundance ratios derived from optical
recombination lines, and from UV, optical and far-IR collisionally excited
lines. Data points with identical x-axis values share the same label. The
dashed lines have a slope of unity.}
\end{center}
\end{figure}

Tables 10 and 11 respectively present the ({\cpp}/{\opp}, C/O) and
({\npp}/{\opp}, N/O) abundance ratios derived from pure ORL and pure CEL
line ratios. In both tables the last column may contain two values: the
one
before the vertical dash is computed with the total oxygen CEL abundance
found if we use the {\opp}($\lambda$4959+$\lambda$5007)/{\hp} ionic ratio
in the \emph{icf} method; the value after the dash adopts the oxygen
abundance found using the {\opp}($\lambda$1663)/{\hp} ratio. In the case
of the CEL N/O ratio the value after the dash is for nitrogen and oxygen
CEL abundances computed by adopting the {\npp}($\lambda$1751)/{\hp} and
{\opp}($\lambda$1663)/{\hp} ionic ratios, respectively, in the \emph{icf}
method.

The results presented in Tables 10 and 11 address an issue that we have
noted in earlier papers: although CELs and ORLs can often yield very
different abundances for the same ion relative to hydrogen, ORL-based and
CEL-based abundance ratios for two different ions are usually quite
comparable. This is because ORL/CEL abundance discrepancy factors are not
very different from one ion to the next for a given nebula (Table~9). It
appears that fairly reliable abundance ratios for ions (e.g.
{\cpp}/{\opp}) and for elements (e.g. C/O) may be obtained, provided that
both abundances are based on the same type of line (e.g. ORL/ORL or
CEL/CEL). It should be noted however that any residual differences between
such ORL/ORL and CEL/CEL elemental ratios may be real and could yield
clues about the relative nucleosynthetic histories of the ORL and CEL
emitting media. We urge further study of this issue. In Fig.\,~2 we show
the CEL {\cpp}/{\opp} and {\npp}/{\opp} ratios compared with the
corresponding ratios derived from ORLs.

The criterion of Kingsburgh \& Barlow (1994) states that Type I PNe are
those objects that have experienced envelope-burning conversion to
nitrogen of dredged-up primary carbon, i.e. PNe with a N/O ratio greater
than the (C + N)/O ratio of {\hii} regions in the host galaxy, the latter
being 0.8 for our own Galaxy. According to this criterion three objects in
our sample, NGC\,2440, NGC\,6302 and My\,Cn\,18, all bipolar, qualify as
Type I PNe by virtue of their ORL N/O ratios; but only the first two
qualify based on their CEL N/O ratios.

\setcounter{table}{9}
\begin{table*}
%\begin{center}
\centering
\begin{minipage}{130mm}
\caption{Comparison of C/O abundance ratios from CELs and ORLs.}
\label{TCORATIOPN}
\begin{tabular}{lcllccl}
\noalign{\vskip3pt} \noalign{\hrule} \noalign{\vskip3pt}
PN         &    {\cpp}/{\opp}  &    {\cpp}/{\opp}               &    {\cpp}/{\opp}     &   {\cpp}/{\opp}       &ORL C/O    &CEL C/O        \\
 &{\scriptsize$\lambda$4267/{\oii}} &{\scriptsize$\lambda$1908/$\lambda$1663}    &{\scriptsize$\lambda$1908/$\lambda$4959+$\lambda$5007}  &{\scriptsize$\lambda$1908/52+88~$\mu$m}            &          &                \\
\noalign{\vskip3pt} \noalign{\hrule}
\noalign{\vskip3pt}                                                                                                                                                                                 %C/(UV) O
NGC\,2022  &0.68               &1.52                            &1.08\,(1.21)$^b$  &0.60       &0.82
&0.47               \\   %&0.84         $\mid$0.49
NGC\,2440  &0.85               &1.93\,(1.23)$^a$                &1.16\,(0.94)      &*        &0.99
&0.96               \\   %&(.48)        $\mid$1.43
NGC\,3132  &0.81               &0.64\,(0.63)                    &0.44\,(1.68)      & 0.51   &0.81       &0.48               \\   %&0.55         $\mid$0.57
NGC\,3242  &0.98               &1.06\,(0.66)                    &0.42\,(0.62)      & 0.37   &1.01       &0.42$\mid$1.05$^c$ \\   %&0.43   $\mid$1.05
NGC\,3918  &0.94               &1.03\,(1.15)                    &0.83\,(0.90)      & 0.61   &0.60       &0.60               \\   %&0.93         $\mid$0.63
NGC\,5315  &0.77               &0.32\,(0.31)                    &0.45\,(0.27)      & 0.39   &0.58       &0.34               \\   %&0.55         $\mid$0.27
NGC\,5882  &0.39               &0.30                            &027               & 0.27   &0.44       &0.27               \\   %&0.29         $\mid$0.34
NGC\,6302  &0.61               &0.33\,(0.38)                    &0.34\,(0.39)      & 0.31   &0.75       &0.31               \\   %&(.12)        $\mid$0.29
NGC\,6818  &0.64               &1.35\,(1.45)                    &0.81\,(1.28)      &0.60    &0.70      &0.49$\mid$0.64     \\
IC\,4191   &0.40               &*                               &*                 &*        &0.45      &*                  \\
IC\,4406   &1.09               &0.55                            &0.58              & 0.54   &1.25       &0.62               \\   %&0.57         $\mid$0.59
My\,Cn\,18 &0.59               &*                               &*                 &*        &0.59      &*                  \\
\noalign{\vskip3pt}
SMC\,N87   &2.28               &*                               &3.04              &*        &2.52      &3.49     \\
LMC\,N66   &*                  &0.13                            &0.08              &*        &*         &0.10               \\
LMC\,N141  &1.38               &*                               &0.77              &*        &1.48      &1.02     \\
\noalign{\vskip2pt}
Solar      &*                   & *                             &  *                    & *    &*      &0.50\,$^d$        \\
\noalign{\vskip3pt} \noalign{\hrule} \noalign{\vskip3pt}

\end{tabular}
\begin{description}
\item[$^a$] Parenthesized values are from Rola \& Stasi\'{n}ska (1994);
\item[$^b$] Parenthesized values are from Rola \& Stasi\'{n}ska (1994);
%who used the $\lambda$1908/$\lambda$5007 line ratio;
\item[$^c$] See Section 3.3 for explanation when two values are listed for the CEL C/O ratio;
\item[$^d$] Solar oxygen and carbon abundances taken from Allende Prieto et al. (2001, 2002), respectively.
\end{description}
\end{minipage}
\end{table*}

\setcounter{table}{10}
\begin{table*}
\begin{center}
\caption{Comparison of N/O abundance ratios from CELs and ORLs.}
\begin{tabular}{l@{\hspace{2mm}}c@{\hspace{2mm}}c@{\hspace{2mm}}c@{\hspace{2mm}}c@{\hspace{2mm}}c@{\hspace{2mm}}l@{\hspace{2mm}}}
\noalign{\vskip3pt} \noalign{\hrule} \noalign{\vskip3pt}
PN         &    {\npp}/{\opp}  &    {\npp}/{\opp}                          &    {\npp}/{\opp}                                                   &    {\npp}/{\opp}                       &ORL N/O    &CEL N/O        \\
           &{\scriptsize{\nii}/{\oii}} &{\scriptsize$\lambda$1750/$\lambda$1663}    &{\scriptsize$\lambda$1750/$\lambda$4959+$\lambda$5007}     &{\scriptsize57\,$\mu$m/52+88\,$\mu$m}                        &          &                \\
\noalign{\vskip3pt} \noalign{\hrule}
\noalign{\vskip3pt}                                                                                                                                                                          %4931      UV (N/O)
NGC\,2022  &0.34               &0.23                            &0.16                                     &0.27            &0.38                                              &0.13           \\ %&0.12     $\mid$0.14
NGC\,2440  &0.91               &1.21                            &0.73                                     &*            &0.99                                              &0.59$\mid$1.09$^a$      \\ %&(.30)    $\mid$1.09
NGC\,3132  &0.42               &0.42                            &0.29                                     &0.30         &0.51                                              &0.36$\mid$0.43        \\ %&0.37     $\mid$0.43
NGC\,3242  &0.27               &0.20                            &0.08                                     &0.13         &0.29                                              &0.10$\mid$0.26      \\ %&0.08     $\mid$0.26
NGC\,3918  &0.26               &0.27                            &0.22                                     &0.22         &0.21                                              &0.23           \\ %&0.24     $\mid$0.25
NGC\,5315  &0.40               &0.32                            &0.45                                     &0.56         &0.47                                              &0.54$\mid$0.42        \\ %&0.55     $\mid$0.42
NGC\,5882  &0.23               &0.25                            &0.23                                     &0.22         &0.29                                              &0.33           \\ %&0.25     $\mid$0.34
NGC\,6302  &1.72               &1.10                            &1.12                                     &3.16         &1.53                                              &1.34           \\ %&(.39)    $\mid$1.30
NGC\,6818  &0.58               &0.22                            &0.14                                     &0.34         &0.58                                              &0.16$\mid$0.21 \\ %&0.20     $\mid$0.21
IC\,4191   &0.39               &*                               &*                                        &0.16         &0.56                                              &0.06$\mid$0.17      \\ %& *       $\mid$0.17
IC\,4406   &0.27               &0.28                            &0.29                                     &0.23         &0.34                                              &0.36$\mid$0.34      \\ %&0.10     $\mid$0.34
My\,Cn\,18 &3.16               &*                               &*                                        &*            &$\geq$2.0                                        &0.39 \\           %& *
\noalign{\vskip3pt}
SMC\,N87   &*                  &*                               &*                                        &*            & *                                                &0.33   \\      %& *       --0.44
LMC\,N66   &*                  &0.55                            &0.34                                     &*            & *                                                &0.41 \\           %& *
LMC\,N141  &*                  &*                               &*                                        &*            &*                                                 &0.45 \\      %&*        --0.83
\noalign{\vskip2pt}
Solar$^b$  &*                   & *                             &*                                        &*            &*                                      &0.19   \\
\noalign{\vskip3pt} \noalign{\hrule} \noalign{\vskip3pt}
\end{tabular}
\begin{description}
\item[$^a$] See Section 3.3 for explanation when two values are listed for the CEL N/O ratio;
\item[$^b$] Solar oxygen abundance taken from Allende Prieto et al. (2001); solar nitrogen abundance from Grevesse, Noels and Sauval (1996).
\end{description}
\end{center}
\end{table*}

\section{Correlations}

\subsection{ORL/CEL ADF's versus {\elt} discrepancies}

The O$^{2+}$ ORL/CEL abundance discrepancy factor is known to be strongly
correlated with the difference between the [O~{\sc iii}] forbidden line
and H~{\sc i} Balmer jump electron temperatures (Liu et al. 2001b). In
Fig.\,~3~(top) we plot the ratio of {\cpp}/{\hp} ionic abundances derived
from the {\cii} $\lambda$4267 ORL to those from the C~{\sc iii}]
$\lambda$1908 CEL versus the difference between the {\foiii} nebular to
auroral forbidden line temperature and the Balmer jump temperature, for a
sample of 17 planetary nebulae. 

\noindent{We see that a strong, positive linear correlation exists
between}\\

\noindent{ADF(C$^{2+}$/H$^+$) $\equiv$ log(C$^{2+}$/H$^+$)$_{\rm ORL}$ $-$
log(C$^{2+}$/H$^+$)$_{\lambda1908\,\rm CEL}$,}

\noindent{and}\\
\noindent{$\Delta T$ $\equiv$ {\elt}(\foiii) $-$ {\elt}(BJ).
\\

\noindent{A linear fit to the 17 PNe plotted in Fig.\,~3 (top) yields,}
\begin{equation} \label{eq:c2dT}
{\rm ADF}(\frac{\rm C^{2+}}{\rm H^+}) =
(0.419\,\pm\,0.051)\,+\,(14.6\,\pm\,1.8)\,\times\,10^{-5}\,\Delta \emph{T},
\end{equation}
\noindent{with a linear correlation coefficient of 0.91.}

\setcounter{figure}{2}
\begin{figure}
\begin{center} \epsfig{file=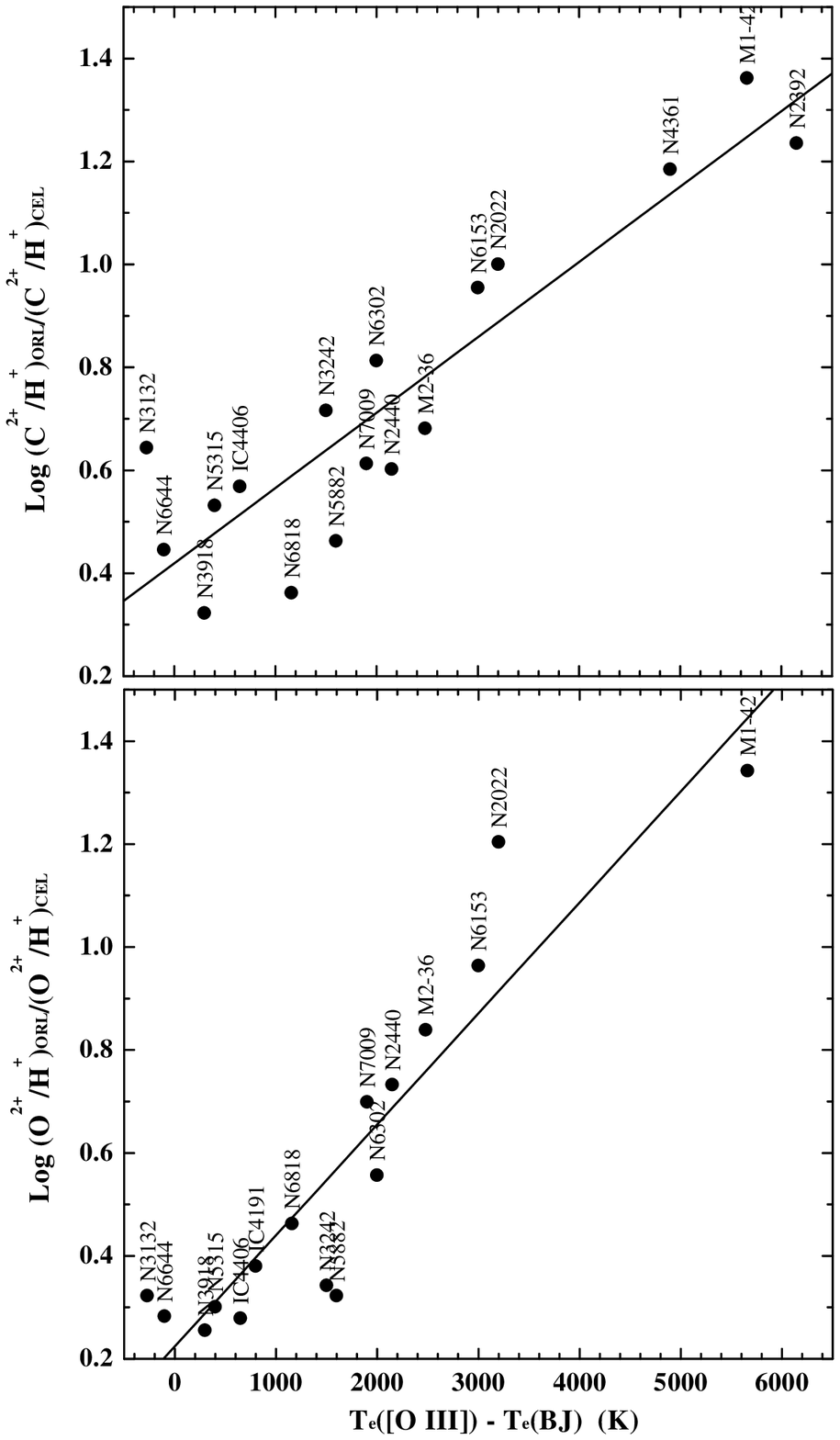, width=11.5 cm, bbllx=90pt,
bblly=190pt, bburx=581pt, bbury=829pt, clip=, angle=0} \caption {
(\emph{top}): ORL/CEL ionic abundance discrepancy factors for {\cpp} plotted
against the difference between the {\foiii} and BJ electron temperatures;
(\emph{bottom}): the same for the {\opp} ratios; the solid lines are the
linear fits of Eqs.~1 and~2; see text for details.}
\end{center}
\end{figure}

The data include those for seven PNe from other studies:  NGC\,7009
(LSBC), NGC\,4361 (Liu 1998; who quotes {\cpp}($\lambda$1908)/{\hp} values
from Torres-Peimbert, Peimbert \& Pe\~{n}a 1990), NGC\,6153 (Liu et al.
2000), M\,1-42 and M\,2-36 (Liu et al. 2001b), NGC\,6644 (our unpublished
observations), and NGC\,2392 [ADF({\cpp}) from Barker 1991;  {\elt}(BJ)
from Liu \& Danziger (1993b)]. The electron temperatures for the PNe in
the current sample are taken from Table~7 of Paper~I.

Fig.\,~3~(bottom) shows that a very similar linear correlation exists between
ADF({\opp}/{\hp}) $\equiv$ log(O$^{2+}$/H$^+$)$_{\rm ORL}$ $-$
log(O$^{2+}$/H$^+$)$_{\lambda4959+5007\,\rm CEL}$ and $\Delta T$ too (cf. Liu
et al. 2001b for a similar fit to a subset of the nebulae discussed here). A
linear fit to the sixteen planetary nebulae plotted in the bottom panel of
Fig.\,~3, yields,

\begin{equation} \label{eq:o2dT}
{\rm ADF}(\frac{\rm O^{2+}}{\rm H^+}) =
(0.224\,\pm\,0.053)\,+\,(21.6\,\pm\,2.4)\,\times\,10^{-5}\,\Delta \emph{T},
\end{equation}
\noindent{with a linear correlation coefficient of 0.92.}

The correlations of the abundance discrepancy factors for doubly ionized C and
O with the difference between the {\foiii} and Balmer jump nebular
temperatures are significant, since they provide a strong observational
indication that the nebular thermal structure is intimately tied to the
problem of discordant---ORL versus CEL---abundance determinations.

\subsection{Abundance discrepancy factors versus PN intrinsic surface
brightnesses}

Nebular surface brightness can be considered as an evolutionary parameter,
since for an expanding nebula it decreases as the nebula ages and the mean
density drops. Garnett \& Dinerstein (2001a) found that for a sample of about
a dozen PNe the magnitude of the ADF({\opp}/{\hp}) discrepancy factor was
inversely correlated with the mean nebular Balmer line surface brightness,
`suggesting that the abundance problem is a function of PN evolution'. In
Fig.\,~4 we have plotted ADF({\opp}/{\hp}) and ADF({\cpp}/{\hp}) values, taken
from the same sources as used for Fig.\,~3 (described in Section~4.1), against
mean nebular {\Hb} surface brightnesses, for samples of 21 and 20 PNe,
respectively. Here the surface brightness, $S$({\Hb}), is defined as the flux
received per square arcsec of the nebula corrected for interstellar
extinction. For all galactic PNe we use nebular angular radii, {\Hb}
integrated fluxes and logarithmic extinction coefficients,
\emph{c}(H$\beta$)$^\mathrm{{rad}}$, from Cahn, Kaler \& Stanghellini (1992;
CKS92); for the Cloud PNe integrated fluxes are from Meatheringham et al.
(1988); for LMC N66 we adopt an angular radius of 1.5 arcsec (Dopita et al.
1993), for LMC N141 we adopt 0.30 arcsec (Shaw et al. 2001), while for SMC~N87
we adopt 0.23 arcsec (Stanghellini et al. 2003). The ORL/CEL abundance 
ratios for NGC\,6572 were derived from our unpublished ESO 1.52-m 
observations.

\setcounter{figure}{3}
\begin{figure}
\begin{center} \epsfig{file=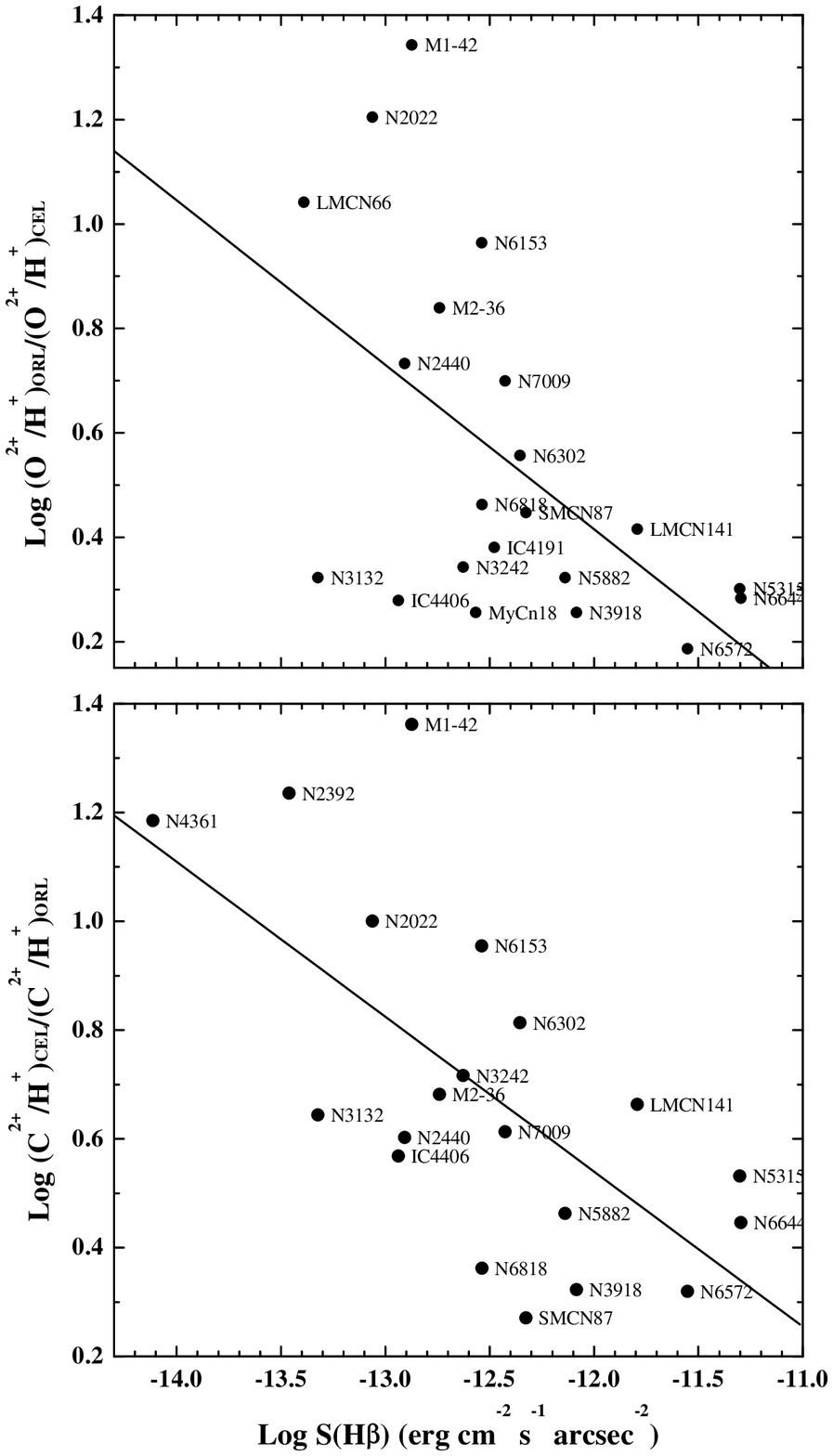, width=11.5 cm, bbllx=90pt,
bblly=190pt, bburx=581pt, bbury=829pt, clip=, angle=0} \caption { {\opp}/{\hp}
(\emph{top}) and {\cpp}/{\hp} (\emph{bottom}) abundance discrepancy factors,
plotted against nebular {\Hb} surface brightness. The solid lines are the
linear fits discussed in the text.}
\end{center}
\end{figure}

From Fig.\,~4, we confirm that the abundance discrepancy factors are
inversely correlated with the nebular surface brightness. A linear fit to the
21 PNe plotted in Fig.\,~4(top) yields,
\begin{equation} \label{eq:o2Sb}
{\rm ADF}(\frac{\rm O^{2+}}{\rm H^+}) =
(-3.52\,\pm\,1.41)\,-\,(0.327\,\pm\,0.113)\,\times\,{\rm log}\,S({\rm
H}\beta),
\end{equation}
\noindent{with a linear correlation coefficient of $-$0.55; while a linear fit
to the 20 PNe plotted in Fig.\,~4(bottom) yields, }
\begin{equation} \label{eq:c2Sb}
{\rm ADF}(\frac{\rm C^{2+}}{\rm H^+}) =
(-2.88\,\pm\,1.00)\,-\,(0.285\,\pm\,0.080)\,\times\,{\rm log}\,S({\rm
H}\beta),
\end{equation}
\noindent{with a linear correlation coefficient of $-$0.64, i.e. the fit and
correlation coefficient is not very strong for ADF({\opp}/{\hp}) (Fig.\,~4,
upper), being quite better for ADF({\cpp}/{\hp}) (Fig.\,~4, lower).}

\subsection{Abundance discrepancy factors versus PN absolute radii}

\setcounter{figure}{4}
\begin{figure}
\begin{center} \epsfig{file=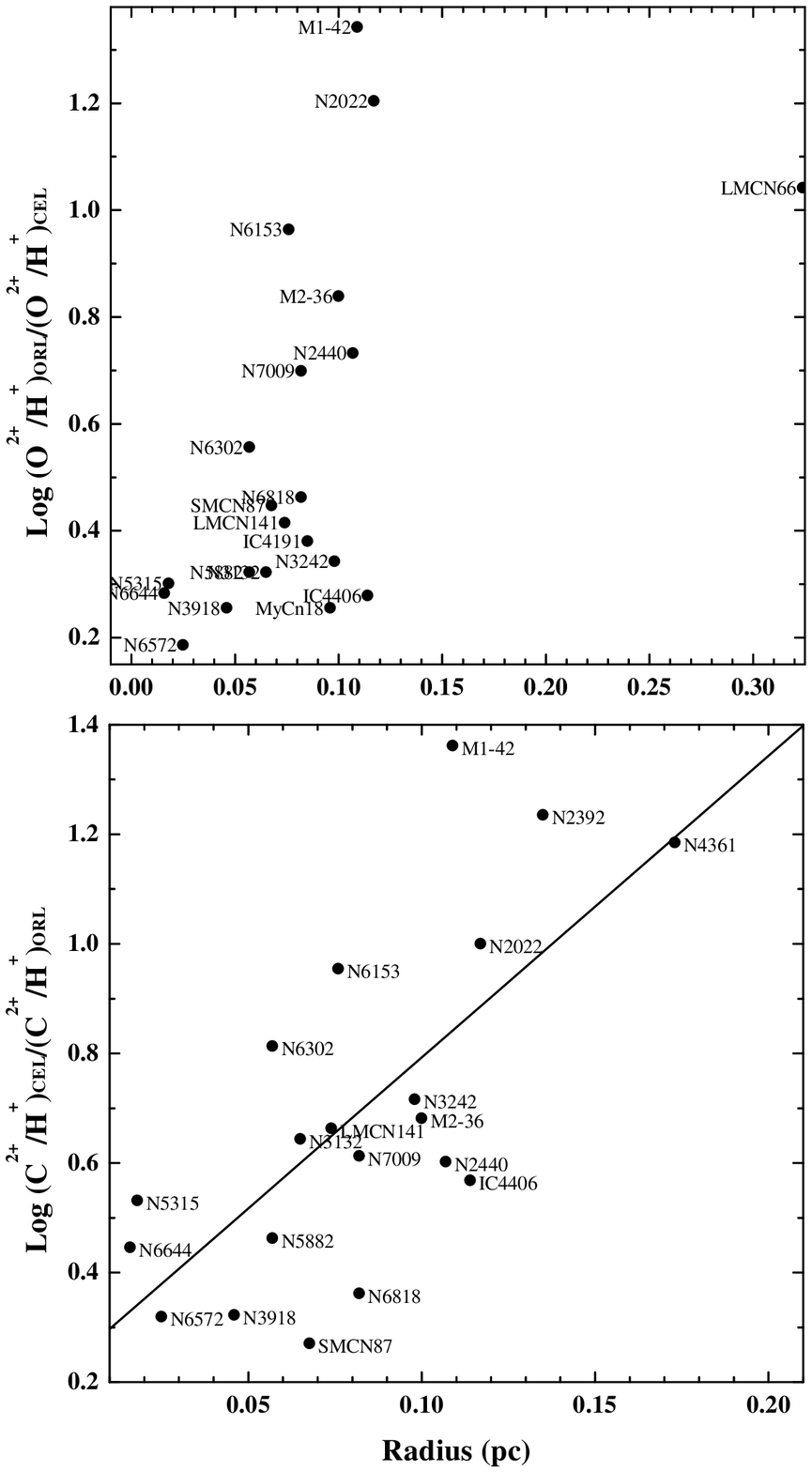, width=11.5 cm, bbllx=90pt,
bblly=190pt, bburx=581pt, bbury=829pt, clip=, angle=0} \caption{ 
{\opp}/{\hp}
(\emph{top}) and {\cpp}/{\hp} (\emph{bottom}) abundance discrepancy 
factors,
plotted against nebular radius. The solid line in the lower plot is the 
linear
fit discussed in the text.}
\end{center}
\end{figure}

Fig.\,~5 plots the ADF({\opp}/{\hp}) and ADF({\cpp}/{\hp}) abundance
discrepancy factors versus nebular absolute radii. For all galactic PNe we
have used the absolute radii quoted by CKS92, except for NGC\,3132 where
the radius was derived assuming a distance to the nebula of 600\,pc (Sahu
\& Desai 1986) and an angular radius of 22.5\,arcsec (CKS92). For the
Cloud PNe we adopted the following: for LMC N66 a radius of 0.324\,pc
(Dopita et al. 1993); for LMC N141 a radius of 0.074\,pc, adopting a
nebular angular radius of 0.30\,arcsec (Shaw et al. 2001) and assuming the
same distance to the LMC (50.6\,kpc) as adopted by Dopita et al. (1996);
finally for SMC~N87 a radius of 0.067\,pc, based on the (photometric)  
optical {\it HST} diametre of 0.45 arcsec (Stanghellini et al. 2003) and
assuming a distance to the SMC of 60\,kpc.

We see that a positive linear correlation exists between ADF({\cpp}/{\hp}) and
the absolute nebular radius for 20 PNe [Fig.\,~5(bottom)], which can be fitted
by,
\begin{equation} \label{eq:c2rad}
{\rm ADF}(\frac{\rm C^{2+}}{\rm H^+}) =
(0.242\,\pm\,0.121)\,+\,(5.51\,\pm\,1.35)\,\times\,\emph{R},
\end{equation}
\noindent{with a linear correlation coefficient of 0.69.}

No such correlation is obvious regarding ADF({\opp}/{\hp}) [Fig.\,~5(top)].

\subsection{Temperature differences versus PN radii and surface
brightnesses}

So far we saw that abundance discrepancy factors are: i) positively correlated
with the difference between the {\foiii} forbidden line and BJ temperatures,
ii) weakly correlated with decreasing intrinsic nebular surface brightness,
and iii) positively correlated with the absolute PN radii [for the case of
ADF({\cpp}/{\hp}) mostly]. Therefore it is not surprising to find that the
aforementioned temperature difference is also inversely correlated with
$S$({\Hb}) and positively correlated with the absolute nebular radii
(Fig.\,~6).

A linear fit to the 25 nebulae plotted in Fig.\,~6~(left) yields,
\begin{equation} \label{eq:dTSb}
\Delta \emph{T} =
(-17.6\,\pm\,6.1)\,\times\,10^3\,-\,(1517\,\pm\,481)\,\times\,{\rm log}\,S({\rm
H}\beta),
\end{equation}
\noindent{with a linear correlation coefficient of $-$0.55.

The relation between $\Delta T$ and the nebular radii for the same 25 objects
[Fig.\,~6~(right)] can be fitted by,
\begin{equation} \label{eq:dTrad}
\Delta \emph{T} =
(-853\,\pm\,744)\,+\,(28.1\,\pm\,7.6)\,\times\,10^3~~\emph{R},
\end{equation}
\noindent{which has a linear correlation coefficient of 0.61.}

In Fig.\,~6, apart from the 11 nebulae whose BJ temperatures were measured
in the context of this study, we also included data for 6 objects
published previously (see Section~4.1 for details on these), along with
data for a further 8 nebulae whose BJ and {\foiii} temperatures were
presented by Liu \& Danziger (1993b).

\setcounter{figure}{5}
\begin{figure*}
\begin{center} \epsfig{file=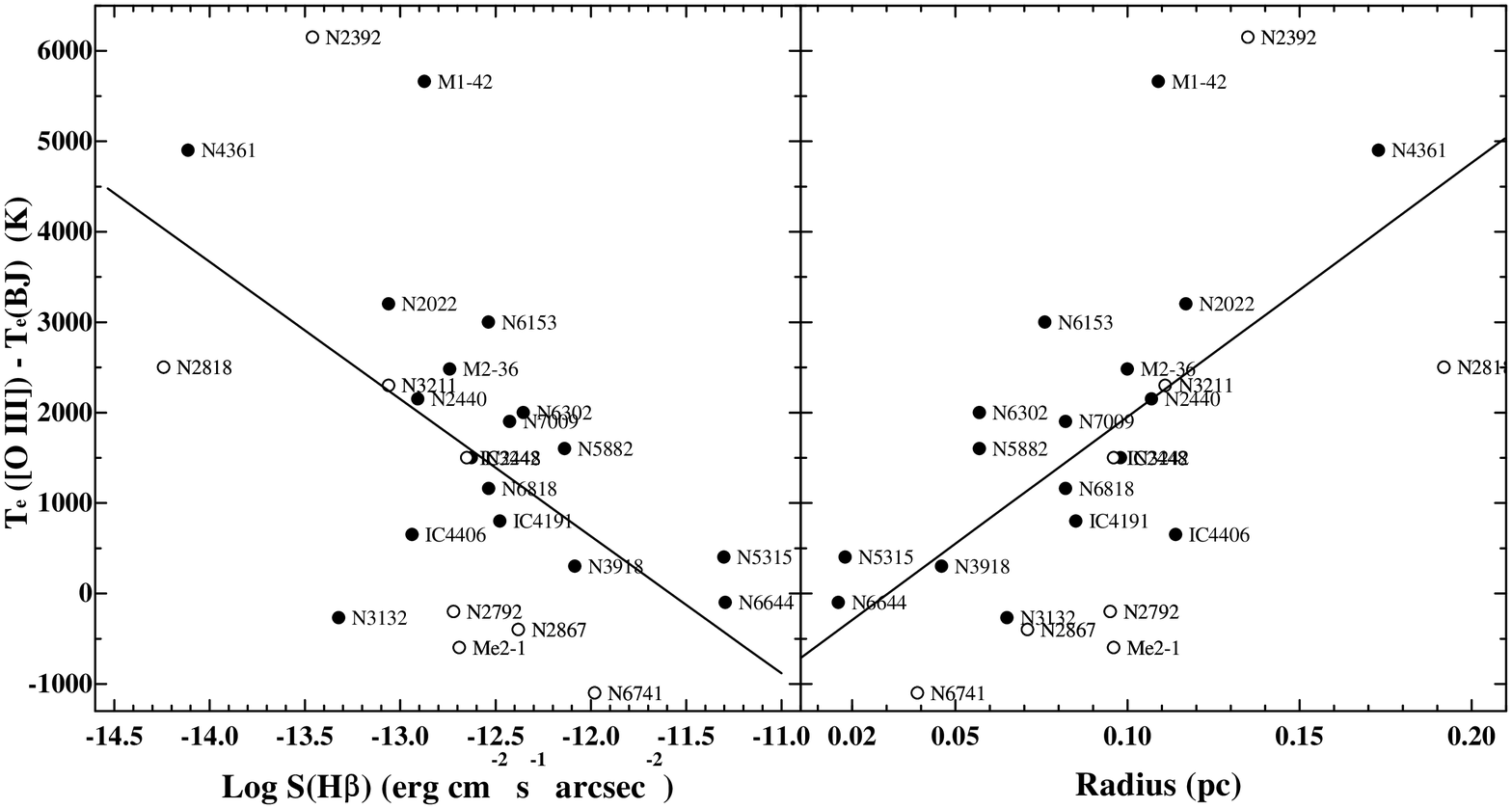, width=16.5 cm, bbllx=14pt, 
bblly=90pt, bburx=850pt, bbury=550pt, clip=}
\caption{The difference between the nebular electron temperatures derived
from the {\foiii} nebular to auroral forbidden line ratio,
{\elt}({\foiii}), and from the nebular Balmer continuum discontinuity,
{\elt}(BJ), is plotted against the intrinsic nebular {\Hb} surface
brightness (\emph{left}) and absolute radius (\emph{right}). Open circles
data are from LD93b. The solid lines are the linear fits discussed in the
text.}
\end{center}
\end{figure*}

\section{The effects of nebular density and temperature variations}

\subsection{Density inhomogeneities}

In Paper~I nebular electron densities derived from various diagnostic ratios
were presented. It was shown that the low critical density {\foiii} 52- and
88-$\mu$m lines yield electron densities that are, on average, a factor of 6
lower than those derived from the optical {\fariv} and {\fcliii} doublets,
which have much higher critical densities. Given that the {\foiii, {\fariv}
and {\fcliii} lines arise from regions of similar ionization
degree,\footnote{The ionization potential of {\op} (= 35.1\,eV), falls
between those of Cl$^+$ (= 23.8\,eV) and Ar$^{2+}$ (= 40.7\,eV).}~this result
points to the presence of strong density variations within the nebulae. The
effects of such variations on the {\opp}/{\hp} and {\npp}/{\hp} abundance
ratios derived from the far-IR lines were discussed in Paper~I in the context
of Rubin's (1989) examination of the effects of varying {\eld} on nebular
abundances; the general conclusion from Rubin's work is that when there are
variable density conditions, the intensity ratios of forbidden lines with low
critical densities relative to {\Hb} (or any recombination line with a volume
emissivity $\propto$\,{\eld}$^2$), will generally underestimate the true
ionic abundances. We saw that in accord with these theoretical predictions,
abundances from far-IR lines \emph{are}, in general, underestimated relative
to UV and optical values (that are subject to smaller bias due to their
higher {\crd}), when the `low' {\foiii} 52$\mu$m/88$\mu$m densities are
adopted for the calculations. On the contrary, we saw that when higher
{\eld}'s were adopted (intermediate or equal to the {\fariv} and {\fcliii}
densities), the inferred abundances from the far-IR lines become compatible
with those derived from the UV and optical indicators (see Fig.\,~1 of this
paper and Fig.\,~3 of Paper~I); with the exception of the high critical
density {\fneiii} 15.5-$\mu$m line which is not affected by such modest
density variations.

In all cases, however, and even after accounting for modest density
variations, the CEL abundances derived from UV, optical or IR lines remain
consistently lower than the ORL values (Fig.\,~1). Viegas \& Clegg (1994)
showed that dense clumps with {\eld}\,$>$\,10$^6$\,{\cmt} can have a
substantial effect on the derived {\foiii} forbidden-line temperature, via
collisional suppression of the nebular $\lambda\lambda$4959, 5007 lines; in
this way the observationally derived {\elt}({\foiii}) could be significantly
overestimated and CEL abundances accordingly underestimated. In their analysis
of the extreme nebula NGC\,6153, Liu et al. (2000) discussed the case for the
existence of dense condensations with a very small filling factor as a
potential explanation for the large ORL/CEL abundance discrepancies. They
concluded that such dense clumps would have a different effect on different
CEL lines. In particular, they found that in the increased density environment
of the condensations, most IR and optical CELs would be collisionally
suppressed, accounting for the abundance ratios, thus covering \emph{part} of
the distance from the ORL values; on the other hand however, the abundances
derived from the high critical density UV CELs, using the downward corrected
electron temperatures that result from allowing for the collisional
suppression of the $\lambda\lambda$4959, 5007 lines, would become so enhanced
as to \emph{exceed} the ORL abundance results. The conclusion in this case is
that the existence of dense condensations is not in itself a sufficient
solution to the problem, since it requires the discrepancies to be correlated
with the critical densities of the various CELs, something not indicated by
the observations.

\subsection{Temperature fluctuations}

Peimbert (1967) first proposed that in the
presence of temperature fluctuations within nebulae, adoption of the
{\foiii} ($\lambda$4659 + $\lambda$5007)/$\lambda$4363 line ratio as a
standard thermometer would result in the underestimation of elemental
abundances as derived from CELs, since that line ratio would be biased
towards high-temperature nebular regions. Generally speaking, if
variations in electron temperature exist along the line of sight through a
nebula, abundance ratios derived from the ratio of a collisionally excited
line to a recombination line (e.g. the most commonly used recombination
line, {\Hb}) will be underestimated, while those derived from a pure
recombination line ratio should be almost insensitive to {\elt} and thus
largely unaffected by any incorrect assumptions regarding the nebular
temperature distribution.

Peimbert (1971) found possible evidence for the existence of such
temperature fluctuations, by comparing for three PNe the temperature
derived from the ratio of the Balmer continuum jump to the intensity of
{\Hb} with that obtained from the {\foiii} nebular to auroral CEL ratio;
he found that the latter temperatures were higher than {\elt}(BJ). This
result has been supported by further observations of more planetary
nebulae (e.g. Liu \& Danziger 1993b). The Balmer jump temperatures that we
presented for 12 PNe in Table~7 of Paper~I were also systematically lower
than the corresponding {\elt}(\foiii) values presented there.

The concept of temperature fluctuations in a chemically homogeneous nebula has
been invoked several times as a promising cause of the ORL/CEL abundance
discrepancy problem, pertaining to both discrepant {\cpp}($\lambda$4267)/{\hp}
and {\opp}/{\hp} ORL abundance ratios. For instance, Peimbert et al. (1993)
proposed that spatial temperature fluctuations with an amplitude of
$\sim$\,20\,per cent around a mean value would reconcile the factor of 1.55
discrepancy between the ORL and CEL {\opp}/{\hp} abundances in the planetary
nebula NGC\,6572. Peimbert, Luridiana \& Torres-Peimbert (1995) derived
electron temperatures for a sample of nebulae from the C~{\sc iii}]
$\lambda$1908/{\cii} $\lambda$4267 ratio and showed that they are generally
lower than {\elt}(\foiii). On account of this they argued that in order to
explain the discrepancy between the
{\cpp}($\lambda$4267)/{\opp}($\lambda$5007) and
{\cpp}($\lambda$1908)/{\opp}($\lambda$5007) values in planetary nebulae
published by Rola \& Stasi\'{n}ska (1994), large temperature fluctuations were
needed; they also recommended the adoption of {\elt}(\cpp) rather than
{\elt}(\foiii) when deriving ionic abundances from CELs and concluded that due
to their insensitivity to temperature variations, abundances derived from pure
recombination line ratios are more reliable.

However, the validity of the C~{\sc iii}] $\lambda$1908/{\cii}
$\lambda$4267 ratio (or even that of the C~{\sc iv} $\lambda$1549/{\ciii}
$\lambda$4650 ratio) as a reliable nebular thermometer is ambiguous, since
there is evidence that, at for PNe which exhibit non-uniform,
inhomogeneous abundances, the {\cii} $\lambda$4267 ORL and C~{\sc iii}]
$\lambda$1908 CEL do not originate in the same gaseous component.
Harrington \& Feibelman (1984) argued that such is the case for the
hydrogen-deficient material which exists at the centre of the planetary
nebula A\,30 in the form of condensations. Subsequent, detailed
photoionization modelling of {\it HST} and ground-based spectra of several
of the knots in A\,30 (Borkowski et al. 1993, Ercolano et al. 2003) has
confirmed that the ORLs and CELs originate from different regions of the
knots.

We have investigated our results for evidence that temperature
fluctuations may contribute to the observed abundance discrepancies. The
major premise of this scenario is that ionic abundances derived from the
intensity ratio of two lines with very different temperature
dependencies---e.g. $I$({\foiii} $\lambda$1663)/$I$({\Hb})---will be
severely biased if an incorrect {\elt} is adopted in calculating line
emissivities. Since, on average, {\elt}(BJ)  $\leq$ {\elt}(\foiii), and
since the standard practice is to use the {\foiii} optical line ratio
temperature in order to calculate both the emissivities of CELs and of
{\Hb}, the possibility of systematically underestimating CEL abundances is
there; should then ORL abundances that are insensitive to temperature bias
be trusted more and if so why do they point towards heavy element
overabundances in most nebulae?  One way to try to answer this question is
to examine the effects of changing the adopted {\elt} on ionic abundances
derived from CELs that have different temperature dependencies.

In the current study the standard electron temperature from the {\foiii}
($\lambda$4659 + $\lambda$5007)/$\lambda$4363 line ratio was adopted in order
to derive all ionic abundances of doubly ionized species. If we suppose
however that the {\elt} pertinent to the nebular zones where doubly ionized
species exist is lower than that, e.g. that it may be as low as the H\,{\sc i}
BJ temperature, then the ionic abundances derived from CELs would all be
underestimated by our adoption of {\elt} = {\elt}(\foiii), but not all by the
same factor. Abundances derived from high excitation energy transitions, such
as O~{\sc iii}] $\lambda$1663 ({\exe}\,$\simeq$\,86\,kK), N~{\sc iii}]
$\lambda$1751 ({\exe}\,$\simeq$\,82\,kK), or C~{\sc iii}] $\lambda$1908
({\exe}\,$\simeq$\,75\,kK), would be underestimated more relative to those
derived from transitions that have a lower temperature sensitivity, such as
the {\foiii} $\lambda\lambda$4959, 5007 nebular lines
({\exe}\,$\simeq$\,29\,kK). Our results show however (Table~9), that all CEL
{\cpp}, {\npp} and {\opp} abundances are systematically lower than those
derived from ORLs, by rather similar factors.

For instance, the \emph{average} {\cpp} $\lambda$4267/$\lambda$1908 ADF for
eighteen nebulae is 5.4, while the {\opp}(ORL/$\lambda$4959 + $\lambda$5007)
ADF is 5.1; also the {\opp} ORL/$\lambda$1663 ADF for ten nebulae is 5.7
(Table~9). This apparent uniformity cannot be reconciled with the different
behaviour that one would expect amongst the discrepancy ratios in the case of
temperature fluctuations of the type envisaged by Peimbert. We would expect
the {\cpp} ORL/CEL ADF's to be greater than that for {\opp}, since the 75~kK
excitation energy of the upper levels of the $\lambda$1908 doublet exceeds the
62~kK excitation energy of the upper level of the $\lambda$4363 transition.
However, in the extreme nebula NGC\,2022 the ADF for {\cpp} is 10, but that
for {\opp} is 16, i.e. the opposite of that expected; in the other extreme
nebulae NGC\,6153 (Liu et al. 2000) and M\,1-42 (Liu et al. 2001b) the
corresponding factors are almost equal. Similarly, we would expect the {\cppp}
ORL/CEL ADF to be greater than that for {\cpp}, due to the higher excitation
energy of the C~{\sc iv} $\lambda$1549 resonant doublet
({\exe}\,$\simeq$\,93\,kK), relative to that of C~{\sc iii}] $\lambda$1908
({\exe}\,$\simeq$\,75\,kK); however, this is not the case for the nine PN with
both {\cpp} $\lambda$4267/$\lambda$1908 and {\cppp} ORL/$\lambda$1549 ADF's
documented, where on average, these abundance discrepancy factors are 5.3 and
3.9, respectively.

Having just seen that in the presence of temperature fluctuations we would
expect ADF({\cpp}/{\hp}), as defined in the previous section, to be affected
more by errors in {\elt} than in ADF({\opp}/{\hp}), our fits (Eqs.~1 and 2)
show that actually it may well be that the opposite is true, judging from the
slightly steeper ADF({\opp}/{\hp}) function (Fig.\,~3). This behaviour of the
two discrepancy factors conflicts with that expected for temperature
fluctuations of the standard type.

Probably, however, the most serious obstacle to the classical temperature
fluctuation hypothesis lies with the IR abundance results. Inspection of
Fig.\,~1 and Table~9 shows that the ORL/CEL ADF's for CEL abundances obtained
from the far-IR {\foiii} and {\fniii} lines are in line with those from the
optical and UV CELs, i.e. there is no correlation whatsoever with the very low
{\exe} ($\lesssim$\,1000\,K) of IR transitions. As we mentioned in
Section~3.2, very satisfactory agreement is reached amongst UV, optical and IR
CEL {\opp}/{\hp} ADF's, while the same is also true for {\npp}/{\hp} and
{\nepp}/{\hp}; this would not be the case in the presence of strong
temperature fluctuations in a chemically homogeneous nebula.

Ruiz et al. (2003) and Peimbert et al. (2004) have presented deep optical
echelle spectra of NGC\,5307 and NGC\,5315, respectively, deriving O$^{2+}$
ADF's of 1.9 and 1.7 for these two PNe (we obtained an optical O$^{2+}$ ADF of
2.0 for NGC\,5315; see Table~9). They argued that these ADF's could be
explained by classical temperature fluctuations, corresponding to {\it t}$^2$
values of 0.051 and 0.056, respectively. NGC\,5307 has no measurements
available for its far-IR FS lines, but {\it ISO} spectra exist for NGC\,5315.
Peimbert et al. (2004) chose not to use the the 52- and 88-$\mu$m [O~{\sc
iii}] {\it ISO} LWS line fluxes presented by Liu et al. (2001a), on the
grounds that the electron density of 2290~cm$^{-3}$ yielded by the intensity
ratio of these two lines lies between the respective critical densities of
3800 and 1800~cm$^{-3}$. However in Paper~I we found that for NGC\,5315 (and
for all the other PNe in our sample with electron densities exceeding one or
both of these critical densities), the adoption of an electron density given
by the mean of the densities from the [Cl~{\sc iii}] and [Ar~{\sc iv}] optical
doublet ratios gives an O$^{2+}$ abundance from the far-IR fine-structure
lines that agrees with the value obtained from the optical forbidden lines.

With regard to this issue, two PNe in our current sample, NGC\,3132
[ADF(O$^{2+}$) = 2.4] and IC\,4406 [ADF(O$^{2+}$) = 1.9], are of key
importance, since their far-IR [O~{\sc iii}] doublet ratio densities of 355
and 540~cm$^{-3}$ (Paper~I) are well below the critical densities of the two
far-IR lines (as are the electron densities derived for these two nebulae from
the various optical doublet ratios; see Table~6 of Paper~I). Thus there should
be no objection to the derivation of O$^{2+}$ abundances from the far-IR
[O~{\sc iii}] lines for these nebulae. Inspection of Table~9 of Paper~I shows
that for NGC\,3132 the O$^{2+}$ abundance derived for the far-IR FS lines,
using the far-IR doublet ratio density, differs by less than 14\% from the
O$^{2+}$ abundance derived from the optical forbidden lines, while for IC~4406
they differ by less than 3\%.

Expressing this another way, the electron temperature of 9900\,K derived for
both NGC\,3132 and IC\,4406 from the ratio of their IR to optical [O~{\sc
iii}] doublet fluxes agrees with the temperatures of 9530\,K and 10000\,K
derived from their [O~{\sc iii}] $\lambda$4363/($\lambda$5007+$\lambda$4959)
ratios (Table~7 of Paper~I). If the rather typical ADF's of $\sim$2 derived
for these two PNe were caused by temperature fluctuations in a chemically
homogeneous medium, then the IR to optical [O~{\sc iii}] doublet flux ratio
should yield significantly lower temperatures than the much more
temperature-sensitive $\lambda$4363/($\lambda$5007+$\lambda$4959) ratio. This
is not the case, so we conclude that classical temperature fluctuations cannot
explain the ADF's of $\sim$2 found for those two nebulae.

\section{Evidence for cold plasma from ORL electron temperatures}

\setcounter{table}{11}
\begin{table*}
\centering
\begin{minipage}{150mm}
\caption{Comparison of CEL and ORL electron temperatures (in K).}
\begin{tabular}{lrccrcrccc}  \\
\noalign{\vskip3pt} \noalign{\hrule}\noalign{\vskip3pt}

\multicolumn{1}{l}{Nebula}& \multicolumn{1}{c}{$N_e$}&
\multicolumn{1}{c}{$T_e$([O~{\sc iii}])}& \multicolumn{1}{c}{$T_e$(BJ)}&
\multicolumn{1}{l}{$I$($\lambda$6678)}& \multicolumn{1}{c}{$T_e$(He~{\sc i})}&
\multicolumn{1}{l}{$I$($\lambda$4649)}&
\multicolumn{1}{r}{$I$($\lambda$4089)}& \multicolumn{1}{c}{$T_e$(O~{\sc ii})}&
%\multicolumn{1}{l}{I(V1)}&
%\multicolumn{1}{c}{$T_e$(O~{\sc ii})}&
\multicolumn{1}{c}{ADF(O$^{2+}$)}\\

\multicolumn{1}{l}{}& \multicolumn{1}{l}{(cm$^{-3}$)}&
\multicolumn{1}{c}{CEL}& \multicolumn{1}{c}{BJ/H11}&
\multicolumn{1}{l}{/$I$($\lambda$4471)}&
\multicolumn{1}{c}{$\lambda$6678/$\lambda$4471}& \multicolumn{1}{c}{O~{\sc
ii}}& \multicolumn{1}{l}{/$I$(4649)}&
\multicolumn{1}{c}{$\lambda$4089/$\lambda$4649}&
%\multicolumn{1}{l}{/4089}&
%\multicolumn{1}{c}{V1/4089}&
\multicolumn{1}{c}{ORL/CEL}\\
\cline{1-10} \\\\
NGC 2022   &  1500 & 15000 & 13200 & 0.693 & 15900 & 0.333 & 0.607 & $<$300  & 16  \\
NGC 2440   &  6000 & 16150 & 14000 & --    &  --   & 0.100 & 0.420 & $<$300  & 5.4 \\
NGC 3132   &   600 & 9530  & 10800 & 0.725 & 12800 & 0.195 & --   &  --     & 2.4 \\
NGC 3242   &  2000 & 11700 & 10200 & 0.774 & 10000 & 0.216 & 0.307$^c$ & 2600 & 2.2 \\
NGC 3918   &  5000 & 12600 & 12300 & 0.740 & 12000 & 0.211 & 0.288 & 3650    & 1.8 \\
NGC 5315   & 10000 &  9000 &  8600 & 0.775 & 10000 & 0.344 & 0.261 & 5750    & 2.0 \\
NGC 5882   &  4000 &  9400 &  7800 & 0.763 & 10700 & 0.362 & 0.230$^c$ & 8700 & 2.1 \\
NGC 6302   & 14000 & 18400 & 16400 & 0.671 & 15100 & 0.100 &  --   & --      & 3.6 \\
NGC 6818   &  2000 & 13300 & 12100 & 0.841 &  5000 & 0.182 & 0.300$^c$ & 2900 & 2.4 \\
IC 4191$^a$ & 10000 & 10700 & 10500 & 0.908 & 3100 & 0.725 & 0.284 & 3900    & 2.4 \\
IC 4191$^b$ & 10000 & 10000 &  9200 & 0.916 & 2900 & 0.572 & 0.316 & 2150    & 2.4 \\
IC 4406    &  1000 & 10000 &  9350 & 0.797 &  8000 & 0.284 &  --   &  --     & 1.9 \\

%\cline{1-11}

\noalign{\vskip3pt} \noalign{\hrule} \noalign{\vskip3pt}
\end{tabular}
\begin{description}
\item[$^a$] Fixed-slit observation;
\item[$^b$] Scanning-slit observation;
\item[$^c$] Adopted O~{\sc ii} 4089.29~\AA\ intensity is after correction
for Si~{\sc iv} 4088.85~\AA, using the measured intensity of Si~{\sc iv}
$\lambda$4116.10~\AA\ and adopting a 2:1 flux ratio for the
4s$^2$S--4p$^2$P$^{\rm o}$ Si~{\sc iv} $\lambda$4089/$\lambda$4116 doublet
components.
\end{description}
\end{minipage}
\end{table*}
\normalsize

\subsection{Electron temperatures from He~{\sc i} ORL ratios}

Liu (2003) used the weak temperature sensitivity of several He~{\sc i}
recombination lines to derive He$^+$ electron temperatures for five PNe with
particularly high O$^{2+}$ ADF's. He found that both the value of
{\elt}(He~{\sc i}) and the ratio of {\elt}(He~{\sc i})/{\elt}([O~\sc iii}])
decreased systematically with increasing nebular ADF, with {\elt}(He~{\sc i})
ranging from 5380~K for NGC~7009 (ADF = 4.7) down to 2310~K for M~1-42 (ADF =
22) and as low as 775~K for Hf~2-2 (ADF = 84). This provided observational
support for the presence of regions of cold plasma within these high-ADF
nebulae.

It is of interest to investigate what patterns might be shown by the lower-ADF
PNe in the current sample, so we have derived He~{\sc i} electron temperatures
using the curves presented by Liu (2003; based on the data of Smits 1996). We
used his $\lambda$6678/$\lambda$4471 curves only, since the
$\lambda$5876/$\lambda$4471 curves have minima between 6000-10000~K, which
allow matches to a given line intensity ratio at both low and high
temperature. Column 5 of Table~12 presents the dereddened
$\lambda$6678/$\lambda$4471 flux ratios (from Paper~I) and the derived values
of {\elt}(He~{\sc i}) are listed in column 6. Also presented in Table~12 are
the nebular electron densities (column 2; adopted from Table~6 of Paper~I)
that were used to interpolate between the density-dependent
$\lambda$6678/$\lambda$4471 curves, as well as the nebular [O~\sc iii}] and
hydrogen Balmer-jump electron temperatures (columns 3 and 4), which were taken
from Table~7 of Paper~I.  Very accurate relative flux calibration over the
wavelength range spanning the two He~{\sc i} lines is required, since the
$\lambda$6678/$\lambda$4471 ratio varies by only 55-60\% between electron
temperatures of 1000~K and 16000~K. Despite the stringent requirement on the
accuracy of the relative flux calibration over such a wide wavelength range,
six of the nine PNe listed in Table~12 show He~{\sc i} electron temperatures
within 2000~K of the nebular [O~{\sc iii}] electron temperatures, one
(NGC~3132) has a He~{\sc i} electron temperature which is 3300~K larger than
the [O~{\sc iii}] electron temperature, and two of nebulae, NGC~6818 and
IC~4191, yield He~{\sc i} electron temperatures that are respectively 8300~K
and 7300~K {\em lower} than the corresponding [O~{\sc iii}] electron
temperatures.

\subsection{Electron temperatures from O~{\sc ii} ORL ratios}

In this subsection we present observational evidence that the {\oii} ORLs
detected from several of our PNe are emitted from ionized gas which is at
temperatures \emph{much lower} than those derived from either the {\hi}
Balmer discontinuity or the [O~{\sc iii}] optical forbidden lines, for
which values were derived in Paper~I (see also the current Table~12).
We will make use of the temperature sensitivity of the ratio of the
intensities of the $\lambda$4089 and $\lambda$4649 lines of O~{\sc ii},
a technique that was first used by Wesson, Liu \& Barlow (2003), who
derived very low electron temperatures for two hydrogen-deficient knots
in the planetary nebula Abell~30.

The intensities of recombination lines originating from states of
different valence orbital angular momentum have different dependences on
electron temperature. By comparing the intensity of a line in the O~{\sc
ii} 4f--3d transition array with one from the 3p--3s array, for example,
it is possible to deduce a recombination line temperature. There are
several potential difficulties in using such lines to measure the electron
temperature.  Firstly, the variation of the intensity ratios with
temperature is weak, typically a factor of two or three when the
temperature changes from 300~K to 10000~K, meaning that it is important to
ensure that the intensities of the weak recombination lines are determined
as accurately as possible. Secondly, the relative intensities may be
affected by the distribution of population in the recombining ion ground
state, O$^{2+}$\,$^3$P in this case. The available O~{\sc ii}
recombination coefficients assume that the levels of this term are
populated according to their statistical weight (Storey 1994). If the
electron density is sufficiently low, this may not be the case, with the
energetically higher levels of the O$^{2+}$\,$^3$P state being relatively
underpopulated; this has been observed for the low density {\hii} regions
30 Doradus, LMC~N11B and SMC~N66 (Tsamis et al. 2003a). Equally, a very
low electron temperature would have the same effect.

To attempt to circumvent this second problem, we consider the ratio of
intensities of $\lambda$4089 from the 4f--3d transition array and
$\lambda$4649 from the 3p--3s array. These lines originate from the
state of highest total angular momentum, $J$, in each case and should
therefore both be mainly populated from the O$^{2+}$\,$^3$P$_2$
level. The published recombination coefficients for these two lines
(LSBC and Storey 1994, respectively) are only valid for temperatures
{\elt} $\geq$ 5000~K. Since we wish to investigate the very low temperature
regime we have extended the recombination coefficient calculations to
temperatures {\elt} $\geq$ 300~K. Figure\,~7 shows the corresponding
intensity ratio for temperatures between 300~K and 15000~K. The
calculations could not easily be taken to even lower temperatures due
to numerical problems in the recombination coefficient codes.

Measurements of the intensities of $\lambda$4089 and $\lambda4649$ were given
in Paper~I. The $\lambda$4089 line is isolated and measurement of the
intensity is relatively straightforward.\footnote{Apart from the cases of NGC
3242, 5882 and 6818, where the detection of weak emission in the 4116.1~\AA\
component of the Si~{\sc iv} 4s$^2$S--4p$^2$P$^{\rm o}$
$\lambda\lambda$4089,4116 doublet (with relative intensities of 0.022, 0.0308
and 0.022, respectively, on a scale where H$\beta$ = 100) led us to correct
the measured 4089~\AA\ line flux for the presence of Si~{\sc iv} 4088.85~\AA ,
by subtracting twice the measured intensity of the Si~{\sc iv} 4116.10~\AA\
line}~The $\lambda$4649 line, on the other hand is part of a complex blend in
which we identify five components, two lines of the O~{\sc ii} multiplet V\,1,
$\lambda 4649.13$ and $\lambda 4650.84$, and three lines of the C~{\sc iii}
multiplet V\,1, $\lambda\lambda 4647.42, 4650.25, 4651.47$. We have made new
five-component fits to this blend for a subset of the PNe, using the known
wavelength separations between the lines, assuming that all five lines have
the same width as the C~{\sc ii} $\lambda 4267$ line (the strongest heavy
element ORL in the spectra), and using the expected intensity ratio of 5:3:1
for the three C~{\sc iii} lines.  The resulting $\lambda$4089 to $\lambda$4649
line ratios and derived electron temperatures are given in Table~12 for the
six PNe for which both lines were detected, while the derived relative
intensities for $\lambda 4649.13$ and $\lambda 4650.84$ can be found in
Table~B13. Note that the O~{\sc ii} electron temperatures presented here in
Table~12 supersede those that were listed, without discussion, in Table~7 of
Paper~I.

Inspection of Table~12 shows that two of the eight PNe with derived
$\lambda$4089/$\lambda$4649 O~{\sc ii} electron temperatures show values
below our 300\,K theoretical calculation limit. The fixed-slit and
scanning-slit observations of IC~4191 yield {\elt}'s of 3900\,K and
2150\,K, respectively, while NGC\,5882 has the highest measured
{\elt}({\oii}) of 8700\,K. It can be shown that these values are reduced
{\it further} when one takes into account the small contribution to the
{\oii} $\lambda\lambda$4089, 4649 intensities from the normal nebular gas
at {\elt}({\foiii}) temperatures. Once this contribution is subtracted
from the observed {\oii} intensities, the revised {\oii} ORL ratios (not
shown in Table~12) indicate electron temperatures of less than 300\,K
for six PNe, while for NGC\,5315 and NGC\,5882 the resulting temperatures
are 4350, and 7190\,K respectively. These findings point towards the
presence of ultra-cold plasma regions in a large fraction of our sample
PNe.

\setcounter{figure}{6}
\begin{figure}
\begin{center} \epsfig{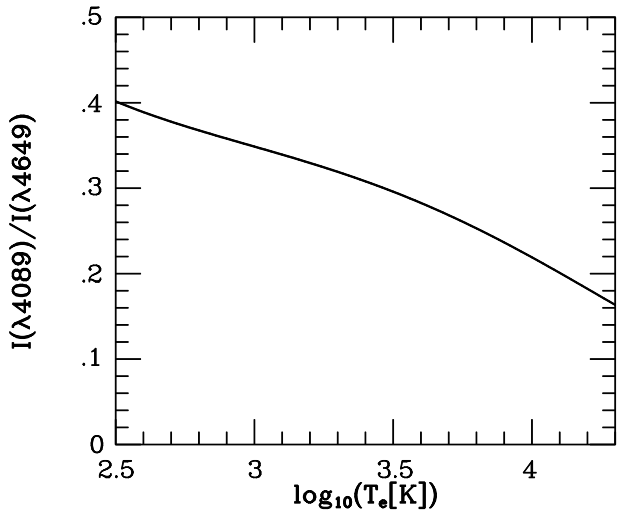}
\caption{The theoretical intensity ratio for O~{\sc ii}
$\lambda$4089/$\lambda$4649 as a function of electron temperature.
See Section~6.2 for details.}
\end{center}
\end{figure}

\section{Discussion}

In Sections~3 and 5.2 it was demonstrated that there is no dependence of
the magnitude of the nebular ORL/CEL abundance discrepancy factors upon
the excitation energy of the UV, optical or IR CEL transition used (see
Table~9), indicating that classical (i.e. in a chemically homogeneous
medium) nebular temperature fluctuations are not the cause of the observed
abundance discrepancies. This reinforces the same conclusion that was
reached in Paper~I, based there upon the fact that [O~{\sc iii}] electron
temperatures derived from the ratio of the 52 and 88-$\mu$m FS lines to
the 4959 and 5007-\AA\ forbidden lines were greater than or comparable to
those derived from the ratio of the higher excitation energy 4363-\AA\
transition to the 4959 and 5007-\AA\ lines---if temperature fluctuations
in the ambient nebular material were the cause of the ORL/CEL abundance
discrepancies then the IR line-based ratio should yield {\em lower}
temperatures. We conclude instead that the main cause of the abundance
discrepancies is enhanced ORL emission from cold ionized gas located in
hydrogen-deficient clumps inside the nebulae, as first postulated by Liu
et al. (2000) for the high-ADF PN NGC~6153.

When nebular heavy element abundances exceed about five times solar,
cooling by their collisionally excited infrared fine-structure (IR FS)
lines is alone sufficient to balance the photoelectric heating from
atomic species. Since cooling by the low-excitation IR FS lines
saturates above a few thousand K, nebular electron temperatures in such
high-metallicity regions will drop to values of this order, the exact
value being determined by the H to heavy element ratios and the input
ionizing spectrum. It is therefore physically plausible and
self-consistent for the heavy element ORLs that yield enhanced abundances
relative to those derived from CELs to also indicate very low electron
temperatures for the regions from which they emit.

Liu et al.  (2000) in their empirical modelling of NGC\,6153 had
postulated the presence of H-deficient ionized clumps within the nebula
that would be cool enough to suppress optical forbidden-line emission and
even some infrared fine-structure line emission but which would emit
strongly in heavy element recombination lines, due to the inverse
power-law temperature dependence of ORL emission. P\'{e}quignot et al.
(2002) have constructed photoionization models of NGC\,6153 and M~1-42
incorporating H-deficient inclusions and found equilibrium {\elt}'s of
$\sim$~10$^3$\,K in the H-deficient clumps and $\sim$~10$^4$\,K in the
ambient gas, with the H-deficient ionized regions being within a factor of
two of pressure equilibrium with the ambient nebular gas. In both cases,
the H-deficient model components contained only $\sim$~1\% of the total
ionized mass, so that the overall metallicity of the whole nebula was
close to that of the `normal' high-temperature component.

Similar, or lower, mass fractions for the postulated H-deficient clumps in
the typical PNe studied here, which have lower ADF's than the extreme
cases discussed above, should ensure that their integrated IR FS line
emission will not represent a significant perturbation to the integrated
IR FS emission from the ambient high-temperature nebular material that
forms the majority of the nebular mass, particularly if the H-deficient
clump electron densities exceed the rather low critical densities of 500
-- 3000~cm$^{-3}$ that correspond to the [O~{\sc iii}] and [N~{\sc iii}]
IR FS lines that have been investigated in this paper. The strong inverse
power-law temperature dependence of ORL emission means that material at a
temperature of 500\,K will emit an ORL such as O~{\sc ii} $\lambda$4649
eighteen times more strongly than at 10$^4$~K. If 0.5\% of the nebular
mass was located in 500\,K clumps having an electron density ten times
that of the ambient nebular gas, then the integrated ORL emission from the
clumps would exceed that from the ambient gas by a factor of nine.

Hydrogen is not expected to be entirely absent in the postulated
H-deficient clumps. The steep inverse power-law temperature dependence of
its recombination emission from the clumps could be responsible for the
strong correlation, discussed in Section~4.1, between the C$^{2+}$ or
O$^{2+}$ ADF's and the temperature difference, $\Delta T$, between the
[O~{\sc iii}] optical forbidden line and H~{\sc i} Balmer jump electron
temperatures (see Fig.\,~3). The correlation between ionic ADF's and
decreasing nebular surface brightness (Fig.\,~4), or increasing absolute
nebular radius (Fig.\,~5), may indicate that the density contrast between
the H-deficient clumps and the ambient nebular material increases as the
nebula evolves, i.e. that the clump density decreases less than the
ambient nebular density does as the nebula expands. Such a situation might
arise if the clump ionized gas originates from photoevaporation of dense
neutral cores (cometary knots) of the kind found in the Helix Nebula
(Meaburn et al. 1992; O'Dell et al. 2000) and the Eskimo Nebula
(NGC\,2392; O'Dell et al. 2002). Indeed, the location of NGC\,2392 in the
$\Delta T$ vs. $S$(H$\beta$) and $\Delta T$ vs. Radius diagrams (Fig.\,~6)
is consistent with this picture and indicates that NGC\,2392 is a
candidate high-ADF nebula.  In confirmation of this, optical and
ultraviolet large aperture measurements presented by Barker (1991) for six
positions in NGC\,2392 yielded $\lambda$4267/$\lambda$1908 C$^{2+}$ ADF's
ranging from 7 to more than 24.  We thus show NGC\,2392 in the
ADF(C$^{2+}$) vs. $\Delta T$ diagram (Fig.\,~3) and the ADF(C$^{2+}$) 
versus Radius diagram (Fig.\,~5) using a mean C$^{2+}$ ADF
of 18 derived for the positions observed by Barker. It would be of
interest to obtain deep optical spectra of this nebula in order to examine
its heavy element ORL spectrum in detail.

Available spatial analyses of long-slit PN spectra show that ORL/CEL ADF's
peak towards the center of nebulae in the cases of NGC\,6153 (Liu et al.
2000) and NGC\,6720 (Ring Nebula; Garnett \& Dinerstein 2001b). An
examination of the results presented by Barker (1991) for NGC\,2392 shows
that the same trend is present in those data too: the inferred
ADF(C$^{2+}$) increases towards the centre of the PN (along the aligned
positions 4, 2, and 1; cf. Fig.\,~1 of that paper). In NGC\,6720
especially, the location of peak O~{\sc ii} ORL emission does not coincide
with the positions of the {\it HST}-resolved dusty cometary knots, which
populate the main shell of the Ring, but is displaced inwards from that of
the peak [O~{\sc iii}] emission (Garnett \& Dinerstein 2001b). We
speculate that such an effect could be due to the advanced
photo-processing of knots in the Ring Nebula that were overcome by the
main ionization front in the past and whose relic material, rich in heavy
elements, is now immersed in the He$^{2+}$ nebular zone, being subjected
to the intense radiation field of the central star. The question of the
possible relationship between the ORL/CEL abundance discrepancy
problem and the cometary knot complexes observed in many PNe warrants 
further investigation.

Ruiz et al. (2003) and Peimbert et al. (2004) have argued against the
presence of H-deficient knots in NGC\,5307 and NGC\,5315 on the grounds
that their 34~km~s$^{-1}$ resolution echelle spectra did not reveal a
difference between the radial velocities or line widths of the heavy
element ORLs and those of the main nebular lines, of the type exhibited by
the high-velocity H-deficient knots in the born-again PNe A30 and A58. We
note however that the H-deficient knot model that has been invoked to
explain the ORL/CEL ADF's of typical planetary nebulae makes no specific
predictions as to whether the knots should exhibit a different kinematic
pattern from the bulk of the nebula. Scenarios for the origin of such
knots in `normal' PNe include i) the evaporation of primitive material
(comets, planetesimals) left over from the formation of the progenitor
star (e.g. Liu 2003), which would predict ORL C/O and N/O ratios typical
of the unprocessed ISM material out of which the star formed; or ii) that
they originated as incompletely mixed material brought to the surface by
the 3rd dredge-up and ejected along with the rest of the AGB progenitor
star's outer envelope during the PN formation phase. In the latter case,
the knots' ORL C/O and N/O ratios should show the same nucleosynthetic
signatures as the rest of the ejected envelope. Although the postulated
H-deficient clumps may not exhibit different kinematics from the bulk of
the nebular material, the low inferred electron temperatures of the clumps
and the consequent very low thermal broadening means that heavy element
ORL's in high-ADF nebulae should exhibit much narrower line widths than do
the strong forbidden lines. For example, a C~{\sc ii} $\lambda$4267 line
originating from 1000\,K material should have a FWHM of 2~km~s$^{-1}$ if
thermal broadening dominates, i.e.  if velocity broadening of the line is
minimal, as when the emission comes from the edge of the nebula, where
material is moving in the plane of the sky, or comes from well-separated
approaching and receding velocity components, e.g. near the centre of the
nebula. Thus observations of suitably chosen nebular sub-regions at a
resolving power of 1.5$\times10^5$ may be capable of confirming the
presence of the postulated cold plasma clumps.

\section*{Acknowledgments}

We thank Daniel P\'equignot for illuminating discussions. YGT gratefully
acknowledges the award of a Perren Studentship while at UCL and of a Peter
Gruber Foundation/International Astronomical Union Fellowship while at the
Observatoire de Paris-Meudon.

{}

\begin{appendix}

\section{The calculation of elemental abundances from recombination lines}

In this Appendix we discuss the derivation of total C, N and O abundances
for our PN sample from ORL ionic abundances. We discuss each PN
individually in order to highlight differences in the adopted \emph{icf}
scheme. \\

\noindent{\bf NGC\,3242}: the C~{\sc ii} $\lambda$4267, C~{\sc iii}
$\lambda\lambda$4069, 4187 and $\lambda$4650 lines are detected; we adopt
C$^{3+}$/H$^+$ = 2.03\,$\times$\,{\tmf}, as an intensity weighted mean from the
$\lambda\lambda$4187, 4650 lines only. The unseen {\cp} is corrected for using
the KB94 \emph{icf} formula: \emph{icf}(C)\,$\equiv$\,1 + {\op}/{\opp} =
1.01. We
calculate the C/H
fraction using,

\begin{equation}\label{eq:icforlc}
{\rm C/H} = icf({\rm C})\,\times\,({\rm C}^{2+} + {\rm C}^{2+} + {\rm
C}^{4+})/{\rm H}^{+}.
\end{equation}

This yields C/H = 8.45$\times${\tmf}; this value is only 2\,per cent more than
the one found if {\cfp}/{\hp} = 1.88\,$\times$\,10$^{-5}$, as derived from
C~{\sc iv} $\lambda$4658 is not added in. The latter value is an upper limit to
the {\cfp}/{\hp} fraction, since $\lambda$4658 is blended with [Fe~{\sc iii}]
$\lambda$4658.10.

From an intensity weighted mean of {\npp}/{\hp} ratios from seven {\nii} lines
(cf. Table~4), we find {\npp}/{\hp} = 1.71\,$\times$\,{\tmf}. From
the N~{\sc iii} $\lambda$4379 line, {\nppp}/{\hp} = 6.85\,$\times$\,10$^{-5}$.
We use the CEL ratio {\np}/{\npp} = 0.0145 (cf. Paper~I) to
account for the unseen {\np}; the error introduced should be negligible, since
{\np}/N = 0.008. Hence, summing three ionic stages, we find N/H =
2.42\,$\times$\,{\tmf}.

From our extensive {\oii} recombination-line survey, we derive
{\opp}/{\hp} = 6.28\,$\times$\,{\tmf}, from five {\oii} multiplets
(Table~6). An ORL {\op}/{\hp} abundance is not available, so the unseen
{\op} is corrected for assuming that the Paper~I CEL ratio of {\op}/{\opp}
= 0.0091 holds for the ORLs too.  In view of the minor concentration of
{\op} in this object (about 1\,per cent), the errors introduced should be
negligible. We then employ the KB94 standard ionization correction factor
to account for the unseen {\oppp}, \emph{icf}(O)\,$\equiv$\,(1 +
{\hepp}/{\hep})$^{2/3}$ = 1.17, and calculate the total O/H abundance,
using,

\begin{equation}\label{eq:icforlo}
{\rm O/H} = icf({\rm O})\,\times\,({\rm O}^{+} + {\rm O}^{2+})/{\rm H}^{+}.
\end{equation}

This yields, O/H = 7.41\,$\times$\,{\tmf}. Alternatively, we can adopt
{\oppp}/{\hp} = 2.05\,$\times$\,{\tmf}, as derived by LD93a from the {\oiii}
V\,8 multiplet at 3265\,\AA; hence, summing all three ionic stages we find O/H
= 8.39\,$\times$\,{\tmf}, i.e. 13\,per cent larger than
yielded by the \emph{icf} method. We adopt this latter value.\\

\noindent{\bf NGC\,5882}: the C~{\sc ii} $\lambda$4267, C~{\sc iii}
$\lambda\lambda$4069, 4187 and $\lambda$4650 lines are detected; we adopt
C$^{3+}$/H$^+$ = 5.27\,$\times$\,10$^{-5}$, as an intensity weighted mean from
the $\lambda\lambda$4187, 4650 lines only. No {\cfp} is expected to exist,
since only 2\,per cent of He is in the form of He$^{2+}$. Using
Eq.~A1 with \emph{icf}(C) = 1.03, we find, C/H = 4.43\,$\times$\,{\tmf}.

From an intensity weighted mean of five {\npp}/{\hp} ratios (Table~4),
{\npp}/{\hp} = 2.24\,$\times$\,{\tmf} is found. From the N~{\sc iii}
$\lambda$4379 line, {\nppp}/{\hp} = 5.75\,$\times$\,10$^{-5}$. As for
NGC\,3242, we use the Paper~I CEL ratio to account for the unseen {\np},
{\np}/{\npp} = 0.0278 in this case; we obtain N/H =
2.88\,$\times$\,{\tmf}.

From eight {\oii} ORL multiplets we derive {\opp}/{\hp} =
9.70\,$\times$\,{\tmf} (Table~6). We account for unseen {\op} as
previously, using the Paper~I CEL ratio of {\op}/{\opp} = 0.029; we have
\emph{icf}(O) = 1.01, so Eq.~A2 yields O/H = 10.08\,$\times$\,{\tmf}.\\

\noindent{\bf NGC\,5315}: the C~{\sc ii} $\lambda$4267, C~{\sc iii}
$\lambda$4187 lines are detected; no {\cfp} exists since the He$^{2+}$/{\hp}
fraction is negligible. We correct for unseen {\cp} using the CEL ratio of
{\cp}/{\cpp} = 0.08 (given the small concentration of {\cp}, about 7\,per cent,
no significant uncertainty is introduced); hence we sum three ionic stages and
find C/H = 7.29\,$\times$\,{\tmf}.

From an intensity weighted mean of four {\npp}/{\hp} ratios (Table~4),
{\npp}/{\hp} = 3.43\,$\times$\,{\tmf} is found; the N~{\sc iii}
$\lambda$4379 line is not detected. If we assume that {\np}/{\npp} = 0.163
and {\nppp}/{\npp} = 0.565 ratios, as given by the CELs (Paper~I), are
also valid for the ORLs, we can account for both {\np} and {\nppp}, hence,
N/H = 5.93\,$\times$\,{\tmf}. This result may be quite uncertain, since
probably the {\nppp}/{\npp} ratio derived from CELs \emph{is not equal} to
the same ratio as derived from ORLs (cf. the case of NGC\,3918 below).

From five {\oii} ORL multiplets, {\opp}/{\hp} = 8.57\,$\times$\,{\tmf};
although this is a relatively low excitation object, a substantial {\oppp}
concentration has however been derived from the $\lambda$1401 CEL line
(Paper~I). If we
account for unseen {\op} using the CEL ratio {\op}/{\opp} = 0.098, we find O/H
= 9.41\,$\times$\,{\tmf}, ignoring {\oppp}/{\hp}; however, if we further assume
that {\oppp}/{\opp} = 0.364, as given by CELs, holds for ORLs too, then this
brings the total ORL abundance
up to O/H = 12.53\,$\times$\,{\tmf}, i.e. 33\,per cent higher. We adopt this latter value.\\

\noindent{\bf NGC\,3918}: the C~{\sc ii} $\lambda$4267, C~{\sc iii}
$\lambda\lambda$4187, 4650 lines are detected; C$^{3+}$/H$^+$ =
1.44\,$\times$\,10$^{-4}$ is derived, as an intensity weighted mean from
the $\lambda\lambda$4187, 4650 lines. Using the [Fe~{\sc iii}]
$\lambda$4702 line, we estimate that 67\,per cent of the detected
$\lambda$4658 intensity is due to {\cfp}, i.e. \emph{I}(C~{\sc iv}
$\lambda$4658) = 0.1093; hence, {\cfp}/{\hp} = 2.62$\times$10$^{-5}$. An
ORL abundance is not available for C$^+$, so it was estimated using the
Paper~I CEL ratio of C$^+$/{\cpp} = 0.155 (given the small ionic
concentration of C$^+$, 9\,per cent maximum, the error introduced should
be negligible); adding this to the remaining three ionic stages we find,
C/H = 7.50\,$\times$\,{\tmf} (if instead we correct for C$^+$ using an
\emph{icf}(C) = 1.09, we arrive at a value which is smaller by only 3\,per
cent).

From an intensity weighted mean of four {\npp}/{\hp} ratios (Table~4),
{\npp}/{\hp} =
1.42\,$\times$\,{\tmf} is found; from the N~{\sc iii} $\lambda$4379 line,
{\nppp}/{\hp} = 6.41\,$\times$\,10$^{-5}$; all of the measured intensity of the
$\lambda$4606 line is attributed to N~{\sc iv} (no contribution from N~{\sc ii}
$\lambda$4607.2 is expected, since the stronger predicted N~{\sc ii}
$\lambda$4601 line is absent); therefore, {\nfp}/{\hp} =
1.90\,$\times$\,10$^{-5}$. We account for the missing {\np} using the
Paper~I CEL ratio of
{\np}/{\npp} = 0.203; summing all four ionic stages, we have N/H =
2.56\,$\times$\,{\tmf}. For this PN, we find that just as is the case for
relative ionic fractions of carbon, the CEL {\nppp}/{\npp} ratio (= 1.22)
\emph{is not equal} to the ORL {\nppp}/{\npp} ratio (= 0.45).

We have derived {\opp}/{\hp} and {\ofp}/{\hp} abundances from five {\oii}
multiplets (Table~6) and the O~{\sc iv} $\lambda$4632 line (Table~7),
respectively; {\opp}/{\hp} = 5.36\,$\times$\,{\tmf} and {\ofp}/{\hp} =
4.74\,$\times$\,{\tmfi}. We account for unseen {\op} and {\oppp}, using
the Paper~I CEL ratios of {\op}/{\opp} = 0.086 and {\oppp}/{\opp} = 1.21;
hence, summing all four ionic stages we have O/H =
12.32\,$\times$\,{\tmf}.\\

\noindent{\bf NGC\,2022}: the C~{\sc ii} $\lambda$4267, C~{\sc iii}
$\lambda\lambda$4069, 4187 and $\lambda$4650 lines as well as the C~{\sc
iv} $\lambda$4658 line are detected; C$^{3+}$/H$^+$ =
3.10\,$\times$\,10$^{-4}$ is the intensity weighted mean from the
$\lambda\lambda$4187, 4650 lines only. We attribute all of the
$\lambda$4658 intensity to C~{\sc iv}, since other [Fe~{\sc iii}] lines
are not present; furthermore this is a very high excitation object
(He$^{2+}$/He = 0.88) and we expect a substantial concentration of {\cfp};
we find {\cfp}/{\hp} = 2.62\,$\times$\,{\tmf}. Using \emph{icf}(C) = 1.02,
to correct for unseen {\cp} and summing the three higher ionic stages, we
obtain C/H = 14.84\,$\times$\,{\tmf}.

We derive {\npp}/{\hp} = 4.38\,$\times$\,\,10$^{-4}$ from an
intensity-weighted mean of three {\npp}/{\hp} ratios (Table~4); from
N~{\sc iii} $\lambda$4379, {\nppp}/{\hp} = 2.08\,$\times$\,10$^{-4}$ and
from N~{\sc iv} $\lambda$4606, {\nfp}/{\hp} = 3.68\,$\times$\,10$^{-5}$
(Table~5; as with NGC\,3918, no contribution by N~{\sc ii} $\lambda$4607.2
is found for the $\lambda$4606 line). We correct for unseen {\np} using
the Paper~I CEL ratio {\np}/{\npp} = 0.013 and sum a total of four ionic
stages to obtain N/H = 6.88\,$\times$\,{\tmf}.

From four {\oii} multiplets (Table~6) and the O~{\sc iv} $\lambda$4632
line (Table~7), respectively, we derive {\opp}/{\hp} =
13.01\,$\times$\,{\tmf} and {\ofp}/{\hp} = 1.10\,$\times$\,{\tmf}; we also
adopt {\oppp}/{\hp} = 3.62\,$\times$\,{\tmf}, as derived by LD93a from the
{\oiii} V\,8 multiplet at 3265\,\AA. The unseen {\op} is accounted for
using the Paper~I CEL ratio of {\op}/{\opp} = 0.020;  hence we obtain O/H
= 17.99\,$\times$\,{\tmf}.\\

\noindent{\bf NGC\,6818}: the C~{\sc ii} $\lambda$4267, C~{\sc iii}
$\lambda\lambda$4069, 4187 and $\lambda$4650 and C~{\sc iv} $\lambda$4658
lines are detected; C$^{3+}$/H$^+$ = 1.37\,$\times$\,10$^{-4}$, is the
intensity-weighted mean from the $\lambda\lambda$4187, 4650 lines only
(Table~2). Using exactly the same argument as in the case of NGC\,2022, we
attribute the $\lambda$4658 line to {\cfp} only; we obtain {\cfp}/{\hp} =
4.47\,$\times$\,10$^{-5}$. If we assume that the ratio C$^+$/{\cpp} =
0.129 derived from the CELs in Paper~I is also valid for the ORLs, we
deduce a total abundance of C/H = 7.71\,$\times$\,{\tmf}.

The N~{\sc ii} $\lambda\lambda$5676, 5679 lines from multiplet V3 are
detected; they yield {\npp}/{\hp} = 4.27\,$\times$\,{\tmf} (Table~4); from
N~{\sc iii} $\lambda$4379 we find {\nppp}/{\hp} =
8.87\,$\times$\,10$^{-5}$, while from N~{\sc iv} $\lambda$4606 we deduce
{\nfp}/{\hp} = 1.46\,$\times$\,10$^{-5}$ (Table~5); the unseen {\np} is
corrected for using the Paper~I CEL ratio {\np}/{\npp} = 0.241. Summing a
total of four ionic stages, N/H = 6.33\,$\times$\,{\tmf} is obtained.

We have derived {\opp}/{\hp} = 7.33\,$\times$\,{\tmf} and {\ofp}/{\hp} =
6.41\,$\times$\,{\tmfi} from four {\oii} multiplets (Table~6) and the
O~{\sc iv} $\lambda$4632 line (Table~7), respectively; we further adopt
{\oppp}/{\hp} = 2.37\,$\times$\,{\tmf}, as derived by LD93a from the
{\oiii} V\,8 line at 3265\,\AA. The unseen {\op} is accounted for using
the Paper~I CEL ratio of {\op}/{\opp} = 0.090; hence, O/H =
11.00\,$\times$\,{\tmf}.\\

\noindent{\bf NGC\,3132}: only C~{\sc ii} $\lambda$4267 is detected. If we
then use the standard KB94 \emph{icf} to account for unseen {\cppp},
\emph{icf}(C)\,$\equiv$\,O/{\opp} = 1.93, and calculate the carbon
abundance as C/H = \emph{icf}(C)\,$\times$\,{\cpp}/{\hp}, we then find C/H
= 12.74\,$\times$\,{\tmf}. If instead, we assume that the ratio
{\cppp}/{\cpp} = 0.111 given by the CELs in Paper~I is also valid for the
ORLs and use it to estimate the ORL {\cppp}/{\hp} abundance, we find, C/H
= 13.86\,$\times$\,{\tmf}, i.e. 9\,per cent larger. However, from
observations of 7 PNe it is found that, in general, ORL
{\cppp}/{\cpp}\,$\neq$\,CEL {\cppp}/{\cpp}; hence, we adopt the former C/H
value.

From three {\nii} ORL lines and the N~{\sc iii} $\lambda$4379 line we
obtain, {\npp}/{\hp} = 3.40\,$\times$\,{\tmf} and {\nppp}/{\hp} =
2.47\,$\times$\,10$^{-5}$, respectively. We correct for the unseen {\np}
using the Paper~I CEL ratio {\np}/{\npp} = 1.296 and sum a total of three
ionic stages to find N/H = 8.05\,$\times$\,{\tmf}.

From the {\oii} V\,1 and V\,10 multiplets (Table~6) we have derived
{\opp}/{\hp} = 8.15\,$\times$\,{\tmf}; we account for unseen {\oppp} using
an \emph{icf}(O) = 1.02 from Eq.~A2; we account for unseen {\op} using the
Paper~I CEL ratio of {\op}/{\opp} = 0.894, hence, O/H =
15.74\,$\times$\,{\tmf}. The concentration of {\op} in this PN is about
46\,per cent and it may be that some uncertainty has been introduced by
assuming that the CEL {\op}/{\opp} ratio equals the ORL {\op}/{\opp}
ratio.\\

\noindent{\bf NGC\,2440}: the C~{\sc ii} $\lambda$4267, C~{\sc iii}
$\lambda\lambda$4187, 4650 and C~{\sc iv} $\lambda$4658 lines are
detected; we attribute the $\lambda$4658 intensity solely to {\cfp}; thus
{\cppp}/{\hp} = 1.66\,$\times$\,{\tmf} and {\cfp}/{\hp} =
1.82\,$\times$\,{\tmf} (Table~2). Using the KB94
\emph{icf}(C)\,$\equiv$\,1 + {\op}/{\opp} = 1.28 to account for unseen
{\cp} and summing the three higher ionic stages, we find C/H =
10.21\,$\times$\,{\tmf}.

Four N~{\sc ii} lines are detected (Table~4); their intensity weighted
mean yields {\npp}/{\hp} = 4.77\,$\times$\,{\tmf}; from N~{\sc iii}
$\lambda$4379 we find {\nppp}/{\hp} = 1.43\,$\times$\,{\tmf}; both N~{\sc
iv} $\lambda\lambda$4606, 4707 lines are detected, yielding a mean
{\nfp}/{\hp} = 1.70\,$\times$\,{\tmf}. No {\np} abundance is available
either from ORLs or CELs (our wavelength coverage did not include the
[N~{\sc ii}] lines). Thus, we adopt the KB94 CEL ratio of {\np}/{\npp} =
0.475, and use it to correct for the unseen {\np}; summing four ionic
stages we obtain N/H = 10.17\,$\times$\,{\tmf}.

From four {\oii} multiplets and the O~{\sc iv} $\lambda$4632 line,
respectively, we derive {\opp}/{\hp} = 5.23\,$\times$\,{\tmf} and
{\ofp}/{\hp} = 1.01\,$\times$\,{\tmf}; we further adopt {\oppp}/{\hp} =
2.56\,$\times$\,{\tmf}, as derived by Liu \& Danziger (1993) from the
{\oiii} V\,8 line at 3265\,\AA, together with the CEL ratio {\op}/{\opp} =
0.285 from the same authors (as reported for their PA = 270$^0$ slit); we
assume that this ratio is valid for ORLs too and hence derive O/H =
10.29\,$\times$\,{\tmf}.\\

\noindent{\bf NGC\,6302}: only C~{\sc ii} $\lambda$4267 is reliably
detected; using the same \emph{icf} scheme as for NGC\,3132, yields
\emph{icf}(C)  = 2.72, from Eq.~A1 we obtain C/H = 5.45\,$\times$\,{\tmf}.

Three N~{\sc ii} lines are detected (Table~4); their intensity weighted mean
yields {\npp}/{\hp} = 5.64\,$\times$\,{\tmf}; N~{\sc iii} $\lambda$4379 yields
{\nppp}/{\hp} = 8.07\,$\times$\,{\tmf}; N~{\sc iv} $\lambda$4606 yields
{\nfp}/{\hp} = 2.28\,$\times$\,10$^{-4}$ (no contribution is expected from
N~{\sc ii} $\lambda$4607). {\np} is accounted for using the CEL ratio
{\np}/{\npp} = 0.435; summing four ionic stages we find N/H =
11.18\,$\times$\,{\tmf}.

We obtain {\opp}/{\hp} = 3.28\,$\times$\,{\tmf} from the {\oii} V\,1
multiplet; {\ofp}/{\hp} =
7.84\,$\times$\,{\tmfi} is derived from O~{\sc iv} $\lambda$4632; we
assume that the Paper~I CEL ratios of {\op}/{\opp} = 0.092 and
{\oppp}/{\opp} = 0.895 are valid for ORLs as well, and deduce O/H =
7.30\,$\times$\,{\tmf}.\\

\noindent{\bf IC\,4406}: C~{\sc ii} $\lambda$4267, C~{\sc iii}
$\lambda\lambda$4187, 4650 are detected (Table~2); the He$^{2+}$
concentration is only
6\,per cent, so it is assumed that the {\cfp}/{\hp} abundance is
negligible. Using
Eq.~A1, with \emph{icf}(C) = 1.31 to account for unseen {\cp}, we
find C/H = 11.27\,$\times$\,{\tmf} (Table~2).

From the sole detected N~{\sc ii} $\lambda$4630 line, {\npp}/{\hp} =
1.90\,$\times$\,{\tmf}; from N~{\sc iii} $\lambda$4379, {\nppp}/{\hp} =
7.31\,$\times$\,10$^{-5}$. {\np} is accounted for using the CEL ratio
{\np}/{\npp} = 0.536; summing three ionic stages we find N/H =
3.65\,$\times$\,{\tmf} (Table~5).

We obtain {\opp}/{\hp} = 7.06\,$\times$\,{\tmf} and use an \emph{icf}(O) =
1.04 to correct for unseen {\oppp}; the missing {\op} is corrected for
using the Paper~I CEL ratio of {\op}/{\opp} = 0.471 (some error may be
introduced since in this object, the {\op} concentration is about 31\,per
cent of all oxygen). Hence, from Eq.~A2 we obtain O/H =
10.80\,$\times$\,{\tmf}.\\

\noindent{\bf IC\,4191}: C~{\sc ii} $\lambda$4267, C~{\sc iii}
$\lambda\lambda$4187, 4650 are detected (Table~2); {\cppp}/{\hp} =
1.07\,$\times$\,{\tmf} is found. No significant {\cfp} is present. Using
Eq.~A1 with \emph{icf}(C) = 1.04 to account for unseen {\cp}, we find C/H
= 6.48\,$\times$\,{\tmf}.

Six {\nii} lines have been detected (Table~4); an intensity weighted mean
yields {\npp}/{\hp} = 5.00\,$\times$\,{\tmf}; from N~{\sc iii}
$\lambda$4379, we derive {\nppp}/{\hp} = 1.15\,$\times$\,10$^{-4}$
(Table~5). {\np} is accounted for using the Paper~I CEL ratio of
{\np}/{\npp} = 0.393, where {\npp}/{\hp} = 2.45\,$\times$\,10$^{-5}$ from
the [N~{\sc iii}] 57-$\mu$m line, as given by Liu et al. (2001) and
{\np}/{\hp} from our optical observations; hence, summing three ionic
stages we obtain N/H = 8.12\,$\times$\,{\tmf}.

We derive {\opp}/{\hp} = 12.92\,$\times$\,{\tmf} from four {\oii}
multiplets (Table~6) and use an \emph{icf}(O) = 1.07 to correct for unseen
{\oppp};  the missing {\op} is corrected for using the CEL ratio of
{\op}/{\opp} = 0.041 (the error introduced in this case should be
negligible, since the {\op} concentration in this PN is only about 4\,per
cent). Hence we deduce O/H = 14.39\,$\times$\,{\tmf}.\\

\noindent{\bf My\,Cn\,18}: only C~{\sc ii} $\lambda$4267 is seen; no
significant amounts of {\cppp} or {\cfp} are expected; using Eq.~A1 with
\emph{icf}(C) = 1.59 to account for unseen {\cp}, we find C/H =
6.04\,$\times$\,{\tmf}.

Three N~{\sc ii} lines of multiplet V5 have been detected (Table~4); from
an intensity weighted mean {\npp}/{\hp} = 20.35\,$\times$\,{\tmf} is
derived. Apart from optical CELs no other information on N exists for this
PN and the standard KB94 \emph{icf} scheme cannot account for unseen
ionization stages. Therefore, as a lower limit we adopt
N/H\,$\geq$\,20.35\,$\times$\,{\tmf}.

We derive {\opp}/{\hp} = 6.43\,$\times$\,{\tmf} from the {\oii} V\,1
$\lambda\lambda$4649, 4650 lines emitted by this low excitation nebula;
unseen {\op} is corrected for using the Paper~I CEL ratio of {\op}/{\opp}
= 0.59 ({\op} represents a significant amount of the total oxygen, about
37\,per cent). The resulting O/H = 10.24\,$\times$\,{\tmf} may be somewhat
uncertain.\\

\noindent{\bf SMC\,N87}: C~{\sc ii} $\lambda$4267, C~{\sc iii}
$\lambda\lambda$4187, 4650 are detected (Table~3); {\cppp}/{\hp} =
6.75\,$\times$\,10$^{-5}$ is found. No {\cfp} is present; using
Eq.~A1 with \emph{icf}(C) = 1.02 to account for the unseen {\cp}
only, we find C/H = 7.55\,$\times$\,{\tmf}. If instead we assume that the
ratio of C$^+$/{\cpp} = 0.095 derived from CELs in Paper~I is also valid
for the ORLs, we arrive at the same total C/H value.

We derive {\opp}/{\hp} = 2.95\,$\times$\,{\tmf} from the {\oii} V\,1
multiplet (Table~3); unseen {\op} is corrected for using the Paper~I CEL
ratio {\op}/{\opp} = 0.018 (this is a medium excitation PN and the
concentration of {\op} is only about 2\,per cent); we obtain O/H =
3.00\,$\times$\,{\tmf}.\\

\noindent{\bf LMC\,N66}: this is a highly ionized nebula reminiscent of the
galactic Type-I PN NGC\,6302; no ORLs of carbon are detected, consistent with
the carbon-poor, nitrogen-rich nature deduced for this nebula from the CEL
analysis of Paper~I, hence no ORL C/H value is available for this object.

From an intensity weighted mean of {\oii} $\lambda\lambda$4649, 4650
(Table~3), we derive {\opp}/{\hp} =
8.53\,$\times$\,{\tmf};  we adopt \emph{icf}(O) = 3.77 [from eq. (A7) of
KB94 coupled with our Paper~I CEL ionic nitrogen abundances for this PN];
we account for unseen {\op} using the CEL ratio {\op}/{\opp} = 0.076 and
obtain O/H = 3.46 \,$\times$\,10$^{-3}$.\\

\noindent{\bf LMC\,N141}: C~{\sc ii} $\lambda$4267, C~{\sc iii}
$\lambda\lambda$4187, 4650 are detected (Table~3); {\cppp}/{\hp} =
5.82\,$\times$\,10$^{-5}$ is found. No {\cfp} is present; using Eq.~A1,
with \emph{icf}(C) = 1.02 to account for the unseen {\cp}, we find C/H =
7.54\,$\times$\,{\tmf}. If instead, we assume that the ratio C$^+$/{\cpp}
= 0.203 derived from the CELs in Paper~I is also valid for the ORLs, we
obtain C/H = 8.75\,$\times$\,{\tmf}. In view of the error this may
introduce, we adopt the former value.

We derive {\opp}/{\hp} = 4.96\,$\times$\,{\tmf} from the {\oii} V\,1 and
V\,10 multiplets; unseen {\op} is corrected for using the Paper~I CEL
ratio of {\op}/{\opp} = 0.023; we obtain O/H = 5.08\,$\times$\,{\tmf}. \\

Nitrogen ORLs have not been detected from either SMC\,N87, LMC\,N66 or
LMC\,N141, hence no inference can be made about their N abundances from optical
recombination lines.

\section{A comparison of observed and predicted intensities of {\oii} ORLs}

In Table~B13 we present a comparison of the dereddened O~{\sc ii} line
intensities detected from all nebulae, against the predicted intensities from
recombination theory. The comparison is relative to the strongest expected line
within each multiplet for the 3--3 transitions, while for the collection of
3d--4f transitions it is relative to the strongest expected 3d--4f line at
4089.3\,\AA. The bracketed figures are the estimated \emph{absolute} errors in
the values, arising from the line profile fitting method only, and do not
include any
possible systematic errors (for instance, those arising from the calibration
process or the corrections for atmospheric and interstellar extinction).
In the case of some of the 4f-3d lines, additional, weaker, lines from the
same multiplet contribute, as listed in Table~4(a) of LSBC.
The $I_{\rm obs}$ values listed for some of the 4f-3d lines in Table~B13
therefore include contributions from weaker components of the multiplet,
but the contributions from such blending lines were corrected for using the
theoretical branching ratios listed in Table~4(a) of LSBC in
order to arrive at the $I_{\rm obs}$/$I_{\rm pred}$ values in the final
column of Table~B13, which refer to the listed multiplet component alone.

\setcounter{table}{12}
\begin{table}
\begin{minipage}{75mm}
\centering \caption{Comparison of observed and predicted relative
intensities of O~{\sc ii} lines.}
\begin{tabular}{lcccc}
\noalign{\vskip3pt} \noalign{\hrule} \noalign{\vskip3pt}
$\lambda_{0}$(\AA) &Mult. &g$_{l}$--g$_{u}$ &$I_{\rm obs}$ &$I_{\rm obs}$/$I_{\rm pred}$\\
\noalign{\vskip3pt} \noalign{\hrule} \noalign{\vskip3pt}
\multicolumn{5}{c}{\bf{NGC\,3242}}\\
\multicolumn{5}{l}{\bf{3s--3p}}\\

4638.86   & V1   &2--4      &0.78[.08] &3.8[0.4]\\
4641.81   & V1   &4--6      &1.00[.07] &1.9[0.1]\\
4649.13   & V1   &6--8      &1.00      &1.0     \\
4650.84   & V1   &2--2      &0.48[.05] &2.3[0.2]\\
4661.63   & V1   &4--4      &0.40[.04] &1.5[0.1]\\
4673.73   & V1   &4--2      &0.14[.01] &3.4[0.3]\\
4676.24   & V1   &6--6      &0.27[.02] &1.2[0.1]\\
\noalign{\vskip2pt}
4317.14   & V2  &  2--4    &0.64[.08]&1.5[0.2]  \\
4319.63   & V2  &  4--6    &0.32[.06]&0.7[0.1]  \\
4325.76   & V2  &  2--2    &0.44[.07]&5.5[0.9]  \\
4345.56   & V2  &  4--2    &1.24[.25]&3.0[0.6]  \\
4349.43   & V2  &  6--6    &1.00     &1.0       \\
4366.89   & V2  &  6--4    &1.28[.21]&2.8[0.5]  \\
\noalign{\vskip2pt}
4414.90   & V5   &4--6    &1.00     &1.0        \\
4416.97   & V5   &2--4    &0.95[.15]&1.7[0.3]   \\
4452.37   & V5   &4--4    &0.51[.17]&4.6[1.5]   \\

\multicolumn{5}{l}{\bf{3p--3d}}\\

4069.89   & V10    &4--6     &1.10[.22] &1.5[0.4]\\
4072.16   & V10    &6--8     &1.29[.21] &1.9[0.3]\\
4075.86   & V10    &8--10    &1.00      &1.0     \\
4078.84   & V10    &4--4     &0.13[.03] &1.2[0.3]\\
4085.11   & V10    &6--6     &0.21[.04] &1.6[0.3]\\
4092.93   & V10    &8--8     &0.14[.04] &1.6[0.4]\\
\noalign{\vskip2pt}
4121.46   & V19   &2--2     &0.90[.59]&2.5[1.6]  \\
4129.32   & V19   &4--2     &1.70[.24]&21.[3.0]  \\
4132.80   & V19   &2--4     &0.69[.15]&1.0[0.2]  \\
4153.30   & V19   &4--6     &1.00     &1.0       \\
4156.53   & V19   &6--4     &1.23[.20]&7.7[1.3]  \\
4169.22   & V19   &6--6     &0.41[.07]&1.2[0.2]  \\

\multicolumn{5}{l}{\bf{3d--4f}}\\

4083.90   & V48b &6--8    &0.19[.03]&0.7[0.1] \\
4087.15   & V48c &4--6    &0.18[.03]&0.7[0.1] \\
4089.29   & V48a &10--12  &1.00     &1.0      \\
4275.55   & V67a &8--10   &0.28[.05]&0.4[0.1] \\
4276.75   & V67b &6--8    &0.25[.04]&1.0[0.1] \\
4277.43   & V67c &2--4    &0.15[.02]&0.7[0.1] \\
4282.96   & V67c &4--6    &0.08[.01]&0.5[0.1] \\
4283.73   & V67c &4--4    &0.09[.01]&0.9[0.1] \\
4285.69   & V78b &6--8    &0.11[.01]&0.6[0.1] \\
4288.82   & V53c &2--4    &0.05[.02]&0.5[0.2] \\
4291.25   & V55  &6--8    &0.15[.04]&0.9[0.2] \\
4292.21   & V78c &6--6    &0.08[.03]&0.9[0.3] \\
4294.78   & V53b &4--6    &0.28[.03]&1.0[0.1] \\
4303.83   & V53a &6--8    &0.52[.09]&1.1[0.2] \\
4307.23   & V53b &2--4    &0.08[.02]&0.7[0.2] \\
4313.44   & V78a &8--10   &0.08[.02]&0.7[0.2] \\
4353.59   & V76c &6--8    &0.09[.02]&0.9[0.2] \\
4357.25   & V63a &6--8    &0.16[.02]&2.7[0.3] \\
4466.42   & V86b &4--6    &0.23[.06]&2.3[0.6] \\
4477.90   & V88  &4--6    &0.13[.03]&1.4[0.3] \\
4489.49   & V86b &2--4    &0.05[.02]&0.7[0.3] \\
4491.23   & V86a &4--6    &0.21[.02]&1.5[0.1] \\
4609.44   & V92a &6--8    &0.20[.23]&0.5[0.4] \\
4669.27   & V89b &4--6    &0.11[.02]&2.8[0.5] \\

\noalign{\vskip3pt} \noalign{\hrule} \noalign{\vskip3pt}

\end{tabular}
\end{minipage}
\end{table}

\setcounter{table}{12}
\begin{table}
\begin{minipage}{75mm}
\centering \caption{{\it --continued}}
\begin{tabular}{lcccc}
\noalign{\vskip3pt} \noalign{\hrule} \noalign{\vskip3pt}
$\lambda_{0}$(\AA) &Mult. &g$_{l}$--g$_{u}$ &$I_{\rm obs}$ &$I_{\rm obs}$/$I_{\rm pred}$\\
\noalign{\vskip3pt} \noalign{\hrule} \noalign{\vskip3pt}
\multicolumn{5}{c}{\bf{NGC\,5882}}\\
\multicolumn{5}{l}{\bf{3s--3p}}\\

4638.86   & V1    &2--4  &0.47[.05]&2.2[0.1]   \\
4641.81   & V1    &4--6  &0.70[.05]&1.3[0.1]   \\
4649.13   & V1    &6--8  &1.00     &1.0        \\
4650.84   & V1    &2--2  &0.33[.02]&1.6[0.1]   \\
4661.63   & V1    &4--4  &0.33[.02]&1.2[0.1]   \\
4673.73   & V1    &4--2  &0.05[.01]&1.2[0.3]   \\
4676.24   & V1    &6--6  &0.25[.02]&1.1[0.1]   \\
\noalign{\vskip2pt}
4317.14   & V2   &2--4  &0.70[.20]&1.8[0.5]    \\
4319.63   & V2   &4--6  &0.37[.13]&0.9[0.3]    \\
4345.56   & V2   &4--2  &1.21[.50]&3.0[1.3]    \\
4349.43   & V2   &6--6  &1.00     &1.0         \\
\noalign{\vskip2pt}
4414.90   & V5   &4--6   &1.00     &1.0         \\
4416.97   & V5   &2--4   &0.91[.22]&1.6[0.4]    \\

\multicolumn{5}{l}{\bf{3p--3d}}\\

4069.89   & V10    &4--6  &0.49[.16]&0.7[0.2]    \\
4072.16   & V10    &6--8  &0.87[.14]&1.3[0.2]    \\
4075.86   & V10    &8--10 &1.00     &1.0         \\
4085.11   & V10    &6--6  &0.11[.04]&0.9[0.3]    \\
4092.93   & V10    &8--8  &0.08[.02]&0.9[0.2]    \\
\noalign{\vskip2pt}
4121.46   & V19   &2--2    &0.30[.23]&0.8[0.6]   \\
4132.80   & V19   &2--4    &0.56[.08]&0.8[0.1]   \\
4153.30   & V19   &4--6    &1.00     &1.0        \\
4156.53   & V19   &6--4    &0.76[.05]&4.8[0.3]   \\
4169.22   & V19   &6--6    &0.29[.06]&0.9[0.2]   \\
\noalign{\vskip2pt}
4110.78   & V20   &4--2    &0.39[.07]&1.4[0.3]   \\
4119.22   & V20   &6--8    &1.00     &1.0        \\
\noalign{\vskip2pt}
4890.86   & V28   &4--2    &0.13[.07]&0.5[0.3]   \\
4906.83   & V28   &4--4    &0.55[.17]&0.9[0.3]   \\
4924.53   & V28   &4--6    &1.00     &1.0        \\
\noalign{\vskip2pt}
\multicolumn{5}{l}{\bf{3d--4f}}\\

4083.90  &V48b &6--8    &0.15[.05]&0.5[0.2]       \\
4087.15  &V48c &4--6    &0.25[.04]&0.9[0.1]       \\
4089.29  &V48a &10--12  &1.00     &1.0            \\
4275.55  &V67a &8--10   &0.40[.07]&0.6[0.1]       \\
4276.75  &V67b &6--8    &0.16[.07]&0.7[0.3]       \\
4277.43  &V67c &2--4    &0.18[.04]&0.8[0.2]       \\
4291.25  &V55  &6--8    &0.15[.03]&0.9[0.2]       \\
4292.21  &V18c &6--6    &0.05[.03]&0.6[0.4]       \\
4294.78  &V53b &4--6    &0.11[.02]&0.4[0.1]       \\
4303.83  &V53a &6--8    &0.39[.04]&0.8[0.1]       \\
4353.59  &V76c &6--8    &0.10[.01]&1.0[0.1]       \\
4366.53  &V75a &6--8    &0.42[.05]&0.6[0.1]       \\
4466.42  &V86b &4--6    &0.21[.05]&2.3[0.6]       \\
4491.23  &V86a &4--6    &0.14[.03]&1.0[0.2]       \\
4609.44  &V92a &6--8    &0.42[.08]&1.0[0.2]       \\

\noalign{\vskip3pt} \noalign{\hrule} \noalign{\vskip3pt}

\end{tabular}
\end{minipage}
\end{table}

\setcounter{table}{12}
\begin{table}
\begin{minipage}{75mm}
\centering \caption{{\it --continued}}
\begin{tabular}{lcccc}
\noalign{\vskip3pt} \noalign{\hrule} \noalign{\vskip3pt}
$\lambda_{0}$(\AA) &Mult. &g$_{l}$--g$_{u}$ &$I_{\rm obs}$ &$I_{\rm obs}$/$I_{\rm pred}$\\
\noalign{\vskip3pt} \noalign{\hrule} \noalign{\vskip3pt}

\multicolumn{5}{c}{\bf{NGC\,5315}}\\
\multicolumn{5}{l}{\bf{3s--3p}}\\

4638.86 & V1    &2--4  &0.49[.04]&2.3[0.2] \\
4641.81 & V1    &4--6  &0.66[.05]&1.3[0.1] \\
4649.13 & V1    &6--8  &1.00     &1.0      \\
4650.84 & V1    &2--2  &0.25[.03]&1.2[0.1] \\
4661.63 & V1    &4--4  &0.25[.02]&0.9[0.1] \\
4673.73 & V1    &4--2  &0.04[.01]&1.0[0.3] \\
4676.24 & V1    &6--6  &0.22[.01]&1.0[0.1]   \\
\noalign{\vskip2pt}
4317.14 & V2  &  2--4  &1.00     &1.0                  \\
4319.63 & V2  &  4--6  &0.71[.10] &0.7[0.1]            \\
4325.76 & V2  &  2--2  &0.28[.04]&1.5[0.2]   \\
4345.56 & V2  &  4--2  &0.92[.20] &1.0[0.3]    \\
4349.43 & V2  &  6--6  &1.07[.20] &0.5[0.1]    \\
4366.89 & V2  &  6--4  &0.85[.40] &0.8[0.3]     \\
\noalign{\vskip2pt}
4414.90 & V5   &4--6    &1.00     &1.0          \\
4416.97 & V5   &2--4    &0.86[.50]&1.5[0.9]     \\
4452.37 & V5   &4--4    &0.20[.11]&1.8[1.0]     \\

\multicolumn{5}{l}{\bf{3p--3d}}\\

4069.62 & V10    &4--6  &3.02[2.2]&4.1[3.0]       \\
4072.16 & V10    &6--8  &1.33[1.0]&1.9[1.4]       \\
4075.86 & V10    &8--10 &1.00     &1.0            \\
4085.11 & V10    &6--6  &0.31[.21]&2.4[1.6]      \\

\multicolumn{5}{l}{\bf{3d--4f}}\\

4083.90  &V48b&6--8  &0.27[.05]&0.9[0.2]    \\
4087.15  &V48c&4--6  &0.35[.04]&1.3[0.2]    \\
4089.29  &V48a&10--12&1.00     &1.0        \\
4275.55  &V67a&8--10 &1.17[.20]&1.0[0.2]  \\
4303.82  &V53a&6--8  &0.37[.06]&0.8[0.1]   \\
4609.44  &V92a&6--8  &0.53[.04]&1.2[0.1]   \\
\noalign{\vskip2pt}
\multicolumn{5}{c}{\bf{NGC\,3918}}\\
\multicolumn{5}{l}{\bf{3s--3p}}\\
4638.86 & V1    &2--4  &1.04[.23]&5.0[1.1] \\
4641.81 & V1    &4--6  &1.03[.40]&1.9[0.7]  \\
4649.13 & V1    &6--8  &1.00     &1.0       \\
4650.84 & V1    &2--2  &0.26[.07]&1.3[0.4]  \\
4661.63 & V1    &4--4  &0.35[.04]&1.3[0.2]  \\
4673.73 & V1    &4--2  &0.15[.03]&3.5[0.8]   \\
4676.24 & V1    &6--6  &0.31[.02]&1.4[0.1]   \\
\noalign{\vskip2pt}
4317.14 & V2  &  2--4  &1.00     &1.0  \\
4319.63 & V2  &  4--6  &0.41     &0.4  \\
4325.76 & V2  &  2--2  &0.34     &1.9   \\
4349.43 & V2  &  6--6  &1.53     &0.7   \\
\noalign{\vskip2pt}
4414.90 & V5   &4--6     &1.00    &1.0 \\
4416.97 & V5   &2--4     &0.56    &1.0 \\
4452.37 & V5   &4--4     &1.26    &11. \\

\multicolumn{5}{l}{\bf{3p--3d}}\\

4069.62 & V10    &4--6   &8.11[4.1]&11[5.6]  \\
4072.16 & V10    &6--8   &3.62[1.8]&5.3[2.6] \\
4075.86 & V10    &8--10  &1.00     &1.0      \\
4085.11 & V10    &6--6   &0.30[.22]&2.3[1.7] \\
\noalign{\vskip2pt}
4121.46 & V19   &2--2    &3.54   &9.8   \\
4132.80 & V19   &2--4    &0.69   &1.0   \\
4153.30 & V19   &4--6    &1.00   &1.0   \\
4156.53 & V19   &6--4    &1.42   &8.9   \\
4169.22 & V19   &6--6    &0.18   &0.5     \\
\end{tabular}
\end{minipage}
\end{table}

\setcounter{table}{12}
\begin{table}
\begin{minipage}{75mm}
\centering \caption{{\it --continued}}
\begin{tabular}{lcccc}
\noalign{\vskip3pt} \noalign{\hrule} \noalign{\vskip3pt}
$\lambda_{0}$(\AA) &Mult. &g$_{l}$--g$_{u}$ &$I_{\rm obs}$ &$I_{\rm obs}$/$I_{\rm pred}$\\
\noalign{\vskip3pt} \noalign{\hrule} \noalign{\vskip3pt}
\multicolumn{5}{c}{\bf{NGC\,3918}}\\

\multicolumn{5}{l}{\bf{3d--4f}}\\

4083.90&V48b&6--8   &0.50[.18]&1.7[0.6] \\
4087.15&V48c&4--6   &0.43[.17]&1.6[0.6] \\
4089.29&V48a&10--12 &1.00     &1.0     \\
4275.55&V67a&8--10  &1.15     &0.9     \\
4291.25&V55 &6--8   &0.44     &1.7     \\
4294.78&V53b&4--6   &0.45     &1.6     \\
4303.83&V53a&6--8   &0.50     &1.1     \\
4466.42&V86b&4--6   &0.43     &4.3     \\
4610.20&V92c&4--6   &0.67     &1.2     \\
4669.27&V89b&4--6   &0.22     &5.5     \\
\noalign{\vskip2pt}
\multicolumn{5}{c}{\bf{NGC 2022}}\\
\multicolumn{5}{l}{\bf{3s--3p}}\\

4638.86 & V1    &2--4   &0.32[.11]&1.5[0.5] \\
4641.81 & V1    &4--6   &1.10[.16]&2.1[0.3] \\
4649.13 & V1    &6--8   &1.00     &1.0      \\
4650.84 & V1    &2--2   &0.15[.21]&0.7[1.0]  \\
4661.63 & V1    &4--4   &0.33[.06]&1.2[0.2]  \\
4676.24 & V1    &6--6   &0.23[.08]&1.0[0.4] \\

\multicolumn{5}{l}{\bf{3p--3d}}\\

4072.16 & V10    &6--8     &1.07[.17]&1.6[0.3]     \\
4075.86 & V10    &8--10    &1.00     &1.0         \\

\multicolumn{5}{l}{\bf{3d--4f}}\\

4089.29&V48a&10--12    &1.00    &1.0            \\
4276.75&V67b&6--8      &0.92    &1.0            \\
4477.90&V88 &4--6      &0.54    &6.0            \\
4609.44&V92a&6--8      &0.29    &0.7            \\
\noalign{\vskip3pt}
\multicolumn{5}{c}{\bf{IC\,4406}}\\
\multicolumn{5}{l}{\bf{3s--3p}}\\
4638.86 & V1    &2--4   &0.64[.11]&3.1[0.5]    \\
4641.81 & V1    &4--6   &0.93[.12]&1.8[0.2]    \\
4649.13 & V1    &6--8   &1.00     &1.0         \\
4650.84 & V1    &2--2   &0.46[.09]&2.2[0.4]   \\
4661.63 & V1    &4--4   &0.35[.09]&1.3[0.3]   \\
4676.24 & V1    &6--6   &0.17[.10]&0.8[0.5]   \\

\multicolumn{5}{l}{\bf{3p--3d}}\\
4069.62 &V10   &2--4    &1.62[.60]&2.2[0.8]       \\
4072.16 &V10   &6--8    &1.49[.51]&2.2[0.8]       \\
4075.86 &V10   &8--10   &1.00     &1.0            \\
\noalign{\vskip3pt}
\multicolumn{5}{c}{\bf{NGC\,6818}}\\

\multicolumn{5}{l}{\bf{3s--3p}}\\
4638.86   &V1   &2--4  &1.19[.30]&5.7[1.4]  \\
4641.81   &V1   &4--6  &2.73[.41]&5.2[0.8]  \\
4649.13   &V1   &6--8  &1.00     &1.0       \\
4650.84   &V1   &2--2  &0.29[.11]&1.4[0.6]  \\
4661.63   &V1   &4--4  &0.47[.07]&1.8[0.3]  \\

\multicolumn{5}{l}{\bf{3p--3d}}\\

4069.78   &V10   &2--4  &0.30[.26]&0.4[0.4]    \\
4072.16   &V10   &6--8  &0.40[.14]&0.6[0.2]    \\
4075.86   &V10   &8--10 &1.00     &1.0         \\
4085.11   &V10   &6--6  &0.16[.06]&1.2[0.5]    \\

\noalign{\vskip3pt}
\multicolumn{5}{c}{\bf{NGC\,2440}}\\
\multicolumn{5}{l}{\bf{3s--3p}}\\
4638.86 & V1    &2--4    &2.85[.56]&14.[2.6]  \\
4641.81 & V1    &4--6    &6.54[.37]&9.4[0.7]  \\
4649.13 & V1    &6--8    &1.00     &1.0       \\
4650.84 & V1    &2--2    &0.20[.26]&1.0[1.2]  \\
4661.63 & V1    &4--4    &0.35[.06]&1.3[0.2]  \\
4676.24 & V1    &6--6    &0.71[.08]&3.2[0.4]  \\
\noalign{\vskip2pt}
4414.90 &V5   &4--6     &1.00     &1.0       \\
4416.97 &V5   &2--4     &0.80[.22]&1.4[0.4]  \\
4452.37 &V5   &4--4     &1.70[.37]&15.[3.3]  \\

\noalign{\vskip3pt} \noalign{\hrule} \noalign{\vskip3pt}

\end{tabular}
\end{minipage}
\end{table}

\setcounter{table}{12}
\begin{table}
\begin{minipage}{75mm}
\centering \caption{{\it --continued}}
\begin{tabular}{lcccc}
\noalign{\vskip3pt} \noalign{\hrule} \noalign{\vskip3pt}
$\lambda_{0}$(\AA) &Mult. &g$_{l}$--g$_{u}$ &$I_{\rm obs}$ &$I_{\rm obs}$/$I_{\rm pred}$\\
\noalign{\vskip3pt} \noalign{\hrule} \noalign{\vskip3pt}

\multicolumn{5}{c}{\bf{IC\,4191}}\\
\multicolumn{5}{l}{\bf{3s--3p}}\\

\multicolumn{5}{c}{entire nebula}\\

4638.86 & V1    &2--4    &0.33[.06]&1.6[0.3]  \\
4641.81 & V1    &4--6    &0.56[.06]&1.1[0.1]  \\
4649.13 & V1    &6--8    &1.00     &1.0       \\
4650.84 & V1    &2--2    &0.27[.05]&1.3[0.3]  \\
4661.63 & V1    &4--4    &0.27[.02]&1.0[0.1]  \\
4676.24 & V1    &6--6    &0.22[.02]&1.0[0.1]  \\

\multicolumn{5}{c}{fixed slit}\\
4638.86 & V1    &2--4    &0.34[.04]&1.6[0.2]   \\
4641.81 & V1    &4--6    &0.28[.02]&0.5[0.04] \\
4649.13 & V1    &6--8    &1.00     &1.0        \\
4650.84 & V1    &2--2    &0.24[.03]&1.2[0.2]   \\
4661.63 & V1    &4--4    &0.20[.03]&0.7[0.1]   \\
4676.24 & V1    &6--6    &0.15[.02]&0.7[0.1]   \\
\noalign{\vskip2pt}
4317.14 & V2  &2--4   &1.00     &1.0         \\
4319.63 & V2  &4--6   &0.82[.25]&0.8[0.3]    \\
4345.56 & V2  &4--2   &1.79[.64]&1.9[0.7]    \\
4349.43 & V2  &6--6   &3.44[1.0]&1.5[0.4]    \\
4366.89 & V2  &6--4   &1.58[.36]&1.6[0.4]    \\
\noalign{\vskip2pt}
4414.90 &V5   &4--6   &1.00     &1.0         \\
4416.97 &V5   &2--4   &0.81[.08]&1.4[0.1]    \\

\multicolumn{5}{c}{entire nebula}\\
4414.90 &V5   &4--6   &1.00     &1.0        \\
4416.97 &V5   &2--4   &0.70[.19]&1.3[0.4]   \\
4452.37 &V5   &4--4   &0.53[.14]&4.8[1.3]   \\

\multicolumn{5}{l}{\bf{3p--3d}}\\

\multicolumn{5}{c}{entire nebula}\\
4069.62 & V10    &4--6   &2.45[.66]&3.3[0.9]  \\
4072.16 & V10    &6--8   &1.27[.36]&1.8[0.5]  \\
4075.86 & V10    &8--10  &1.00     &1.0       \\
4085.11 & V10    &6--6   &0.17[.09]&1.3[0.7]  \\

\multicolumn{5}{c}{fixed slit}\\
4069.62 & V10    &4--6   &1.74[.38]&2.4[0.5]  \\
4072.16 & V10    &6--8   &1.03[.25]&1.5[0.4]  \\
4075.86 & V10    &8--10  &1.00     &1.0       \\
4085.11 & V10    &6--6   &0.08[.04]&0.6[0.3]  \\

\multicolumn{5}{l}{\bf{3d--4f}}\\
\multicolumn{5}{c}{entire nebula}\\
4083.90   &V48b &6--8  &0.28[.13]&1.0[0.5]    \\
4087.15   &V48c &4--6  &0.39[.13]&1.3[0.4]    \\
4089.29   &V48a &10--12&1.00     &1.0         \\
4275.55   &V67a &8--10 &1.42[.28]&1.1[0.2]    \\
4282.96   &V67c &4--6  &0.40[.13]&0.9[0.3]    \\
4303.82   &V53a &6--8  &0.63[.16]&1.3[0.3]    \\
4609.44   &V92a &6--8  &0.58[.10]&1.3[0.2]    \\

\multicolumn{5}{c}{fixed slit}\\
4083.90   &V48b&6--8  &0.34[.07]&1.2[0.3]    \\
4087.15   &V48c&4--6  &0.33[.07]&1.1[0.2]    \\
4089.29   &V48a&10--12&1.00     &1.0         \\
4275.55   &V67a&8--10 &1.18[.18]&1.0[0.2]    \\
4303.82   &V53a&6--8  &0.33[.07]&0.7[0.2]    \\
4466.42   &V86b&4--6  &0.19[.03]&1.9[0.3]    \\
4609.44   &V92a&6--8  &0.29[.08]&0.7[0.1]    \\
\noalign{\vskip3pt}  \noalign{\vskip3pt}  \noalign{\vskip3pt}

\end{tabular}
\end{minipage}
\end{table}

\setcounter{table}{12}
\begin{table}
\begin{minipage}{75mm}
\centering \caption{{\it --continued}}
\begin{tabular}{lcccc}
\noalign{\vskip3pt} \noalign{\hrule} \noalign{\vskip3pt}
$\lambda_{0}$(\AA) &Mult. &g$_{l}$--g$_{u}$ &$I_{\rm obs}$ &$I_{\rm obs}$/$I_{\rm pred}$\\
\noalign{\vskip3pt} \noalign{\hrule} \noalign{\vskip3pt}

\multicolumn{5}{c}{\bf{NGC\,3132}}\\
\multicolumn{5}{l}{\bf{3s--3p}}\\
4638.86 & V1    &2--4    &0.53[.47]&2.6[2.2] \\
4641.81 & V1    &4--6    &0.88[.58]&1.7[1.1] \\
4649.13 & V1    &6--8    &1.00     &1.0      \\
4650.84 & V1    &2--2    &0.66[.38]&3.2[1.8] \\
4661.63 & V1    &4--4    &0.60[.34]&2.2[1.3] \\
4676.24 & V1    &6--6    &0.39[.23]&1.7[1.1]  \\
\multicolumn{5}{l}{\bf{3p--3d}}\\
4069.62  &V10   &2--4   &1.14[.78]     &1.5[1.0] \\
4072.16  &V10   &6--8   &0.41[.34]     &0.6[0.5] \\
4075.86  &V10   &8--10  &1.00          &1.0      \\
4085.11  &V10   &6--6  &0.12[.03]     &0.9[0.2] \\
\noalign{\vskip3pt}
\multicolumn{5}{c}{\bf{NGC\,6302}}\\
\multicolumn{5}{l}{\bf{3s--3p}}\\
4649.13 & V1    &6--8    &1.00     &1.0      \\
4650.84 & V1    &2--2    &0.97     &4.7      \\
4661.63 & V1    &4--4    &1.14     &5.1      \\
\noalign{\vskip3pt}
\multicolumn{5}{c}{\bf{My\,Cn\,18}}\\
\multicolumn{5}{l}{\bf{3s--3p}}\\
4649.13 & V1    &6--8    &1.00     &1.0      \\
4650.84 & V1    &2--2    &0.37     &1.8      \\
\noalign{\vskip3pt}
\multicolumn{5}{c}{\bf{LMC~N66}}\\
\multicolumn{5}{l}{\bf{3s--3p}}\\
4649.13 & V1    &6--8    &1.00     &1.0      \\
4650.84 & V1    &2--2    &1.00     &4.8      \\
\noalign{\vskip3pt}
\multicolumn{5}{c}{\bf{LMC N141}}\\
\multicolumn{5}{l}{\bf{3s--3p}}\\
4638.86 & V1    &2--4     &0.20[.10]&1.0[0.5]   \\
4641.81 & V1    &4--6     &0.95[.28]&1.8[0.5]   \\
4649.13 & V1    &6--8     &1.00     &1.0        \\
4650.84 & V1    &2--2     &0.66[.41]&3.1[1.9]   \\
4661.63 & V1    &4--4     &0.41[.14]&1.6[0.6]   \\
4676.24 & V1    &6--6     &0.37[.09]&1.7[0.4]    \\
\multicolumn{5}{l}{\bf{3p--3d}}\\
4069.62  &V10   &2--4    &0.74[.36]&1.0[0.5]  \\
4072.16  &V10   &6--8    &0.49[.18]&0.7[0.3]  \\
4075.86  &V10   &8--10   &1.00     &1.0       \\
\noalign{\vskip3pt}
\multicolumn{5}{c}{\bf{SMC N87}}\\
\multicolumn{5}{l}{\bf{3s--3p}}\\
4638.86 & V1    &2--4    &1.07[.17]&5.1[0.8] \\
4641.81 & V1    &4--6    &1.06[.26]&2.0[0.4] \\
4649.13 & V1    &6--8    &1.00     &1.0             \\
4650.84 & V1    &2--2    &1.00[.15]&4.8[0.7] \\
4661.63 & V1    &4--4    &0.31[.04]&1.2[0.2] \\
4676.24 & V1    &6--6    &0.32[.11]&1.4[0.5] \\

\noalign{\vskip3pt}  \noalign{\vskip3pt}  \noalign{\vskip3pt}

\end{tabular}
\end{minipage}
\end{table}

\end{appendix}

\end{document}